\documentclass[aps,prl,10pt,twocolumn,superscriptaddress,longbibliography]{revtex4}

\usepackage{graphicx}
\usepackage{amssymb, amsmath}
\usepackage{color}
\usepackage{ulem}
\usepackage{bm}
\usepackage{graphicx}
\usepackage{subfigure}
\usepackage{dcolumn}
\usepackage{mathtools}
\usepackage{pifont}
\usepackage[utf8]{inputenc}
\usepackage[T1]{fontenc}
\usepackage{bbm}
\def\I{\uppercase\expandafter{\romannumeral 1}}
\def\II{\uppercase\expandafter{\romannumeral 2}}
\def\III{{\uppercase\expandafter{\romannumeral 3}}}
\def\IV{{\uppercase\expandafter{\romannumeral 4}}}
\def\V{{\uppercase\expandafter{\romannumeral 5}}}
\def\VI{{\uppercase\expandafter{\romannumeral 6}}}
\def\VII{{\uppercase\expandafter{\romannumeral 7}}}
\def\VIII{{\uppercase\expandafter{\romannumeral 8}}}
\def\IX{{\uppercase\expandafter{\romannumeral 9}}}
\def\X{{\uppercase\expandafter{\romannumeral 10}}}
\def\XI{{\uppercase\expandafter{\romannumeral 11}}}
\def\i{\lowercase\expandafter{\romannumeral 1}}
\def\ii{\lowercase\expandafter{\romannumeral 2}}
\def\iii{{\lowercase\expandafter{\romannumeral 3}}}
\def\iv{{\lowercase\expandafter{\romannumeral 4}}}
\def\v{{\lowercase\expandafter{\romannumeral 5}}}
\def\vi{{\lowercase\expandafter{\romannumeral 6}}}
\def\vii{{\lowercase\expandafter{\romannumeral 7}}}

\def\angstrom{\mbox{\normalfont\AA}}

\def\nn{\nonumber\\}

\def\angstrom{\mbox{\normalfont\AA}}
\def\k{\mathbf{k}}
\def\G{\mathbf{G}}

\def\kt{\widetilde{\mathbf{k}}}

\def\Q{\mathbf{Q}}
\def\qt{\widetilde{\mathbf{q}}}
\def\nn{\nonumber\\}

\begin{document}

\title{Spin polarized nematic order, quantum valley Hall states, and field tunable topological transitions in  twisted multilayer graphene systems}

\author{Shihao Zhang}
\affiliation{School of Physical Science and Technology, ShanghaiTech University, Shanghai 200031, China}

\author{Xi Dai}
\affiliation{Department of Physics, Hong Kong University of Science and Technology, Hong Kong, China}

\author{Jianpeng Liu}
\email[]{liujp@shanghaitech.edu.cn}
\affiliation{School of Physical Science and Technology, ShanghaiTech University, Shanghai 200031, China}
\affiliation{ShanghaiTech laboratory for topological physics, ShanghaiTech University, Shanghai 200031, China}

\begin{abstract}
We theoretically study the correlated insulator states, quantum anomalous Hall (QAH) states, and field-induced topological transitions between different correlated states  in  twisted multilayer graphene systems. Taking twisted bilayer-monolayer graphene and twisted double-bilayer graphene as  examples, we show that both systems stay in spin polarized, $C_{3z}$-broken insulator states with zero Chern number at 1/2 filling of the flat bands under finite displacement fields. 
In some cases these spin polarized, nematic insulator states are in the quantum valley Hall phase by virtue of the nontrivial band topology of the systems. The spin polarized insulator state is quasi-degenerate with the valley polarized state if only the dominant intra-valley Coulomb interaction is included.  Such quasi-degeneracy can be split by atomic on-site interactions such that the spin polarized, nematic state become the unique ground state. Such a scenario applies to various twisted multilayer graphene systems at 1/2 filling, thus can be considered as a universal mechanism.  Moreover, under vertical magnetic fields, the  orbital Zeeman splittings and the field-induced change of charge density in twisted multilayer graphene systems would compete with the atomic Hubbard interactions, which can drive transitions from spin polarized zero-Chern-number states to valley-polarized QAH states with small onset magnetic fields.  
\end{abstract}

\pacs{}

\maketitle

\paragraph{Introduction. \textemdash}

The moir\'e graphene system has been a thriving research area  since the discoveries of the correlated insulator states \cite{cao-nature18-mott,efetov-nature19,tbg-stm-pasupathy19,tbg-stm-andrei19,tbg-stm-yazdani19, tbg-stm-caltech19, young-tbg-science19}, unconventional superconductivity \cite{cao-nature18-supercond,dean-tbg-science19,marc-tbg-19, efetov-nature19}, and quantum anomalous Hall states \cite{sharpe-science-19, young-tbg-science19, andrei-tbg-chern-arxiv20, efetov-tbg-chern-arxiv20, yazdani-tbg-chern-arxiv20} in twisted bilayer graphene (TBG) around the magic angle \cite{macdonald-pnas11}. 
In magic-angle TBG, the valley, spin,  and sublattice degeneracy of the eight-fold degenerate flat bands with nontrivial topological character \cite{song-tbg-prl19, yang-tbg-prx19,po-tbg-prb19, origin-magic-angle-prl19, jpliu-prb19} can be lifted by  the strong Coulomb interactions, leading to topologically distinct correlated insulator states at different partial integer fillings \cite{xu-lee-prb18, liu-prl18, kang-tbg-prl19,Uchoa-ferroMott-prl,xie-tbg-2018, huang-tbg-sb19, zaletel-tbg-2019, zaletel-tbg-hf-prx20,jpliu-tbghf-prb21,zhang-tbghf-arxiv20,hejazi-tbg-hf,kang-tbg-dmrg-prb20, kang-tbg-topomott,yang-tbg-arxiv20, meng-tbg-arxiv20,Bernevig-tbg3-arxiv20,Lian-tbg4-arxiv20}.   Moreover, it has been theoretically proposed \cite{senthil-hbn-tbg-prr19, jpliu-prx19, rademaker-prr20, ma-tbmg-sb20, jung-tbmg-prb20} and experimentally realized \cite{chen-trilayer-hbn-qah, young-monobi-nature20,Yankowitz-monobi-np2020,Yankowitz-doublebi-np2020} that topologically nontrivial flat bands  exist in  twisted multilayer graphene systems and moir\'e graphene heterostructures as well.  
Recent  transport experiments reveal orbital Chern insulator states \cite{young-monobi-nature20} and correlated insulator states with zero Chern number \cite{young-monobi-nature20,Yankowitz-monobi-np2020, shi-tbmg-np21} in  twisted bilayer-monolayer graphene (TBMG) system at different partial integer fillings. Recent experiments also report spin polarized insulator states \cite{kim-tdbg-nature20, cao-tdbg-nature20, zhang-tdbg-np20, Tutuc-tdbg-correlatd-prl}  and nematic phase \cite{Pasupathy-nematic-tdbg-arxiv20} in twisted double bilayer graphene (TDBG) system.

In this work,  we theoretically study the correlated insulator states, topological properties, and the topological transitions between different correlated states induced by vertical magnetic fields in various twisted multilayer graphene systems, which are exemplified by TBMG and TDBG.
We first focus at the zero-Chern number insulator states  observed at  $1/2$  filling \cite{comment_filling} of TBMG under finite displacement fields\cite{young-monobi-nature20,Yankowitz-monobi-np2020,shi-tbmg-np21} and those observed in TDBG\cite{kim-tdbg-nature20, cao-tdbg-nature20, zhang-tdbg-np20, Tutuc-tdbg-correlatd-prl}.
Using unrestricted Hartree-Fock calculations in the band basis, we show that the ground state of TBMG  at $1/2$ filling under finite displacement field is a spin polarized, nematic insulator with spontaneously broken $C_{3z}$ symmetry. 
In  TDBG system, the calculated ground states at 1/2 filling also exhibits nematicity, consistent with scanning tunneling microscopy (STM) measurements \cite{Pasupathy-nematic-tdbg-arxiv20}.
These zero-Chern-number states at 1/2 filling are driven by the exchange part of inter-site Coulomb interactions, and are  quasi-degenerate with valley polarized states with nonzero Chern numbers.  The inclusion of  atomic on-site Hubbard interactions favors spin polarization, which makes the spin-polarized, nematic state as the unique ground state in both systems. Our calculations indicate that such a  mechanism also applies other twisted multilayer graphene systems at 1/2 filling \cite{supp_info}, thus can be considered as a universal mechanism.

The flat bands in moir\'e graphene superlattices are generally topologically nontrivial \cite{song-tbg-prl19,origin-magic-angle-prl19,jpliu-prb19,zhang-tbg-prb19,jpliu-prx19} with giant and valley-contrasting orbital magnetizations and nonzero valley Chern numbers. 
As a result, under vertical magnetic fields, the  orbital magnetic Zeeman splitting and the field-induced change of charge density would compete with the  atomic on-site interactions: the latter favors a spin polarized insulator state with $C_{3z}$ breaking, while the former favors a valley-polarized QAH state. Thus at 1/2 filling we predict topological transitions with small onset vertical magnetic fields in TMG systems.  

\begin{figure}[!htbp]
\includegraphics[width=0.5\textwidth]{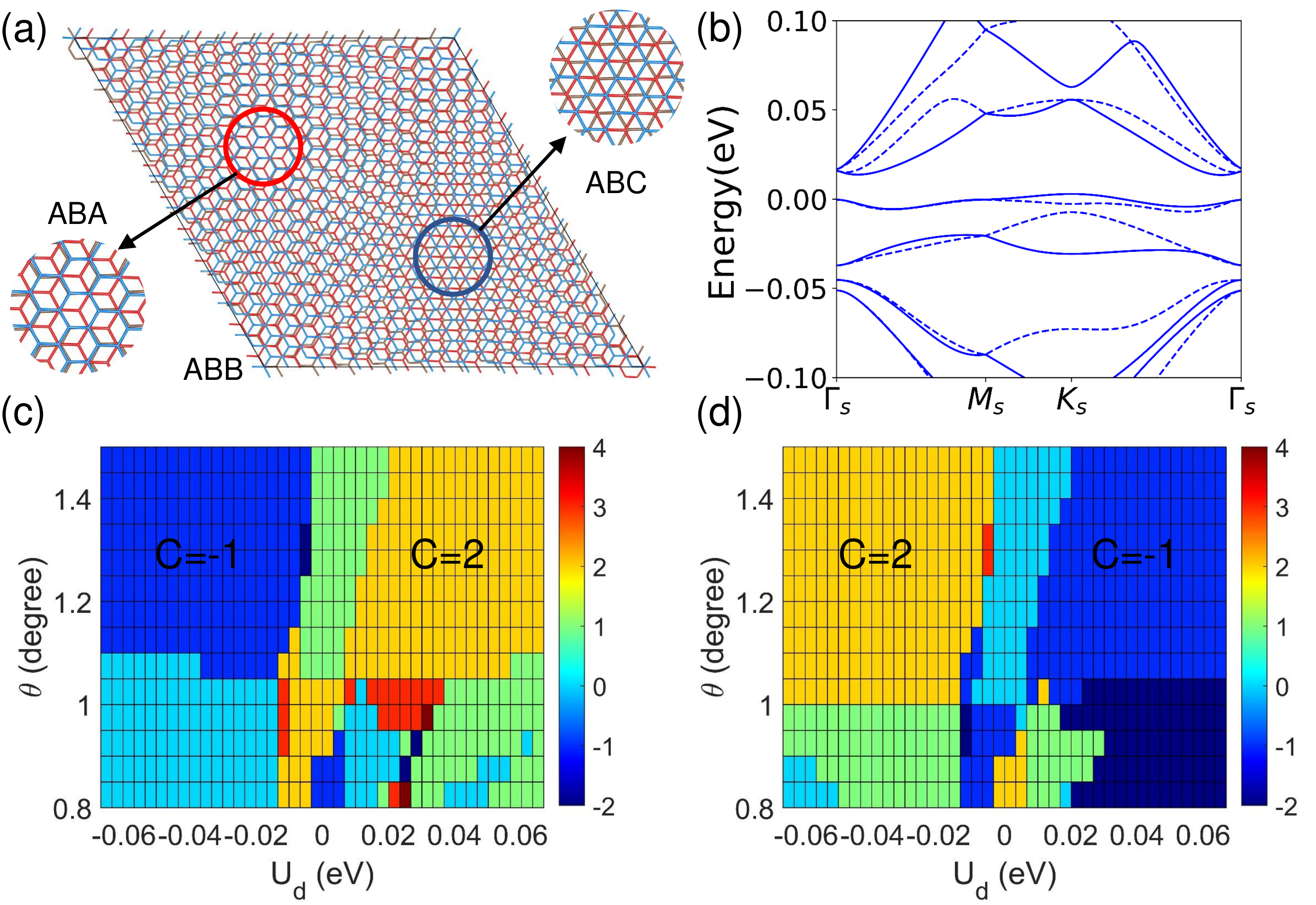}
\caption{~\label{fig1} (a) The illustrations of twisted bilayer-monolayer graphene. The blue and red hexagonal lattice represent bilayer graphene, and brown lattice is monolayer graphene. (b) The non-interacting energy bands of 1.25$^\circ$-twisted bilayer-monolayer graphene under 0.4\,V/nm electric displacement field. The solid (dashed) lines are the energy bands of K (K$^{\prime}$) valley. (c) and (d) show the Chern numbers of the highest valence band and lowest conduction band under different electric displacement field  $U_d$ (-0.067$-$0.067\,eV) and varying twist angle $\theta$.}
\end{figure}

\paragraph{Single-particle picture. \textemdash}
The non-interacting physics of twisted  $(M+N)$-layer graphene system is described by an extension \cite{jpliu-prx19,ashvin-double-bilayer-nc19,koshino-tdbg-prb19,ma-tbmg-sb20} of the Bistritzer-MacDonald continuum model \cite{macdonald-pnas11}
\begin{equation}
H_{\mu}^{0}(M+N)=\begin{pmatrix}
H^{0}_{\mu}(M) & \mathbb{U}_{\mu} \\
\mathbb{U}^{\dagger}_{\mu} & H^{0}_{\mu}(N)
\end{pmatrix}\;,
\label{eq:HMN}
\end{equation}
where $H^{0}_{\mu}(M)$ and $H^{0}_{\mu}(N)$ are the low-energy  Hamiltonians of $\mu$ valley ($\mu=\pm$ denoting the $K'/K$ valley) for the $M$-layer and $N$-layer graphene systems \cite{supp_info}. $\mathbb{U}_{\mu}$ denotes the interlayer coupling between layers near the twisted interface for the $\mu$ valley \cite{supp_info}. 
Vertical displacement field $D$ can be included into the Hamiltonian by adding an on-site energy to the $l$th layer ($l=1, ..., M+N$): $U_l=(l-1)U_d/(M+N-1)$, where $U_d\!=\!-eDd/\epsilon _{\rm{BN}}$, and $d\!=\!(M+N-1)\times3.35\,\angstrom$ denotes the total thickness of the twisted $(M+N)$-layer graphene system. $\epsilon _{\rm{BN}}$ is the dielectric constant of hexagonal boron nitride (hBN) substrate, which is set to 5, e.g., 0.4\,V/nm displacement field corresponds to $U_d\!=\!\--0.0536\,$eV in this work. 

The non-interacting band structures of TBMG  at $1.25^{\circ}$ with $D\sim 0.4\,$V/nm  are presented in Fig.~\ref{fig1}(b), where the solid and dashed lines represent the energy bands from $K$ and $K'$ valleys respectively.  Clearly there are two flat bands that are energetically separated from other dispersive remote bands.
The flat bands possess nonzero valley Chern numbers, the sum of which equals to $\pm 1$ \cite{jpliu-prx19}.  The narrow bandwidth of the conduction flat band implies that the system would be strongly susceptible to Coulomb interactions when it is partially filled.  The  valley Chern numbers of the valence and conduction flat bands at different  displacement fields and twist angles are presented in the Fig.~\ref{fig1}(c) and (d) respectively. 
When the twist angle is $\theta\!\gtrapprox\!1.05^\circ$ and $\vert U_d\vert\!\gtrapprox\!0.015$\,eV, the valley Chern numbers of the highest valence band and lowest conduction band are -1 and 2  for negative $U_d$, and the Chern numbers of the two bands are interchanged for positive $U_d$. This is crucial in determining the nature of the correlated insulators and QAH states at partial integer fillings of the flat bands. The non-interacting band structures and the topological properties of other twisted multilayers including TDBG and hBN-aligned TBG are presented in Supplementary Information \cite{supp_info}.

\paragraph{Coulomb interactions and Hartree-Fock approximation. \textemdash}
We consider the inter-site Coulomb interactions 
\begin{equation}
H_C=\frac{1}{2N_s}\sum _{\alpha \alpha ^{\prime}}\sum _{\mathbf{k} \mathbf{k^{\prime}}\mathbf{q}}\sum _{\sigma \sigma ^{\prime}}\,V(\mathbf{q})\,\hat{c}^{\dagger}_{\mathbf{k+q},\alpha \sigma}\,\hat{c}^{\dagger}_{\mathbf{k^{\prime}-q}, \alpha ^{\prime}\sigma ^{\prime}}
\,\hat{c}_{\mathbf{k^{\prime}},\alpha ^{\prime}\sigma ^{\prime}}\,\hat{c}_{\mathbf{k},\alpha \sigma}
\label{eq:coulomb-project}
\end{equation}
where $\mathbf{k}$ and $\mathbf{q}$ are atomic wavevectors, $\alpha$ represent the the layer and sublattice indices,  and $\sigma$ refers to the spin index. Here $V(\mathbf{q})$ denotes the screened Coulomb interaction $V(\mathbf{q})\!=\!e^2/(\,2\Omega _M\epsilon \epsilon _0\sqrt{q^2+\kappa ^2}\,)$,
where $\Omega _M$ is the area of moir\'e supercell, $\kappa$ is the inverse screening length and $\epsilon$ denotes background dielectric constant. $\epsilon$ and $\kappa$ will be treated as two free parameters in this work. The typical inter-site Coulomb interaction energy on the moir\'e length scale is characterized by $U_M\!=\!e^2/(4\pi\epsilon\epsilon_0 L_s)\!\sim\!25\,$meV for $\epsilon\!=\!5$ and twist angle $1.25^{\circ}$, where $L_s$ is the moir\'e lattice constant. 
At small twist angles one can further decompose the inter-site interaction into the intravalley one and the intervalley one, with the former being two orders of magnitudes larger than the latter \cite{supp_info}, thus we only consider the intravalley part of the inter-site Coulomb interaction.

In addition to the inter-site Coulomb interactions, we also include the on-site Hubbard interaction 
\begin{equation}
H_{\rm{on-site}}= \frac{U_0a^2}{N_M L_s^2}\sum _{\mathbf{k}\mathbf{k}'\mathbf{q}}\sum _{\alpha}  \hat{c}^{\dagger}_{\mathbf{k}+\mathbf{q},\alpha\uparrow}\hat{c}^{\dagger}_{\mathbf{k}'-\mathbf{q},\alpha\downarrow}
\times\hat{c}_{\mathbf{k}',\alpha\downarrow}\hat{c}_{\mathbf{k},\alpha\uparrow}\;,
\label{eq:hubbard}
\end{equation}
%
where $a$ ($L_s$) denotes the atomic (moir\'e) lattice constant,
$U_0$ is the atomic Hubbard $U$ value in
graphene, and $U_0\sim 2.25\textrm{-}9.3\,$eV according to previous first principles studies \cite{Wehling-coulomb-prl11,paiva-prb05}. In this work, $U_0$ is treated as a free parameter  varying from 1\,eV to 5\,eV.
Taking $U_0\!=\!5\,$eV, the characteristic Hubbard interaction energy for the moir\'e system $a^2U_0/L_s^2\!\approx\!2.3\,$meV for 1.25$^{\circ}$.  
We project both the intersite and on-site Coulomb interaction onto the two flat bands of each valley, and solve the interacting Hamiltonian self consistently using Hartree-Fock approximations \cite{supp_info}. 
We construct Hartree-Fock phase diagrams in the ($\epsilon$, $\kappa$) parameter space, and study the nature of the ground states at different partial integer fillings and different displacement fields, with the twist angle fixed at $1.25^{\circ}$ in TBMG and $1.28^{\circ}$ in TDBG.

\paragraph{Correlated insulators with zero Chern number. \textemdash}

\begin{figure}[!htbp]
\includegraphics[width=0.5\textwidth]{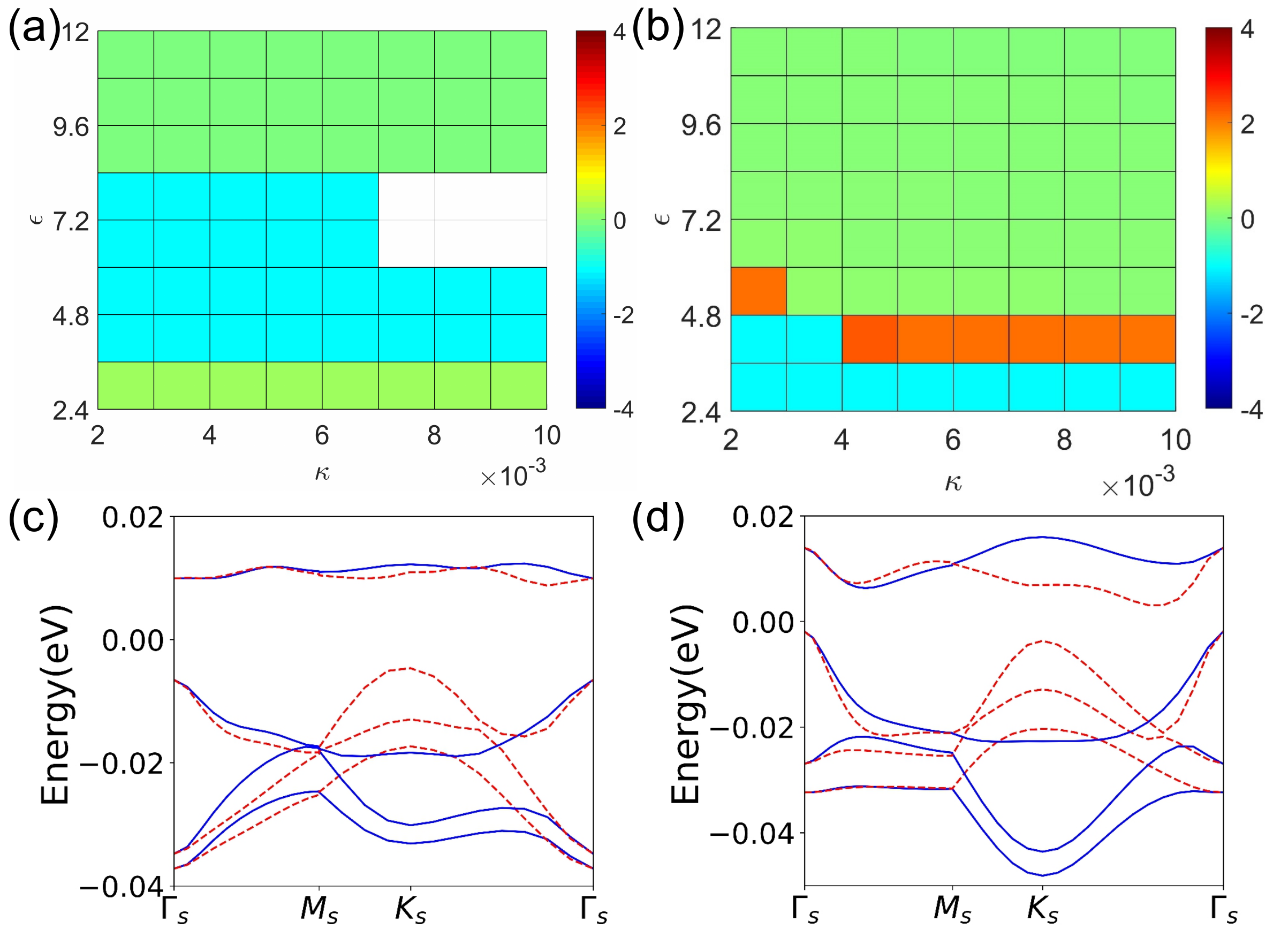}
\caption{~\label{fig2}  Hartree-Fock phase diagrams: (a) $U_d\!=\!\--0.0536$\,eV ($D=0.4\,V/nm$), (b) $U_d$ =0.0536\,eV ($D\!=\!\--0.4\,V/nm$) at 1/2 filling of TBMG with $\theta\!=\!1.25\,^{\circ}$. The color coding indicates Chern numbers, and Hubbard $U_0=2\,$eV in these calculations. The Hartree-Fock bandstructures of TBMG at 1/2 filling with (c) $U_d\!=\!\--0.0536$\,eV  and (d) $U_d\!=\!0.0536$\,eV ($\epsilon\!=\!9.6$ and $\kappa\!=\!0.005$\,\AA{}$^{-1}$). The blue and red lines represent the energy bands from different valleys.}
\end{figure}

The Chern numbers of the Hartree-Fock ground states with $U_d\!=\!- 0.0536\,$eV and $U_d\!=\!0.0536\,$eV at 1/2 filling of TBMG are presented in Fig.~\ref{fig2}(a) and (b) respectively. 
Our calculations indicate that there are two  (quasi-)degenerate ground states (with energy difference $\sim\!0.1\rm{-}1\mu\,$eV) at 1/2 filling for  TBMG: one is valley polarized (VP) state with nonzero Chern number ($\vert C\vert\!=\!2$ in  TBMG under $U_d\!=\!0.0536\,$eV),  and the other is a spin polarized (SP) state with zero Chern number, which is consistent with previous report \cite{rademaker-prr20}. 
Such quasi degeneracy is present whenever there is isolated conduction flat band in a TMG system with finite displacement field, which arises due to the analytic properties of the single-band interaction form factors \cite{supp_info}.
However, the inclusion of atomic Hubbard interactions would lower the energy of the SP state by about $0.4\!\sim\!0.6\,$meV per electron, such that the SP state becomes the unique ground state.
This because the Hartree part of the Hubbard interaction favors a spin polarized state. This is clearly shown in Fig.~\ref{fig2}(a)-(b), where large portions of the phase diagrams are occupied by the $C\!=\!0$  insulator states with the gaps $\sim\!10\,$meV \cite{supp_info}.
The Hartree-Fock bandstructures of the SP zero-Chern-number insulator state at 1/2 filling of TBMG with $U_d\!=\!\mp0.0536\,$eV are presented in Fig.~\ref{fig2}(c)-(d), where the blue and red lines represent the two opposite valleys.  
We have also double checked our results by including more active bands into the Hartree-Fock calculations, and the conclusions are unchanged \cite{supp_info}. Similar spin polarized and nematic insulator states that are energetically stabilized by atomic Hubbard interactions are obtained for other twisted multilayer graphene systems at 1/2 filling under finite displacement fields including TDBG and hBN-aligned TBG \cite{supp_info}. 

\begin{figure}[!htbp]
\includegraphics[width=0.5\textwidth]{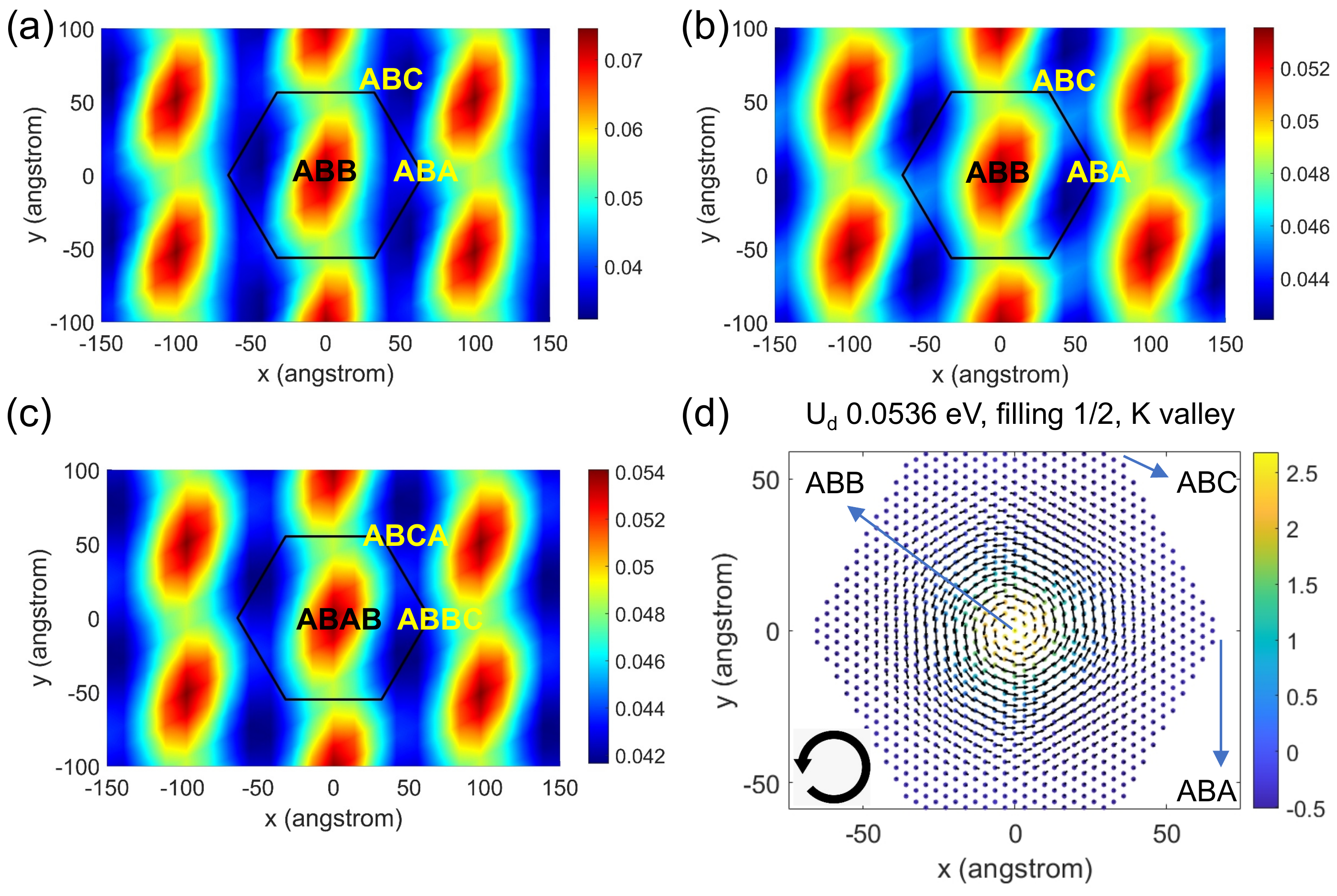}
\caption{~\label{fig3} The charge density distribution in real space at half filling for (a)  $U_d\!=\!\--0.0536$\,eV,  (b) $U_d\!=\!0.0536$\,eV in  TBMG, and (c) $U_d\!=\!0.04$\,eV   in  TDBG. (d) present the current loops from K valley at the valley Hall phase in  TBMG. The color coding represents the strength of  magnetic field generated by the current loops, in units of Gauss. The current loops from K$^{\prime}$ valley are opposite due to time-reversal symmetry.}
\end{figure}

 
The dominant order parameters of the zero-Chern-number states at 1/2 filling in TBMG and TDBG are $\tau _{0,z}s_{0,z}\sigma _k$ ($k=x,y$), where $\mathbf{\tau}$, $\mathbf{s}$, and $\mathbf{\sigma}$ represent Pauli matrices in the valley, spin, and sublattice spaces.  
The order parameters $s_{0,z}\sigma _y$ and $\tau _z s_{0,z}\sigma _x$ satisfy the so-called ``Kramers time-reversal symmetry"  $\mathcal{T}^{\prime}\!=\!i\tau _y \mathcal{K}$ \cite{zaletel-tbg-hf-prx20}, and $s_{0,z}\sigma _x$ and $\tau _zs_{0,z}\sigma _y$ obey time-reversal symmetry $\mathcal{T}\!=\!\tau _x \mathcal{K}$, both of which would enforce vanishing total orbital magnetization in the ground states.
The order parameters ($\sigma _y$, $\tau _z\sigma _x$) and ($\sigma _x$, $\tau _z\sigma _y$) form two-dimensional representations of the $C_{3z}$ operation, and detailed analysis reveal that the order parameters at 1/2 filling actually break $C_{3z} $ symmetry \cite{supp_info}. 
The charge densities of the Hartree-Fock ground states at $U_d\!=\!\mp 0.0536\,$eV in  TBMG 
are shown in Fig.~\ref{fig3}(a)-(b). It is clearly shown that the system forms $C_{3z}$-breaking, stripe-like charge patterns. 
In Fig.~\ref{fig3}(c) we show the charge density of TDBG at 1/2 filling at $\vert U_d\vert\!=\!0.04\,$eV, which shows similar stripe-like charge pattern, and it is consistent with recent STM measurements \cite{Pasupathy-nematic-tdbg-arxiv20}.

As shown in Fig.~\ref{fig1}(c) and (d),  the Chern numbers $C$ for the highest valence band and  lowest conduction band of TBMG for the $K^{\prime}$ valley are $-1$ and $2$ at $U_d\!=\!\--0.0536\,$eV, and are interchanged with each other for $U_d\!=\!+0.0536\,$eV ($\theta\!=\!1.25\,^{\circ}$). Therefore, the spin polarized $C\!=\!0$ state for negative $U_d$ consists of one $C\!=\!2$ and two $C\!=\!-1$ occupied bands, with zero total Chern number for each valley; for positive $U_d$, however, for each valley two $C=\pm 2$ bands and one $C\!=\!\mp 1$ band are occupied , leading to a $C\!=\!3 (-3) $ state for $K'$ ($K$) state, which is a QVH state.  Similarly, the $C\!=\!0$ nematic insulator state in TDBG is also a QVH state with valley Chern numbers $\pm 2$ \cite{supp_info}. The $C\!=-\!1$ phase in Fig.~\ref{fig2}(a)-(b) are states with both spin and valley polarizations \cite{supp_info}. The $C\!=\!0$ QVH states are associated with opposite chiral current loops in real space  for the two opposite valleys. The calculated current patterns for  TBMG system contributed by $K$ valley are shown in  Fig.~\ref{fig3}(d), where the black arrows denote the real-space current density vector circulating around the $ABB$ region. The color coding refers to the magnetic fields generated by these current loops, in units of Gauss.  
Such QVH states would give rise to helical edge states, which can be probed through nonlocal transport measurements \cite{sinha-tdbg-nonlocal-nc20}. 

\begin{figure}[!htbp]
\includegraphics[width=3.5in]{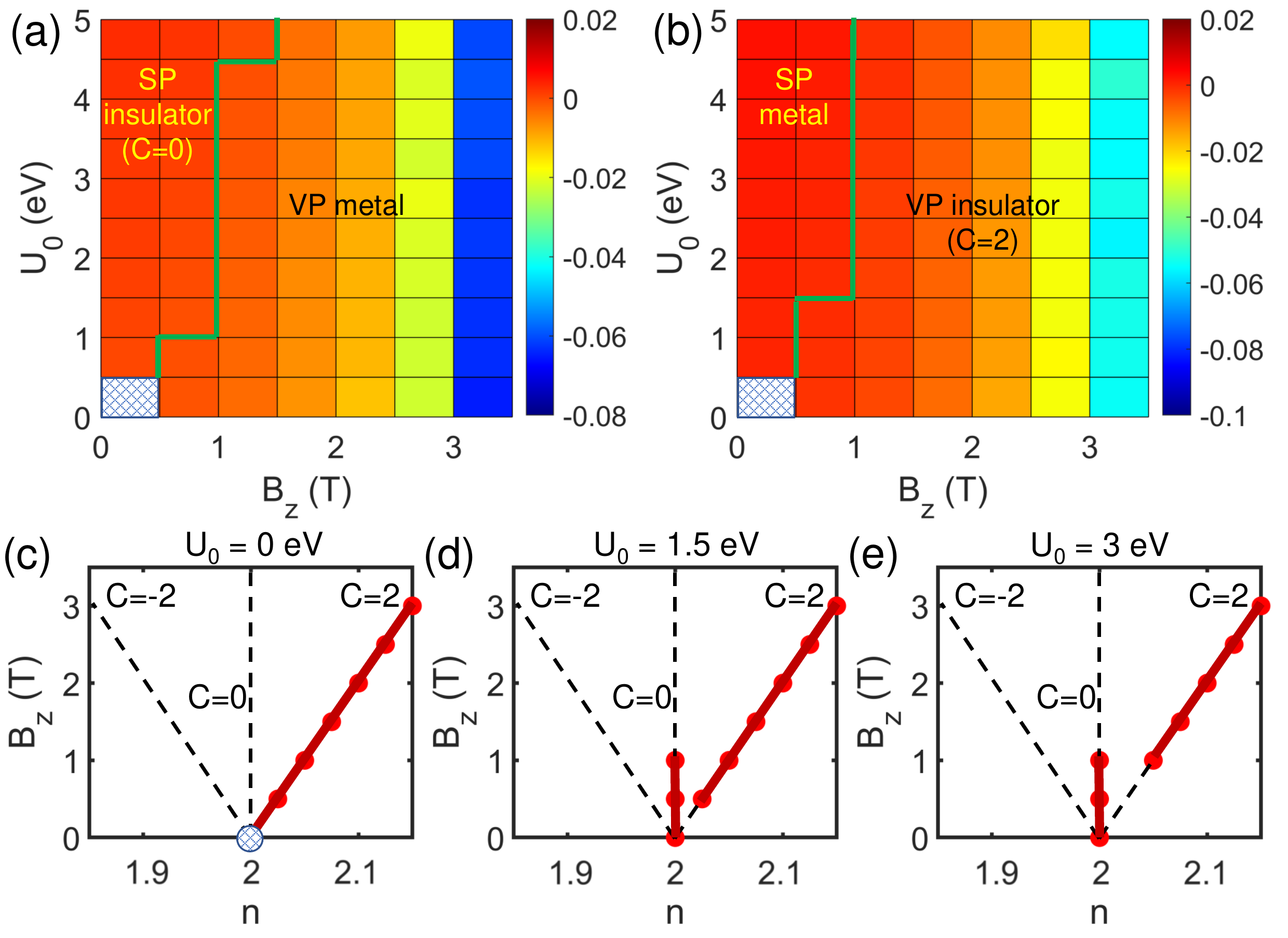}
\caption{~\label{fig4}  The Hartree-Fock phase diagram around  filling 2 in the parameter space of on-site Hubbard interaction and out-of-plane magnetic fields ($U_0$, $B_z$) in  TBMG with $\epsilon$ = 9.6, $\kappa$ = 0.005\,\AA{}$^{-1}$, $\theta=1.25^{\circ}$, and $U_d\!=\!0.0536\,$eV: (a) the occupation is fixed at $n(C=0)=2$, and (b) with the occupation  fixed at  $n(C\!=\!2)=2+2B\Omega_M e/h$ . ``SP" and ``VP" stand for ``spin polarized" and ``valley polarized" respectively. The color coding denotes the energy difference between VP and SP states $\Delta E=E_{VP}-E_{SP}$ in (a)-(b). The shaded block in (a) and (b) indicate that the SP and VP states are nearly degenerate at the origin. We also present the calculated gapped states remarked by red lines under magnetic fields with (c) $U_0\!=\!0$\,eV, (d) $U_0\!=\!1.5$\,eV, and (e) $U_0\!=\!3$\,eV in the TBMG system.}
\end{figure}

We have also studied the ground states at 1/4 filling with $D\!=\!0.5\,$V/nm ($U_d\!=\!-0.067\,$eV) and at 3/4 filling with $D\!=\!0.3\,$V/nm, and find that the systems stays in QAH states with $C\!=\!2$ when $\epsilon\gtrapprox\!6$ \cite{supp_info}, which are consistent with experimental observations \cite{young-monobi-nature20} and previous calculations \cite{rademaker-prr20}.  If the direction of the displacement field is reversed, the Chern numbers of the Hartree-Fock ground states at 1/4 and 3/4 fillings become $\mp 1$ due to the change of the band topology \cite{supp_info}.


\paragraph{Field tunable topological transitions \textemdash}
We further consider effects of external magnetic fields, which can be separated into two parts:  the spin Zeeman effects and the orbital magnetic effects.  The former can be trivially described by the spin Zeeman splitting, the latter deserves careful discussions. First, the vertical magnetic field tends to recombine the flat bands into a series of recurring Landau levels (LLs), i.e., the  Hofstadter butterfly spectra \cite{hofstadter-prb76},  which are dependent on the number of magnetic fluxes in each moir\'e primitive cell.  Second,
the magnetic field also induces splitting between the flat bands from the opposite valleys due to the large, valley-contrasting orbital magnetizations ($\sim 10\,\mu_{\rm{B}}$ for TBMG). Moreover, vertical magnetic field ($B_z$) also changes the density of the Chern bands as characterized by St$\check{\rm{r}}$eda formula \cite{streda} $\delta\rho=\delta n/\Omega_M=C B_z e/h$ ($\Omega_M$ is the area of moir\'e primitive cell), which also induces valley polarizations due  to the valley-contrasting Chern numbers in TMG systems. Such effects can be well captured by imposing a Berry-curvature correction to the density matrix \cite{xiao-prl05, dai-tbg}.      
Therefore, the vertical magnetic field tends to drive the system into a valley polarized, time-reversal breaking state. 
For weak vertical magnetic fields, the magnetic flux for each moir\'e supercell is small in TBMG, e.g., for $B\!=\!2\,$T, the flux per supercell $\Phi/\Phi_0\!=\!0.053\!\approx\!7/132$ ($\Phi_0\!=\!h/e$), which is far from forming notable Hofstadter bands. Thus we  neglect the effects of LL quantization for $B\!\lessapprox\!2\,$T, and only consider the orbital magnetic Zeeman effect \cite{song-zeeman-arxiv15,Koshino-orbital-prb11,ashvin-double-bilayer-nc19, wu-tdbg-arxiv20, Sun-zeeman-prb20} and the field-induced Berry-curvature correction to the charge density, the details of which are presented in Supplementary Information \cite{supp_info}. 
The orbital Zeeman splitting and the field-induced density variation for the Chern bands would compete with the Hubbard interaction: the former mechanism  favors a valley  polarized QAH state, while the latter favors a nematic, spin polarized state. Thus we expect to see a topological transition from the nematic spin polarized insulator to the QAH state with the increase of magnetic field.  

In Fig.~\ref{fig4}(a) and (b) we present the Hartree-Fock phase diagrams of TBMG around 1/2 filling with $U_d\!=\!0.0536\,$eV  including both orbital magnetic Zeeman effects and the field-induced Berry-curvature correction to the charge density. In particular, in Fig.~\ref{fig4}(a)  we fix the filling number  as $n(C=0)=2$, and we obtain a transition from a SP insulator (with Chern number 0) to a VP metal; while in Fig.~\ref{fig4}(b) we fix the filling number as $n(C\!=\!2)=2+ 2\Omega_M e B_z/h$ , and we find that there is a transition from a SP metallic state to a VP  insulator with Chern number 2 for $B_z\gtrapprox 0.5\rm{-}1\,$T. In Fig.~\ref{fig4}(c), (d) and (e) we also present the calculated insulator states with different Chern numbers in the parameter space of filling number $n$ and magnetic field $B_z$,  where the insulator states are marked by the red lines for different Hubbard interaction $U_0\!=\!0$, 1.5\,eV, and 3\,eV. Clearly we see the onset of field-driven $C\!=\!2$ Chern insulator states with small onset magnetic field $B_z\!\gtrapprox\!0.5\rm{-}1\,$T. For the $C\!=\!-2$ occupation $n(C\!=\!-2)=2-2\Omega_M e B_z/h$, the transition  from SP to VP states still exists but the system cannot open up a global gap in the VP phase up to $B_z\!=\!3\,$T.

\paragraph{Conclusion. \textemdash}
To summarize, we have theoretically studied the correlated  and topological states in twisted multilayer graphene systems.  Taking TBMG and TDBG as examples, we find that at 1/2 filling the Hartree-Fock ground states of both systems under finite displacements are  spin polarized, nematic insulators with zero Chern numbers, which are energetically stabilized by atomic on-site Hubbard interactions. In some cases, these spin polarized, nematic states can also be QVH states. Such a scenario generically applies to other twisted multilayer graphene systems.  Moreover, by virtue of the  orbital magnetic Zeeman effects and the field-induced change of charge density in the systems, the spin polarized, nematic insulator states at 1/2 filling can be driven into valley polarized QAH states with small  onset  magnetic fields.   
The competition between the valley polarized and spin polarized states and the resulted field-driven topological transitions may be a universal phenomena for moir\'e graphene superlattices.

\acknowledgements
This work is supported by  the National Key R \& D program of China (grant no. 2020YFA0309601) and the start-up grant of ShanghaiTech University. We would like to thank Yue Zhao, Jian Kang, and Yi Zhang for valuable discussions. 


\bibliography{tmg}

\begin{thebibliography}{77}
\bibitem[{com()}]{comment_filling}
\bibinfo{note}{In this paper, $\nu/4$ ($-4\!\leq\!\nu\!\leq\!4$) filling means
  that there are $\vert\nu\vert$ charge carriers per moir\'e supercell, where
  positive and negative $\nu$ corresponds to electron and hole carriers
  respectively.}
\end{thebibliography}

\widetext
\clearpage

\begin{center}
\textbf{\large Supplementary Information for ``Spin polarized nematic order, quantum valley Hall states, and field tunable topological transitions in twisted multilayer graphene systems"}
\end{center}

\vspace{12pt}
\begin{center}
\textbf{\large \I\ Lattice structures for twisted multilayer graphene systems}
\end{center}
\begin{figure}[!htbp]
\includegraphics[width=3.5in]{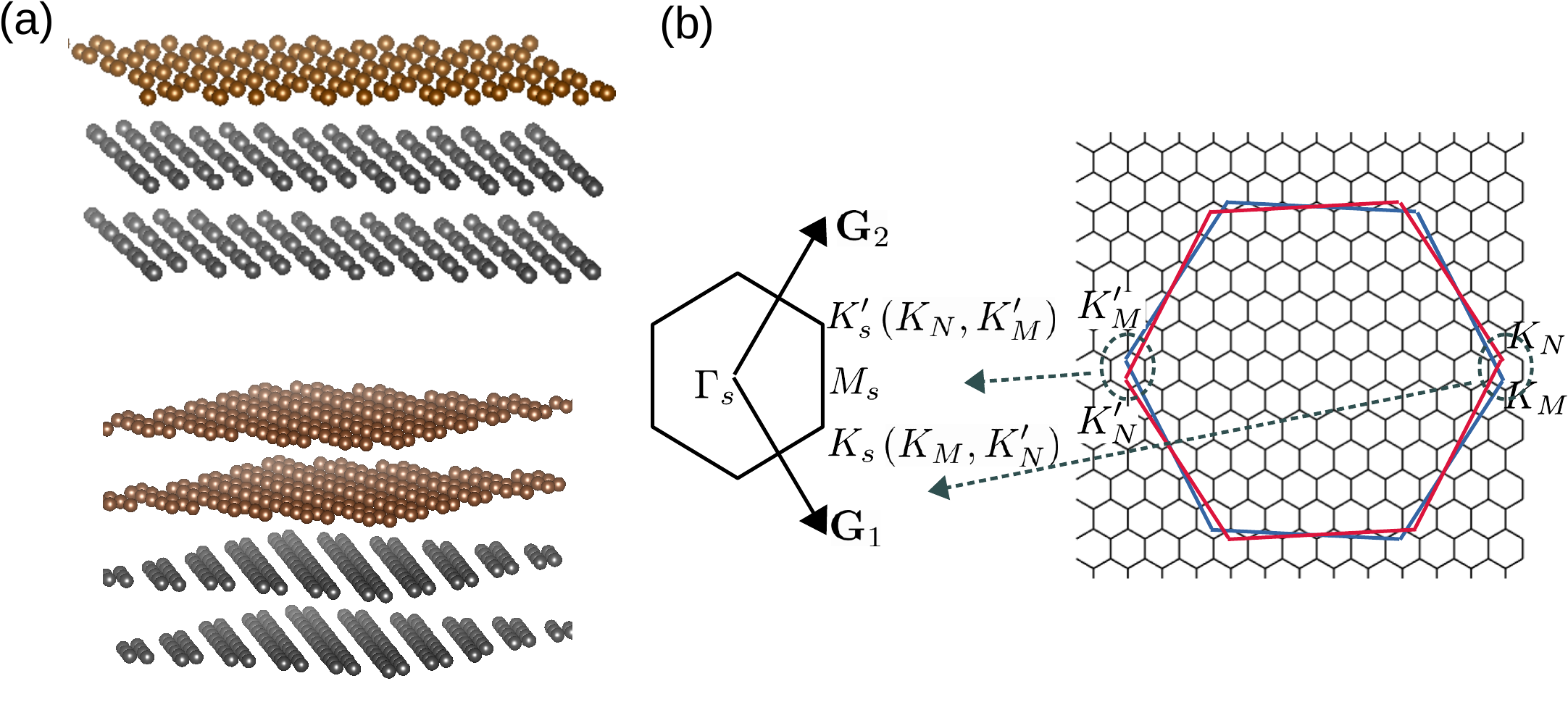}
\caption{~\label{figs-lattice} (a) Schematic illustration of the moir\'e lattice structures of twisted bilayer-monolayer graphene (upper panel) and twisted double bilayer graphene (lower panel).  (b) Brillouin zone of the twisted graphene systems: the blue and red hexagons represent the atomic Brillouin zones of the bottom and top layers, and the small black hexagon denotes the mini Brillouin zone of the moir\'e superlattices. $\mathbf{G}_1$ and $\mathbf{G}_2$ represent two moir\'e reciprocal lattice vectors.}
\end{figure}
We first review the lattice structures of moir\'e superlattices of twisted multilayer graphene systems. Here we consider the generic case of  twisted multilayer graphene with chiral stackings, i.e., we place $N$ chiral graphene multilayers on top of $M$ chiral graphene multilayers, and twist them with respect to each other by an angle $\theta$. This is schematically shown in Fig.~\ref{figs-lattice}(a) for the case of $M\!=\!2,N=1\!$ (upper panel) and $M\!=\!2, N\!=\!2$ (lower panel), i.e., for twisted bilayer-monolayer graphene (TBMG) and twisted double-bilayer graphene (TDBG) systems. Similar to the case of twisted bilayer graphene (TBG), commensurate moir\'e supercells are formed when the twist angle $\theta(m)$ obeys the condition 
$\cos{\theta(m)}=(3m^2+3m+1/2)/(3m^2+3m+1)$ \cite{castro-neto-prb12}, where $m$ is a positive integer. The  lattice vectors of the moir\'e superlattice are expressed as $\mathbf{t}_1=(\sqrt{3}L_s/2,-L_s/2)$, and $\mathbf{t}_2=(\sqrt{3}L_s/2,L_s/2)$, where $L_s=\vert\mathbf{t}_1\vert=a/(2\sin{(\theta/2)})$ is  lattice constant of the moir\'e supercell, and $a=2.46\,$\angstrom\ is the lattice constant of monolayer graphene.  In TBG it is well known that there are atomic corrugations, i.e., the variation of interlayer distances on the moir\'e length scale. In particular, in the $AB (BA)$ region of TBG,  the interlayer distance $d_{AB}\!\approx\!3.35\,$\angstrom\, while in the $AA$-stacked region the interlayer distance
$d_{AA}\!\approx\!3.6\,$\angstrom\ \cite{graphite-AA}. Such atomic corrugations  may be modeled as \cite{koshino-prx18}
\begin{equation}
d_z(\mathbf{r})=d_{0}+2d_1\sum_{j=1}^{3}\cos{(\,\mathbf{g}_j\!\cdot\!\mathbf{r}\,)}\;,
\label{eq:dz-1}
\end{equation}
where $\mathbf{G}_1$, $\mathbf{G}_2$, and $\mathbf{G}_3\!=\!\mathbf{G}_1+\mathbf{G}_2$ are the three reciprocal lattice vectors of the moire supercell. We take $d_0=3.433\,$\angstrom\ and $d_1=0.0278\,$\angstrom\ in order to reproduce the interlayer distances in $AA$- and $AB$-stacked bilayer graphene. Such atomic corrugations lead to different interlayer coupling parameters at the twisted interface. 
At a small twist angle $\theta$, the Brillouin zone (BZ) of the moir\'e supercell is much smaller  compared with those of the untwisted multilayers as shown in Fig.~\ref{figs-lattice}(b). In particular, the $K$ points of the bottom $M$-layer graphene ($K_M$) and  the top $N$-layer graphene ($K_N$) are mapped to the moir\'e $K_s$ and $K_s'$ points respectively, while the $K'_M$ and $K_N'$ points are mapped to $K_s'$ and $K_s$ points in the moir\'e BZ. 

\vspace{12pt}
\begin{center}
\textbf{\large \II\ Continuum Hamiltonian for twisted multilayer graphene systems}
\end{center}
The low-energy effective Hamiltonian of the twisted $(M+N)$-layer TMG of the $K$ valley is expressed as 
\begin{equation}
H^{0}_{\mu}(M+N)=\begin{pmatrix}
H^{0}_{\mu}(M) & \mathbb{U}_{\mu} \\
\mathbb{U}^{\dagger}_{\mu} & H^{0}_{\mu}(N)
\end{pmatrix}\;,
\label{eq:HMN}
\end{equation}
where $H^{0}_{\mu}(M)$ and $H^{0}_{\mu}(N)$ are the low-energy effective Hamiltonians for the $M$-layer and $N$-layer graphene with chiral stackings, and $\mu=\pm$ denotes the $K'/K$ valley. In particular, 
\begin{equation}
H^{0}_{\mu}(M)=\begin{pmatrix}
h^{0}_{\mu}(\mathbf{k}) & h_{\alpha} & 0 & 0 & ... \\
h_{\alpha}^{\dagger} & h^{0}_{\mu}(\mathbf{k}) & h_{\alpha} & 0 & ...\\
0 & h_{\alpha}^{\dagger} & h^{0}_{\mu}(\mathbf{k}) & h_{\alpha} & ...\\
 & & & ... &   
\end{pmatrix}\;,
\label{eq:HM}
\end{equation}
where  $h^{0}_{\mu}(\mathbf{k})\!=\!-\hbar v_{F}(\mathbf{k}-\mathbf{K}^{\mu}_{M})\cdot\mathbf{\sigma}_{\mu}$ stands for the low-energy effective Hamiltonian for monolayer graphene near the Dirac point $\mathbf{K}_{M}^{\mu}$, with $\mathbf{K}_M^{-}\!\equiv\!K_M$ and $\mathbf{K}_M^{+}\!\equiv\!K_M'$, and $\mathbf{\sigma}_{\mu}=(\mu\sigma_x,\sigma_y)$. $h_{\alpha}$ is the interlayer hopping for the chiral multilayer graphene with stacking chirality $\alpha=\pm$ , with
\begin{equation}
h_{+}=\begin{pmatrix}
t_2f(\mathbf{k}) & t_2 f^*(\mathbf{k})\;\\
t_{\perp}-3t_3 & t_2 f(\mathbf{k}) 
\end{pmatrix}\;,
\label{eq:chiral-hopping}
\end{equation}
where in $t_2\!=\!0.21$\,eV, $t_3\!\approx\!0.05\,$eV are extracted from the Slater-Koster hopping parameters in Ref.~\onlinecite{moon-tbg-prb13}. We set $t_{\perp}\!=\!0.48\,$eV, which is also from the Slater-Koster formula \cite{moon-tbg-prb13}. The phase factor $f(\mathbf{k})\!=\!(e^{-i\sqrt{3}ak_y/3}+e^{i(k_xa/2+\sqrt{3}ak_y/6)}+e^{i(-k_xa/2+\sqrt{3}ak_y/6)}$. The interlayer hopping with $-$ stacking chirality $h_{-}\!=\!h_{+}^{\dagger}$. Since the stacking chiralities of the multilayer graphene can be either $+$ or $-$, representing  $AB$ or $BA$ stacking. Therefore, there are four distinct stacking configurations for twisted $(M+N)$-layer graphene systems, i.e., stacking chiralities for the $M$ layers ($\alpha$) and $N$ layers ($\alpha'$) can be: $(\alpha, \alpha')=(\pm, \pm)$. Previous theory shows that in the chiral limit (i.e., the intrasublattice interlayer hopping vanishes), the magic angle of TBG  also applies to twisted $(M+N)$-layer systems (with chiral stackings), and the total valley Chern numbers of the two flat bands (per valley per spin) for $(\alpha,\alpha')$ stacking configurations obeys the following equation \cite{jpliu-prx19}:
\begin{equation}
C_{\alpha,\alpha'}^{\mu}=-\mu(\alpha(M-1)-\alpha'(N-1))\;.
\label{eq:chern-number}
\end{equation}
For TDBG, obviously, the total valley Chern number is 0 for $AB$-$AB$ stacking, and equals to $\pm 2$ for $AB$-$BA$ stacking \cite{jpliu-prx19,koshino-tdbg-prb19}. In this work we only consider the $AB$-$AB$ stacked TDBG, which is the case that is realized in experiments \cite{kim-tdbg-nature20,cao-tdbg-nature20,zhang-tdbg-np20}.

The off-diagonal term $\mathbb{U}$ represents the coupling between the twisted $M$ layers and $N$ layers. Here we assume that there is only the nearest neighbor interlayer coupling, i.e., the topmost layer of the $M$-layer graphene is only coupled with the bottom-most layer of the $N$-layer graphene, thus   
\begin{equation}
\mathbb{U}_{\mu}=\begin{pmatrix}
0 & ... & 0 \\
\vdotswithin{0} & ... & 0 \\
U_{\mu}(\mathbf{r})e^{i\mu\Delta\mathbf{K}\cdot\mathbf{r}} & ... & 0
\end{pmatrix}\;,
\label{eq:twist-coupling1}
\end{equation}
where the $2\!\times\!2$ matrix $U$ describes the tunneling between the Dirac states of the twisted bilayers \cite{macdonald-pnas11,koshino-prx18}
\begin{equation}
U_{\mu}(\mathbf{r})=\begin{pmatrix}
u_0 g_{\mu}(\mathbf{r}) & u_0'g_{\mu}(\mathbf{r}-\mathbf{r}_{AB})\\
u_0'g_{\mu}(\mathbf{r}+\mu\mathbf{r}_{AB}) & u_0 g_{\mu}(\mathbf{r})
\end{pmatrix}\;,
\label{eq:u}
\end{equation}
where $\mathbf{r}_{AB}\!=\!(\sqrt{3}L_s/3,0)$,  $u_0'$ and $u_0$ denote the intersublattice and intrasublattice interlayer tunneling amplitudes, with $u_0'\!\approx\!0.098\,$eV, and $u_0\!\approx\!0.078$\,eV \cite{koshino-prx18}. $u_0\!$ is  smaller than $u_0'$ due to the effects of atomic corrugations \cite{koshino-prx18, jpliu-prb19}. $\Delta\mathbf{K}=\mathbf{K}_N-\mathbf{K}_M=(0,4\pi/3L_s)$ is the shift between the Dirac points of the $N$ layers and the $M$ layers. The phase factor $g(\mathbf{r})$ is defined as $g_{\mu}(\mathbf{r})=\sum_{j=1}^{3}e^{-i\mu\mathbf{q}_j\cdot\mathbf{r}}$, with $\mathbf{q}_1=(0,4\pi/3L_s)$, $\mathbf{q}_2=(-2\pi/\sqrt{3}L_s,-2\pi/3L_s)$, and $\mathbf{q}_3=(2\pi/\sqrt{3}L_s,-2\pi/3L_s)$. 

In realistic devices the twisted multilayer graphene systems are encapsulated by hexagonal boron nitride (hBN) substrates.  When the hBN substrate is aligned with the twisted graphene system, i.e., in the $AA$ region when the boron/nitrogen atom is exactly below the A/B carbon sublattice, the hBN substrate would impose a staggered sublattice potential to the twisted graphene system.  This staggered sublattice potential is especially important for TBG because  it breaks the $C_{2z}$ symmetry in TBG, which opens gaps at the Dirac points and gives rise to nonzero Berry curvatures and valley Chern numbers.  On the other hand, $C_{2z}$ symmetry is always broken for chiral twisted $(M+N)$-layer graphene for $M+N\!\geq\!3$, therefore the alignment of hBN substrate is not important for twisted multilayers except for TBG.  In this work, we consider twisted graphene systems with broken $C_{2z}$ symmetry. To be specific, we theoretically study the non-interacting bandstructures and Hartree-Fock ground states for TBMG, TDBG, TBG aligned with hBN substrate, twisted (3+1)-layer graphene, and twisted (3+2)-layer graphene systems.

It is also worthwhile to note that when TBG is aligned with hBN substrate, there are also two moir\'e patterns: one from the mutual twist of the two graphene layers, and the other from the lattice mismatch between the hBN and the graphene layers. However, the moir\'e potential from the hBN substrate is one order of magnitude weaker than that from the mutual twist \cite{jung-prb14,koshino-prb14}, therefore the effect of the hBN moir\'e can be neglected in hBN-aligned TBG system, and dominant effects of the hBN substrate is to impose a staggered sublattice potential to the bottom layer graphene which breaks $C_{2z}$ (sublattice) symmetry.

\vspace{12pt}
\begin{center}
\textbf{\large \III\ The noninteracting physics in twisted multilayer graphene}
\end{center}

\begin{figure}[!htbp]
\includegraphics[width=3.5in]{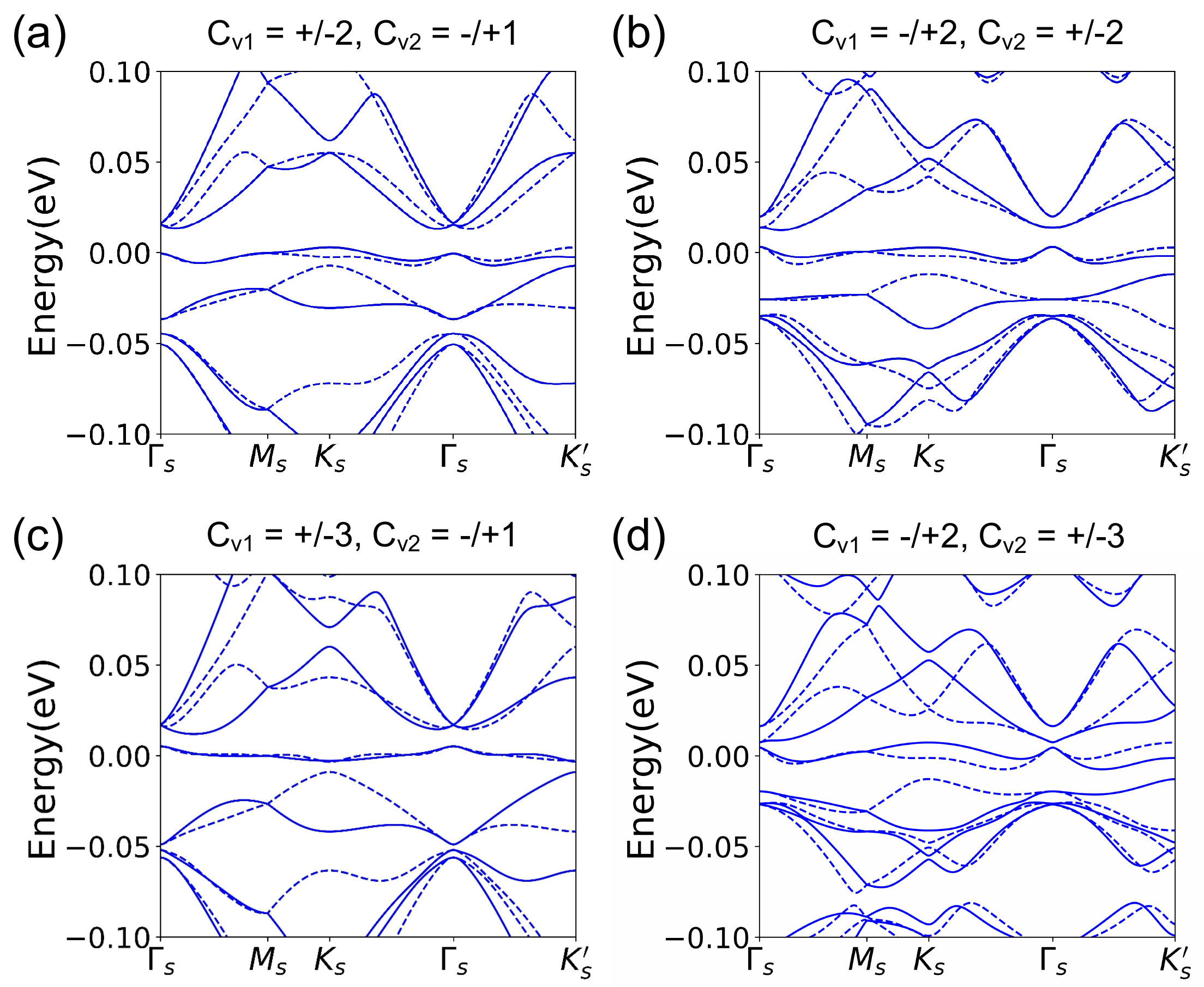}
\caption{~\label{figsband} The non-interacting energy bands: (a)  1.25\,$^\circ$-twisted bilayer-monolayer graphene under $U_d\!=\!-0.0536\,$eV, (b) 1.28\,$^\circ$-twisted double bilayer graphene under $U_d\!=\!0.04\,$eV, (c) twisted $(3+1)$-layer graphene with $\theta\!=\!1.4^{\circ}$ and $U_d=-0.06\,$eV, (d) twisted $(3+2)$-layer graphene with $\theta\!=\!1.28\,^{\circ}$ and $U_d=0.04\,$eV.  The solid (dashed) lines represent the energy bands of K (K$^{\prime}$) valley. The valley Chern numbers of the conduction flat bands and valence flat bands are denoted by $C_{v1}$ and $C_{v2}$ respectively.}
\end{figure}

In this section we first discuss the non-interacting physics of twisted multilyer graphene systems. The typical bandstructures of TBMG, TDBG with $AB$-$AB$ stacking, hBN-aligned TBG, twisted (3+1)-layer graphene (with chiral stacking), and twisted (3+2)-layer graphene (with chiral stacking) systems are presented in Fig.~\ref{figsband}(a)-(d). To be specific, in Fig.~\ref{figsband}, we show the non-interacting bandstructures of (a) TBMG for $\theta\!=\!1.25\,^{\circ}$ with $U_d\!=\!-0.0536\,$eV, (b) TDBG with $\theta=1.28^{\circ}$ and $U_d=0.04\,$eV, (c) twisted $(3+1)$-layer graphene with $\theta\!=\!1.4^{\circ}$ and $U_d=-0.06\,$eV, and (d) twisted $(3+2)$-layer graphene with $\theta\!=\!1.28\,^{\circ}$ and $U_d=0.04\,$eV. We have also marked the valley Chern numbers of the conduction flat bands and valence flat bands in these twisted multilayer graphene systems, which are marked by $C_{v1}$ and $C_{v2}$ in Fig.~\ref{figsband}.
Here $U_d=-e D d/\epsilon_{BN}$ is the vertical electrostatic energy drop across the graphene multilayer with total thickness $d$, $D$ is the displacement field, and $\epsilon_{BN}\approx 5$ is the dielectric constant of BN substrate (see main text). The solid and dashed lines represent the energy bands from the $K$ and $K'$ valleys respectively. In these five systems, there are two flat bands for each valley which are energetically separated from other  remote bands. In TBMG and TDBG under finite displacement fields, the conduction flat band is flatter than the valence flat band, which means the system is more susceptible to Coulomb interactions at electron fillings.
%

\begin{figure}[!htbp]
\includegraphics[width=3.5in]{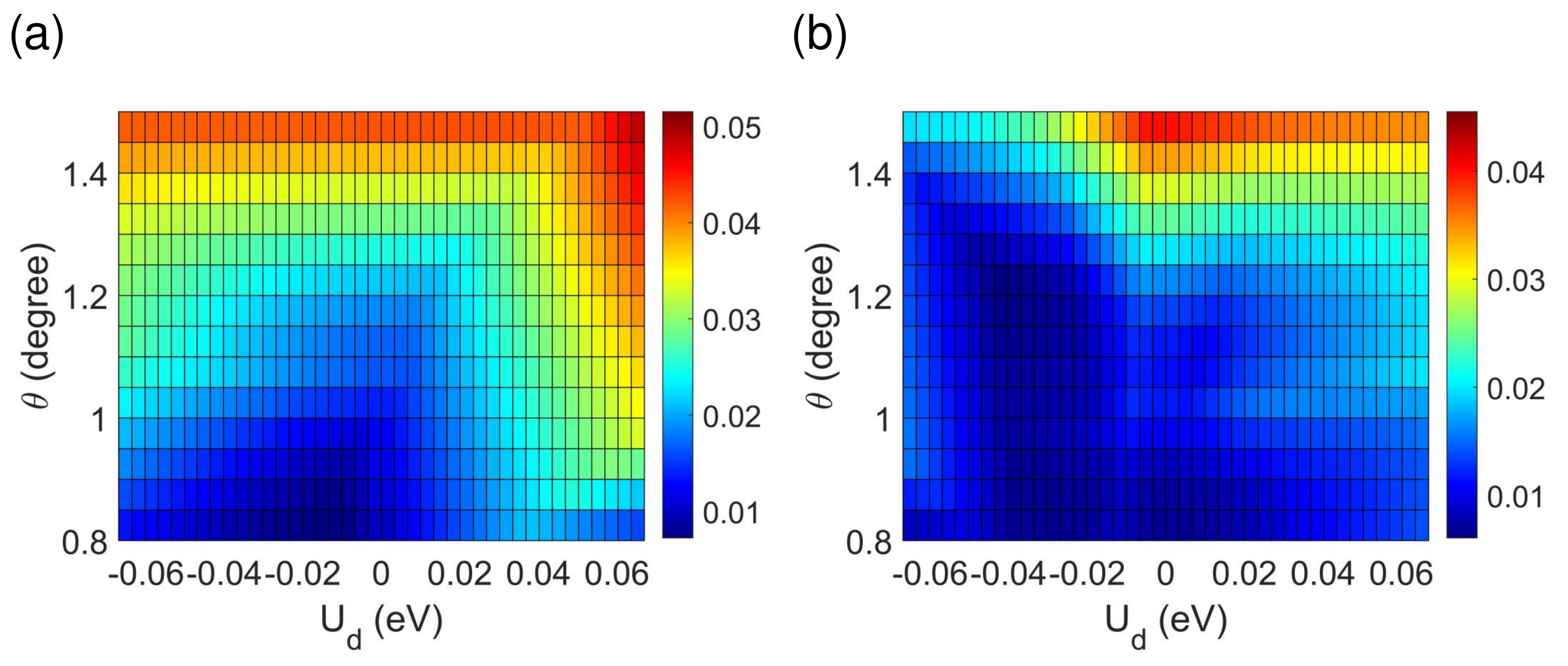}
\caption{~\label{figs1} (a) and (b) show the band widths of the highest valence band and lowest conduction band under different electric displacement field  $U_d$ (--0.067$\sim$0.067\,eV) and varying twist angle $\theta$ in the TBMG.}
\end{figure}

As discussed in Ref.~\onlinecite{jpliu-prx19}, the flat bands in twisted multilayer graphene  have nontrivial topological properties. The topology of the flat bands can be further tuned by vertical displacement fields. We have shown the Chern numbers of the valence and conduction flat bands from the $K^{\prime}$ valley in Fig.~1(c)-(d) of the main text. Here in Fig.~\ref{figs1} (a) and (b) we further show the bandwidths of the valence and conduction flat bands of TBMG under different electrostatic potential energy drop $U_d$ and twist angle $\theta$. We see that the bandwidth of the conduction flat band is $\lessapprox 10\,$meV when $0.8\,^{\circ}\lessapprox\theta\lessapprox 1.3\,^{\circ}$ and $-0.02\,\textrm{eV}\!\lessapprox\!U_d\!\lessapprox\!-0.06\,$eV ($0.15\,\textrm{V/nm}\lessapprox D\lessapprox 0.45\,\textrm{V/nm}$), which is ideal for the realization of various correlation effects. Indeed, the experimentally observed correlated insulators \cite{young-monobi-nature20,Yankowitz-monobi-np2020,shi-tmgUd-arxiv20} and quantum anomalous Hall insulators \cite{young-monobi-nature20} which show up at partial integer electron fillings are all within this parameter regime.  For example, in Ref.~\onlinecite{young-monobi-nature20}, the correlated insulator at 1/2 filling is observed at $\theta\!=\!1.25\,^{\circ}$ for $0.2\,\textrm{V/nm}\lessapprox\!D\!\lessapprox\!0.55\,\textrm{V/nm}$; in Ref.~\onlinecite{Yankowitz-monobi-np2020}, correlated insulator at 1/2 filling is observed at $\theta\!=\!1.08\,^{\circ}$ for $0.3\,\textrm{V/nm}\lessapprox\!D\!\lessapprox\!0.5\,\textrm{V/nm}$; and quantum anomalous Hall states are observed at $\theta\!=\!1.25\,^{\circ}$ for $D\!\approx\!0.5\,V/nm$ at 1/4 filling and $D\!\approx\!0.39\,V/nm$ at 3/4 filling.

\begin{figure}[!htbp]
\includegraphics[width=3.5in]{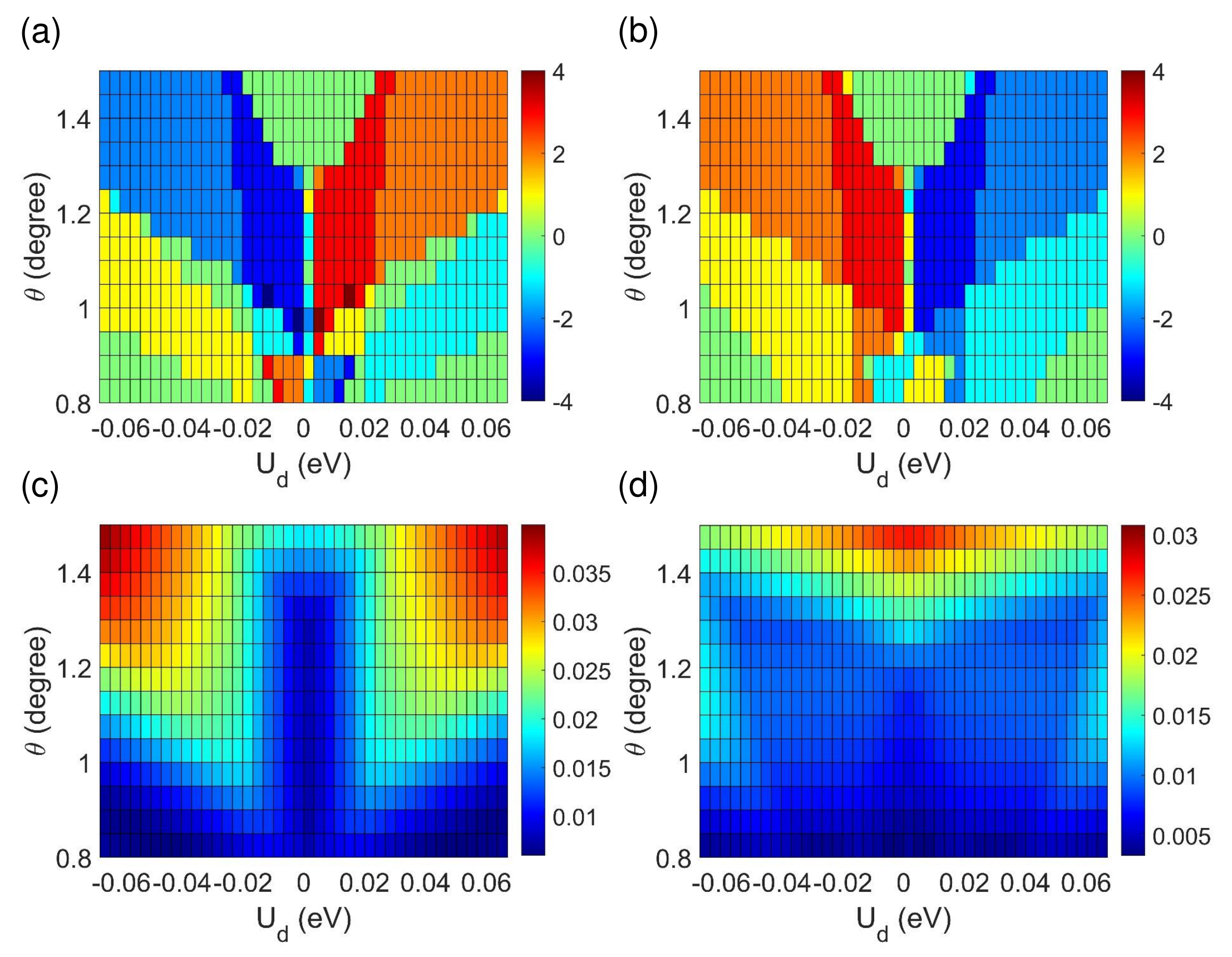}
\caption{~\label{figstdbg} (a) and (b) show the Chern numbers of the highest valence band and lowest conduction band, and (c) and (d) show the band widths of the highest valence band and lowest conduction band, under different electric displacement field  $U_d$ (--0.067$\sim$0.067\,eV) and varying twist angle $\theta$ in the TDBG.}
\end{figure}

In Fig.~\ref{figstdbg} we show the bandwidths and Chern numbers of the valence flat band and conduction flat band of the $K^{\prime}$ valley, in the parameter space of $(U_d, \theta)$. In Fig.~\ref{figstdbg}(a) and (b) we show the Chern numbers of the valence flat band and conduction flat band of the $K$ valley, which can be tuned to large extent by both twist angle and vertical displacement fields. The Chern numbers vary from -4 to 4 for different displacement fields and twist angles. Moreover, the distribution of the Chern number is anti-symmetric with respect to $U_d$, for fixed twist angle, the Chern number is opposite for opposite $U_d$, and the total Chern number of the two flat bands sum up to zero \cite{jpliu-prx19,koshino-tdbg-prb19}.

\vspace{12pt}
\begin{center}
\textbf{\large \IV\ The Coulomb interactions in the twisted graphene system}
\end{center}


We consider the Coulomb interactions
\begin{equation}
H_C=\frac{1}{2}\int d\,\mathbf{r}\,d\,\mathbf{r}^{\prime} \sum _{\sigma, \sigma ^{\prime}}\hat{\psi}_\sigma ^{\dagger}(\mathbf{r})\hat{\psi}_{\sigma ^{\prime}}^{\dagger}(\mathbf{r} ^{\prime})\frac{e^2}{4\pi \epsilon _0\vert \mathbf{r}-\mathbf{r}^{\prime}\vert}\hat{\psi}_{\sigma ^{\prime}}(\mathbf{r} ^{\prime}) \hat{\psi}_{\sigma}(\mathbf{r})
\label{eq:coulomb}
\end{equation}
where $\hat{\psi}_{\sigma}(\mathbf{r})$ is real-space electron annihilation operator at $\mathbf{r}$ site with spin $\sigma$. In the graphene system, this interaction can be written as 
\begin{equation}
H_C=\frac{1}{2}\sum _{i i' j j'}\sum _{\alpha \alpha '\beta \beta '}\sum _{\sigma \sigma '} \hat{c}^{\dagger}_{i\alpha \sigma}\hat{c}^{\dagger}_{i'\alpha '\sigma '}U^{\alpha \beta \sigma , \alpha ' \beta ' \sigma '} _{ij,i'j'}\hat{c}_{j'\beta '\sigma '}\hat{c}_{j\beta \sigma}\;,
\end{equation}
where
\begin{align}
U^{\alpha \beta \sigma , \alpha ' \beta ' \sigma '} _{ij,i'j'}=&\int d \mathbf{r}\,d\mathbf{r}^{\prime}\,\frac{e^2}{4\pi \epsilon _0\vert \mathbf{r}-\mathbf{r}^{\prime}\vert}\,\phi ^*_\alpha (\mathbf{r}-\mathbf{R}_i-\tau _\alpha)\,\phi _\beta (\mathbf{r}-\mathbf{R}_j-\tau _\beta)\;\nn
&\times\phi ^*_{\alpha  '}(\mathbf{r}-\mathbf{R}_i'-\tau _{\alpha '})\phi _{\beta  '}(\mathbf{r}-\mathbf{R}_j'-\tau _{\beta '})\chi ^{\dagger}_\sigma \chi ^{\dagger}_{\sigma '}\chi _{\sigma '}\chi _{\sigma}\;.
\end{align}
Here $i$, $\alpha$, and $\sigma$ refer to Bravis lattice vectors, layer/sublattice index, and spin index. $\phi$ is Wannier function and $\chi$ is the two-component spinor wave function. We further assume that the "density-density" like interaction is dominant in the system, i.e., $U^{\alpha \beta \sigma , \alpha ' \beta ' \sigma '} _{ij,i'j'}\approx U^{\alpha \alpha \sigma , \alpha ' \alpha ' \sigma '} _{ii,i'i'}\equiv U_{i\alpha \sigma ,i'\alpha '\sigma '}$,  then the Coulomb interaction is simplified to
\begin{align}
H_C=&\frac{1}{2}\sum _{i i'}\sum _{\alpha \alpha '}\sum _{\sigma \sigma '}\hat{c}^{\dagger}_{i\alpha \sigma}\hat{c}^{\dagger}_{i'\alpha '\sigma '}U_{i\alpha \sigma ,i'\alpha '\sigma '}\hat{c}_{i'\alpha '\sigma '}\hat{c}_{i\alpha \sigma}\;\nn
=&\frac{1}{2}\sum _{i\alpha \neq i'\alpha '}\sum _{\sigma \sigma '}\hat{c}^{\dagger}_{i\alpha \sigma}\hat{c}^{\dagger}_{i'\alpha '\sigma '}U_{i\alpha,i'\alpha '}\hat{c}_{i'\alpha '\sigma '}\hat{c}_{i\alpha \sigma}\;\nn
&+\sum _{i\alpha}U_0 \hat{c}^{\dagger}_{i\alpha\uparrow}\hat{c}^{\dagger}_{i\alpha\downarrow}\hat{c}_{i\alpha\downarrow}\hat{c}_{i\alpha\uparrow}
\end{align}
Here we can see that the Coulomb interaction can be divided into intersite Coulomb interaction and on-site Coulomb interaction. Given that the electron density in a typical moir\'e graphene system is extremely low, i.e., a few electrons per moir\'e supercell, the chance that two electrons meet at the same atomic site is very low. The Coulomb correlations between two electrons in the moir\'e system are mostly contributed by the inter-site Coulomb interactions.
Therefore,  in most of the previous studies, the on-site Hubbard interaction has been neglected. In this work, we first consider the effects of the dominant inter-site Coulomb interactions,  then we discuss the effects of the atomic on-site Hubbard interactions, which turn out to be crucial in lifting the (quasi-)degeneracy between the valley polarized and spin polarized states.


In order to model the screening effects to the electron-electron Coulomb interactions from the dielectric environment, we introduce two screening parameters: the background dielectric constant $\epsilon$ and the inverse screening length $\kappa$, and we assume the Coulomb interaction takes the Thomas-Fermi screened form:
\begin{equation}
V(\vert\mathbf{r}-\mathbf{r}'\vert)=\frac{e^2}{4\pi \epsilon \epsilon _0\vert \mathbf{r}-\mathbf{r}^{\prime}\vert}e^{-\kappa \vert \mathbf{r}-\mathbf{r}^{\prime}\vert}  
\end{equation} 
Then the Fourier transform is expressed as 
\begin{align}
V(\mathbf{q})&=\frac{1}{\Omega _M}\int d\mathbf{r}\frac{e^2}{4\pi \epsilon \epsilon _0\vert \mathbf{r}\vert}e^{-\kappa \vert \mathbf{r}\vert} e^{-i\mathbf{q}\cdot \mathbf{r}}\;\nn
&=\frac{e^2}{2\Omega _M\epsilon _0\epsilon \sqrt{\vert\mathbf{q}\vert^2+\kappa ^2}}\;,
\end{align}
where $\Omega _M$ is the area of a moir\'e primitive cell.

At small twist angles, the intersite Coulomb interactions can be divided into the intra-valley term and the inter-valley term \cite{ashvin-double-bilayer-nc19}. The intra-valley term $H_{C}^{\rm{intra}}$ can be expressed as
\begin{equation}
H_{C}^{\rm{intra}}=\frac{1}{2N_s}\sum_{\alpha\alpha '}\sum_{\mu\mu ',\sigma\sigma '}\sum_{\mathbf{k}\mathbf{k} '\mathbf{q}}\,V(\mathbf{q})\,
\hat{c}^{\dagger}_{\mathbf{k}+\mathbf{q},\mu \sigma \alpha} \hat{c}^{\dagger}_{\mathbf{k}'-\mathbf{q},\mu '\sigma '\alpha '}\hat{c}_{\mathbf{k}',\mu '\sigma '\alpha '}\hat{c}_{\mathbf{k},\mu \sigma \alpha}\;,
\label{eq:h-intra}
\end{equation}
and the inter-valley term $H_{C}^{\rm{inter}}$ is expressed as
\begin{equation}
H_{C}^{\rm{inter}}=\frac{1}{2N_s}\sum_{\alpha\alpha '}\sum_{\mu ,\sigma\sigma '}\sum_{\mathbf{k}\mathbf{k} '\mathbf{q}}\,V(\vert\mathbf{K}-\mathbf{K}'\vert)\hat{c}^{\dagger}_{\mathbf{k}+\mathbf{q},\mu \sigma \alpha} \hat{c}^{\dagger}_{\mathbf{k}'-\mathbf{q},-\mu \sigma '\alpha '}\,
\hat{c}_{\mathbf{k}',\mu \sigma '\alpha '}\hat{c}_{\mathbf{k},-\mu \sigma \alpha}\;.
\label{eq:h-inter}
\end{equation}
Here $V(\mathbf{q})$ denotes the screened Coulomb interaction $V(\mathbf{q})\!=\!e^2/(\,2\Omega _M\epsilon \epsilon _0\sqrt{q^2+\kappa ^2}\,)$, where $\Omega _M$ is the area of moir\'e supercell, $\kappa$ is the inverse screening length and $\epsilon$ denotes background dielectric constant. $\epsilon$ and $\kappa$ will be treated as two free parameters in this work. 
$H_{C}^{\rm{intra}}$ includes the Coulomb scattering processes of two electrons created and annihilated in the same valley, and $H_{C}^{\rm{inter}}$ includes the processes that two electrons are created in $\mu$ and $-\mu$ and get annihilated in $-\mu$ and $\mu$ valleys. Here the atomic wavevector $\mathbf{k}$ is expanded around the valley $K^{\mu}$ in the big Brillouin zone of the monolayer graphene, which can be decomposed as $\mathbf{k}=\widetilde{\mathbf{k}}+\mathbf{G}$, where $\widetilde{\mathbf{k}}$ is the moir\'e wavevector in the moir\'e BZ, and $\mathbf{G}$ denotes a moir\'e reciprocal lattice vector.  We note that the typical intravalley interaction energy $V_M\approx e^2/(4\pi\epsilon_0\epsilon L_s)\approx 25\,$meV for twist angle $\theta\approx 1.2^{\circ}$ and $\epsilon\approx 5$; while the intervalley interaction $V(\vert\mathbf{K}-\mathbf{K}'\vert)\sim 0.35\,$meV for $\theta\approx 1.2^{\circ}$ and $\epsilon\approx 5$, which is two orders of magnitudes smaller than the intravalley interaction, thus we neglect the intervalley term (Eq.~(\ref{eq:h-inter}) in our calculations.

The electron annihilation operator can be transformed from the original basis to the band basis:
\begin{equation}
\hat{c}_{\mathbf{k},\mu\alpha\sigma}=\sum _n C_{\mu \alpha \mathbf{G},n}(\widetilde{\mathbf{k}})\,\hat{c}_{\mu\sigma , n\widetilde{\mathbf{k}}}\;,
\label{eq:transform}
\end{equation}
where $C_{\mu \sigma \alpha \mathbf{G},n}(\widetilde{\mathbf{k}})$ is the expansion coefficient in the $n$th Bloch eigenstate at $\widetilde{\mathbf{k}}$ of valley $\mu$: 
\begin{equation}
\vert \Psi _{n\widetilde{\mathbf{k}}}^{\mu}\rangle=\sum _{\alpha \mathbf{G}}C_{\mu \alpha \mathbf{G},n}(\widetilde{\mathbf{k}})\,\vert \mu \sigma \alpha \mathbf{G},\widetilde{\mathbf{k}} \rangle\;.
\end{equation}
We note that the non-interacting Bloch functions are spin degenerate due to the separate spin rotational symmetry ($SU(2)\otimes SU(2)$ symmetry) of each valley. Using the transformation given in Eq.~(\ref{eq:transform}), the intravalley Coulomb interaction (Eq.~(\ref{eq:h-intra}) can be re-written in the band basis:
\begin{equation}
H^{\rm{intra}}=\frac{1}{2N_s}\sum _{\widetilde{\mathbf{k}} \widetilde{\mathbf{k}}'\widetilde{\mathbf{q}}}\sum_{\substack{\mu\mu' \\ \sigma\sigma'}}\sum_{\substack{nm\\ n'm'}}\left(\sum _{\mathbf{Q}}\,V(\mathbf{Q}+\mathbf{\widetilde{q}})\,\Omega^{\mu \sigma,\mu'\sigma'}_{nm,n'm'}(\widetilde{\mathbf{k}},\widetilde{\mathbf{k}}',\widetilde{\mathbf{q}},\mathbf{Q})\right)\hat{c}^{\dagger}_{\mu\sigma,n\widetilde{\mathbf{k}}+\widetilde{\mathbf{q}}} \hat{c}^{\dagger}_{\mu'\sigma',n'\widetilde{\mathbf{k}}'-\widetilde{\mathbf{q}}}\,\hat{c}_{\mu'\sigma',m'\widetilde{\mathbf{k}}'}\,\hat{c}_{\mu\sigma,m\widetilde{\mathbf{k}}}
\label{eq:Hintra-band}
\end{equation}
%
where the form factor $\Omega ^{\mu \sigma,\mu'\sigma'}_{nm,n'm'}$ is written as
\begin{equation}
\Omega ^{\mu \sigma,\mu'\sigma'}_{nm,n'm'}(\widetilde{\mathbf{k}},\widetilde{\mathbf{k}}',\widetilde{\mathbf{q}},\mathbf{Q})
=\sum _{\alpha\alpha'\mathbf{G}\mathbf{G}'}C^*_{\mu\sigma\alpha\mathbf{G}+\mathbf{Q},n\widetilde{\mathbf{k}}+\widetilde{\mathbf{q}}}C^*_{\mu'\sigma'\alpha'\mathbf{G}'-\mathbf{Q},n'\widetilde{\mathbf{k}}'-\widetilde{\mathbf{q}}}C_{\mu'\sigma'\alpha'\mathbf{G}',m'\widetilde{\mathbf{k}}'}C_{\mu\sigma\alpha\mathbf{G},m\widetilde{\mathbf{k}}}
\end{equation}
We make Hartree-Fock approximation to Eq.~(\ref{eq:Hintra-band}) such that the two-particle Hamiltonian is decomposed into a superposition of the Hartree and Fock single-particle Hamiltonians, where the Hartree term is expressed as
\begin{equation}
\begin{split}
H_H^{\rm{intra}}=&\frac{1}{2N_s}\sum _{\widetilde{\mathbf{k}} \widetilde{\mathbf{k}}'}\sum _{\substack{\mu\mu'\\ \sigma\sigma'}}\sum_{\substack{nm\\ n'm'}}\left(\sum _{\mathbf{Q}} V(\mathbf{Q})\Omega ^{\mu \sigma,\mu'\sigma'}_{nm,n'm'}(\widetilde{\mathbf{k}},\widetilde{\mathbf{k}}',0,\mathbf{Q})\right)\\
&\times \left(\langle \hat{c}^{\dagger}_{\mu\sigma,n\widetilde{\mathbf{k}}}\hat{c}_{\mu\sigma,m\widetilde{\mathbf{k}}}\rangle \hat{c}^{\dagger}_{\mu'\sigma',n'\widetilde{\mathbf{k}}'}\hat{c}_{\mu'\sigma',m'\widetilde{\mathbf{k}}'} + \langle \hat{c}^{\dagger}_{\mu'\sigma',n'\widetilde{\mathbf{k}}'}\hat{c}_{\mu'\sigma',m'\widetilde{\mathbf{k}}'}\rangle \hat{c}^{\dagger}_{\mu\sigma,n\widetilde{\mathbf{k}}}\hat{c}_{\mu\sigma,m\widetilde{\mathbf{k}}}\right)
\end{split}
\end{equation}
%
and the Fock term is expressed as:
%
\begin{equation}
\begin{split}
H_F^{\rm{intra}}=&-\frac{1}{2N_s}\sum _{\widetilde{\mathbf{k}} \widetilde{\mathbf{k}}'}\sum _{\substack{\mu\mu'\\ \sigma\sigma'}}\sum_{\substack{nm\\ n'm'}}\left(\sum _{\mathbf{Q}} V(\widetilde{\mathbf{k}}’-\widetilde{\mathbf{k}}+\mathbf{Q})\Omega ^{\mu \sigma,\mu'\sigma'}_{nm,n'm'}(\widetilde{\mathbf{k}},\widetilde{\mathbf{k}}',\widetilde{\mathbf{k}}’-\widetilde{\mathbf{k}},\mathbf{Q})\right)\\
&\times \left(\langle \hat{c}^{\dagger}_{\mu\sigma,n\widetilde{\mathbf{k}}'}\hat{c}_{\mu'\sigma',m'\widetilde{\mathbf{k}}'}\rangle \hat{c}^{\dagger}_{\mu'\sigma',n'\widetilde{\mathbf{k}}}\hat{c}_{\mu\sigma,m\widetilde{\mathbf{k}}} + \langle \hat{c}^{\dagger}_{\mu'\sigma',n'\widetilde{\mathbf{k}}}\hat{c}_{\mu\sigma,m\widetilde{\mathbf{k}}}\rangle \hat{c}^{\dagger}_{\mu\sigma,n\widetilde{\mathbf{k}}'}\hat{c}_{\mu'\sigma',m'\widetilde{\mathbf{k}}'}\right)\;.
\end{split}
\end{equation}
%
Since the inter-site interaction $V(\textbf{q})\!=\!e^2/(2\epsilon\epsilon_0\Omega_M\sqrt{q^2+\kappa ^2})$, the intervalley section of the intersite Coulomb interaction is much weaker than intravalley section as discussed above. However, in the atomic on-site Hubbard interaction, the characteristic interaction strength is independent of wavevector $\mathbf{q}$, thus both intervalley and intravalley sections of the on-site Hubbard interaction have to be taken into account. In particular, the intravalley section of on-site Hubbard interaction with Hartree-Fock approximation in the band basis can be expressed as:
%
\begin{equation}
\begin{split}
H_{\rm{on-site}}^{\rm{intra}}=&\frac{U_0a^2}{L_s^2N_s}\sum _{\widetilde{\mathbf{k}} \widetilde{\mathbf{k}}'}\sum _{\substack{\mu\mu'\\ \sigma\sigma'}}\sum_{\substack{nm\\ n'm'}}\left(\sum _{\mathbf{Q}} \sum _{\alpha\alpha'\mathbf{G}\mathbf{G}'}C^*_{\mu\uparrow\alpha\mathbf{G}+\mathbf{Q},n\widetilde{\mathbf{k}}}C^*_{\mu'\downarrow\alpha'\mathbf{G}'-\mathbf{Q},n'\widetilde{\mathbf{k}}'}C_{\mu'\downarrow\alpha'\mathbf{G}',m'\widetilde{\mathbf{k}}'}C_{\mu\uparrow\alpha\mathbf{G},m\widetilde{\mathbf{k}}} \right)\\
&\times \left(\langle \hat{c}^{\dagger}_{\mu\uparrow,n\widetilde{\mathbf{k}}}\hat{c}_{\mu\uparrow,m\widetilde{\mathbf{k}}}\rangle \hat{c}^{\dagger}_{\mu'\downarrow,n'\widetilde{\mathbf{k}}'}\hat{c}_{\mu'\downarrow,m'\widetilde{\mathbf{k}}'} + \langle \hat{c}^{\dagger}_{\mu'\downarrow,n'\widetilde{\mathbf{k}}'}\hat{c}_{\mu'\downarrow,m'\widetilde{\mathbf{k}}'}\rangle \hat{c}^{\dagger}_{\mu\uparrow,n\widetilde{\mathbf{k}}}\hat{c}_{\mu\uparrow,m\widetilde{\mathbf{k}}}\right)
\end{split}
\label{eq:hubbard1}
\end{equation}
%
and the intervalley section of the on-site interaction with Hartree-Fock approximation in the band basis is expressed as 
\begin{equation}
\begin{split}
H_{\rm{on-site}}^{\rm{inter}}=&\frac{U_0a^2}{L_s^2N_s}\sum _{\widetilde{\mathbf{k}} \widetilde{\mathbf{k}}'}\sum _{\substack{\mu\\ \sigma\sigma'}}\sum_{\substack{nm\\ n'm'}}\left(\sum _{\mathbf{Q}} \sum _{\alpha\alpha'\mathbf{G}\mathbf{G}'}C^*_{\mu\uparrow\alpha\mathbf{G}+\mathbf{Q},n\widetilde{\mathbf{k}}}C^*_{-\mu\downarrow\alpha'\mathbf{G}'-\mathbf{Q},n'\widetilde{\mathbf{k}}'}C_{\mu\downarrow\alpha'\mathbf{G}',m'\widetilde{\mathbf{k}}'}C_{-\mu\uparrow\alpha\mathbf{G},m\widetilde{\mathbf{k}}} \right)\\
&\times \left(\langle \hat{c}^{\dagger}_{\mu\uparrow,n\widetilde{\mathbf{k}}}\hat{c}_{-\mu\uparrow,m\widetilde{\mathbf{k}}}\rangle \hat{c}^{\dagger}_{-\mu\downarrow,n'\widetilde{\mathbf{k}}'}\hat{c}_{\mu\downarrow,m'\widetilde{\mathbf{k}}'} + \langle \hat{c}^{\dagger}_{-\mu\downarrow,n'\widetilde{\mathbf{k}}'}\hat{c}_{\mu\downarrow,m'\widetilde{\mathbf{k}}'}\rangle \hat{c}^{\dagger}_{\mu\uparrow,n\widetilde{\mathbf{k}}}\hat{c}_{-\mu\uparrow,m\widetilde{\mathbf{k}}}\right)
\end{split}
\label{eq:hubbard2}
\end{equation}
%
Since the full interacting Hamiltonian of twisted graphene system preserves spin rotational symmetry ($SU(2)$ symmetry), one can fix the spin quantization axis (the spin ``$z$" axis), then the spin-flip density matrix $\langle \hat{c}_{\uparrow}^{\dagger} \hat{c}_{\downarrow}$ vanishes.  Therefore, the Fock term vanishes for the Hubbard interactions as shown in Eq.~(\ref{eq:hubbard1}) and Eq.~(\ref{eq:hubbard2}).

We continue to discuss the symmetries of the full interacting Hamiltonian of twisted multilayer graphene systems.  First, all the twisted graphene systems have time-reversal ($\mathcal{T}$) symmetry and $C_{3z}$ symmetry. Some of the twisted multilayer graphene systems may have additional crystalline symmetries. For example, in free-standing TBG, there is $C_{2z}$ symmetry; in $AB$-$AB$ stacked TDBG there is $C_{2x}$ symmetry. These additional crystalline symmetries are important in the determining the topological properties of the energy bands. For example, when $C_{2z}$ symmetry is present in TBG, the $C_{2z}$ combined with $\mathcal{T}$ symmetry ($C_{2z}\mathcal{T}$) enforces that the Berry curvature has to vanish at every $\widetilde{\mathbf{k}}$ point in the moir\'e BZ for each valley; and the $C_{2x}$ symmetry of $AB$-$AB$ stacked TDBG enforces that the total valley Chern number of the two flat bands have to be zero \cite{koshino-tdbg-prb19,jpliu-prx19}.  

Moreover, if one only considers the non-interacting continuum model Eq.~(\ref{eq:HMN}) and the intravalley inter-site Coulomb interaction Eq.~(\ref{eq:h-intra}), then all the twisted graphene systems have $U(1)\times U_v(1)\times SU(2)\times SU(2)$ symmetry which can be interpreted as follows. The Bloch states of graphene around  $K$ and $K'$ valleys can  be separately folded into the moir\'e BZ.  For small twist angles the separation between the low-energy states around the $K$ and $K'$ valleys $\sim\!\vert\mathbf{K}-\mathbf{K}' \vert=4\pi/(3a)$ is much greater than the size of the moir\'e reciprocal lattice vector $\vert\mathbf{g}_1\vert=4\pi/(\sqrt{3}L_s)$. Therefore the coupling between the low-energy states around the two valleys can be neglected, because the Fourier components of the corresponding moir\'e potential  is vanishingly small given that the potential is smooth on the moir\'e length scale and that $L_s\!\gg\!a$.  As a result, the  charge is separately conserved for each valley, and the low-energy states of the system has an emergent valley $U(1)$ symmetry \cite{po-prx18} (dubbed as $U_v(1)$). Moreover, as spin-orbit coupling is negligible in graphene and at the non-interacting level the two valleys are approximately decoupled for small twist angles, approximately there is separate spin $SU(2)$ symmetry for each valley. Therefore, all the moir\'e graphene systems have the approximate continuous $U(1)\times U_v(1)\times SU(2)\times SU(2)$ symmetry \cite{po-prx18}, where $U(1)$ stands for global charge conservation symmetry. Such a symmetry can be re-written as $U(2)\otimes U(2)$ symmetry, which means that there is separate charge-conservation and spin rotational symmetry for each valley. Such $U(2)\times U(2)$ symmetry is preserved if one only includes the intravalley inter-site Coulomb interaction Eq.~(\ref{eq:h-intra}).  If one includes the atomic Hubbard interaction, the two valleys are coupled by the intervalley component of the Hubbard interaction (Eq.~(\ref{eq:hubbard2})), which reduces the $U(2)\times U(2)$ symmetry to a global $U(2)$ symmetry, i.e., there is only global charge conservation and global spin rotational symmetry if atomic Hubbard interaction is included.

First let us neglect the atomic Hubbard interactions ,then a generic twisted multilayer graphene system \textit{at least} has $U(2)\times U(2)$, $\mathcal{T}$, and $C_{3z}$ symmetry. The generators of the $U(2)\times U(2)$ symmetry are $\{\tau^{0,z}\, s^{a}, a=0,x,y,z\}$, the $C_{3z}$ operator  can be represented as $C_{3z}=e^{-i(2\pi/3)\tau_z\sigma_z}$, and the time-reversal operator $\mathcal{T}=\tau_x\mathcal{K}$, where  $\tau^{i}$, $s^{j}$, and $\sigma^{k}$ denote Pauli matrices in the valley, spin, and sublattice spaces respectively. $\mathcal{K}$ denotes complex conjugation operation.
Under a symmetry operation $g$, the electron annihilation operator $\hat{c}_{\mathbf{k},\mu\sigma\alpha}$ is transformed to:
\begin{equation}
    g\,\hat{c}_{\mathbf{k},\mu\sigma\alpha}\,g^{-1}=\sum_{\mu'\sigma'\alpha'}\,(\hat{O}_g)_{\mu'\sigma'\alpha',\mu\sigma\alpha}\,\hat{c}_{g^{-1}\mathbf{k},\mu'\sigma'\alpha'}\;,
    \label{eq:c-symmetry}
\end{equation}
Then one can easily verify that the intravalley Coulomb interaction Eq.~(\ref{eq:h-intra}) is invariant under $U(2)\otimes U(2)$, $\mathcal{T}$ and $C_{3z}$ symmetries.
Including the atomic Hubbard interactions would break the $U(2)\times U(2)$ symmetry to a global $U(2)$ symmetry, i.e., there is only total charge conservation global spin rotational symmetry.

In our calculations we project the interaction Hamiltonian given onto a few energy bands around the charge-neutrality point (CNP), assuming that interaction effects are negligible for energy bands at higher energies. In particular, we have performed Hartree-Fock calculations with interactions projected onto 2, 4, and 6 low-energy bands around the CNP for each valley each spin. The results are qualitatively consistent for all the twisted graphene systems.

\vspace{12pt}
\begin{center}
\textbf{\large \V\ Symmetry analysis on the order parameters}
\end{center}

In this section we discuss how the order parameters defined in the valley, spin, and sublattice space are transformed under symmetry operations.
First we define the order parameter $\hat{O}^{a b c}=\tau^{a}\, s^{b} \,\mathbb{I}_{N_l\times N_l}\, \sigma^{c}$ in the valley-spin-sublattice space, where $\mathbf{\tau}$, $\mathbf{s}$,  and $\mathbf{\sigma}$ represent the Pauli matrices in the valley, spin, and sublattice spaces, with $a, b, c=0,x,y,z$ (the zeroth component of the Pauli matrix is a $2\times 2$ identity matrix). $\mathbb{I}_{N_l\times N_l}$ is an identity matrix of dimension $N_l=(M+N)$ defined in the layer space ($N_l=M+N$ is the number of layers). Here we only consider the order parameters in the valley, spin, and sublattice space, and take the average over the layer degrees of freedom.
The expectation value of the order parameter $\hat{O}^{a, b, c}$ at moir\'e wavevector $\kt$ can be expressed as
\begin{equation}
\Delta_{a b c}(\kt)=\sum _n \left\langle \Psi _{n\widetilde{\mathbf{k}}}\vert \hat{O}^{a b c} \vert \Psi _{n\widetilde{\mathbf{k}}} \right\rangle\theta(\varepsilon _F-E_{n\widetilde{\mathbf{k}}})\;,
\label{eq:Delta}
\end{equation}
where $\theta(\varepsilon _F-E _{n\widetilde{\mathbf{k}}})$ refers to Fermi-Dirac distribution at zero temperature and 
\begin{equation}
   \vert \Psi _{n\widetilde{\mathbf{k}}} \rangle_{\rm{HF}} = \sum _{\mu \sigma l \alpha \mathbf{G}}\,C_{\mu  \sigma l \alpha \mathbf{G},n\widetilde{\mathbf{k}}}^{\rm{HF}}\,\vert \mu \sigma l\alpha,\mathbf{G}+\widetilde{\mathbf{k}}\rangle 
\end{equation}
is the Bloch eigenstate of a Hartree-Fock Hamiltonian expressed in the original basis of the continuum model, and $\mu$, $\sigma$, $l$, and $\alpha$ refer to the valley, spin, layer, and sublattice indices.
The basis function $\langle \mathbf{r} \vert \mu s l \sigma,\mathbf{G}+\widetilde{\mathbf{k}}\rangle = \chi _{\mu sl\sigma,\kt+\G}(\mathbf{r})\,e^{i(\widetilde{\mathbf{k}}+\mathbf{G})\cdot\mathbf{r}}$ transforms as follows under symmetry operation $g$:
\begin{align}
\chi _{\mu\sigma l\alpha,\kt+\G}(\mathbf{r})e^{i(\widetilde{\mathbf{k}}+\mathbf{G})\cdot\mathbf{r}} \longrightarrow \begin{cases}\sum _{\mu'\sigma' l'\alpha'}\,(O_{\mathit{g}})_{\mu'\sigma' l'\alpha',\mu \sigma l \alpha}\,\chi _{\mu'\sigma' l' \alpha',g^{-1}(\kt+\G)}(\mathbf{r})\,e^{i\mathit{g}^{-1}(\widetilde{\mathbf{k}}+\mathbf{G})\cdot \mathbf{r}}\; & \hbox{if no $\mathcal{T}$ operation involved}\nn
\sum _{\mu'\sigma' l'\alpha'}\,(O_{\mathit{g}})_{\mu'\sigma' l'\alpha',\mu \sigma l\alpha}\,\chi^{*} _{\mu'\sigma' l'\alpha', -g^{-1}(\kt+\G)}(\mathbf{r})\,e^{-i g^{-1}(\widetilde{\mathbf{k}}+\mathbf{G})\cdot \mathbf{r}}\; & \hbox{if $\mathcal{T}$ operation involved}
\end{cases}
\end{align}
%
Therefore, under symmetry operation $g$, the matrix element of the order parameter $\hat{O}^{a b c}$ in the original basis of the continuum model $\langle \mu'\sigma' l'\alpha',\mathbf{G}+\widetilde{\mathbf{k}} \vert\, \hat{O}^{a b c}\, \vert\mu\sigma l\alpha,\mathbf{G}+\widetilde{\mathbf{k}}\rangle$  becomes
\begin{equation}
\langle \mu'\sigma'l'\alpha',\mathbf{G}+\widetilde{\mathbf{k}} \vert \hat{O}^{a b c} \vert \mu\sigma l \alpha,\mathbf{G}+\widetilde{\mathbf{k}}\rangle \longrightarrow (\,O^{\dagger}_g(g^{-1}\widetilde{\mathbf{k}},g^{-1}\mathbf{G})\,\hat{O}^{a b c}\, O_g(g^{-1}\widetilde{\mathbf{k}}, g^{-1}\mathbf{G})\,)_{\mu'\sigma' l'\sigma',\mu\sigma l\sigma}
\end{equation}
where ``$(g^{-1}\kt,g^{-1}\mathbf{G})$" means that the symmetry operation transforms $\kt$ and $\mathbf{G}$ to  $g^{-1}\kt$ and $g^{-1}\mathbf{G}$.

Thus the expectation value $\Delta _{a b c}$ under $\mathit{g}$ operation becomes
\begin{equation}
\Delta _{a b c d}(\widetilde{\mathbf{k}}) \to \begin{cases}
&\Delta ^{\mathit{g}}_{a b c}(\mathit{g}^{-1}\widetilde{\mathbf{k}})=\sum _n\langle\Psi_{n\mathit{g}^{-1}\widetilde{\mathbf{k}}}\vert\, (\,O^{\dagger}_{\mathit{g}}\,\hat{O}^{a b c}\,O_{\mathit{g}})\,\vert \Psi_{n\mathit{g}^{-1}\widetilde{\mathbf{k}}}\rangle\,\theta(\varepsilon _F-E_{n\widetilde{\mathbf{k}}})\;,\hbox{if $\mathcal{T}$ operation is not involved in $g$}\;,\nn
&\Delta ^{\mathit{g}}_{a b c d}(-\mathit{g}^{-1}\widetilde{\mathbf{k}})=\sum _n\,\langle\Psi_{n-\mathit{g}^{-1}\widetilde{\mathbf{k}}}\vert\, (\,O^{\dagger}_{\mathit{g}}\,(\hat{O}^{a b c})^*\, O_{\mathit{g}}\,)\,\vert \Psi_{n-\mathit{g}^{-1}\widetilde{\mathbf{k}}}\rangle\,\theta(\varepsilon _F-E_{n\widetilde{\mathbf{k}}})\;, \hbox{if $\mathcal{T}$ operation is involved in $g$}\;.
\end{cases}
\end{equation}
If $g$ is a symmetry, it is required that $\Delta _{abc}(\widetilde{\mathbf{k}}) = \Delta^{g}_{a b c}(g^{-1}\widetilde{\mathbf{k}})$ or $\Delta _{abc}(\widetilde{\mathbf{k}}) = \Delta^{g}_{a b c}(-\mathit{g}^{-1}\widetilde{\mathbf{k}})$.


The order parameters $\sigma^x$, $\tau^z\sigma^y$, $\sigma_y$ and $\tau_z\sigma_x$  transform as follows under $C_{3z}$ operation:
\begin{eqnarray}
\sigma_x &\longrightarrow &\cos(2\theta)\sigma_x - \sin(2\theta)\tau_z\sigma_y \;\nn
\tau_z\sigma_y &\longrightarrow &\sin(2\theta)\sigma_x+\cos(2\theta)\tau_z\sigma_y \;\nn
\sigma_y &\longrightarrow &\cos(2\theta)\sigma_y + \sin(2\theta)\tau_z\sigma_x \;\nn
\tau_z\sigma_x &\longrightarrow &-\sin(2\theta)\sigma_y+\cos(2\theta)\tau_z\sigma_x
\end{eqnarray}
Here $\theta = 2\pi/3$. Thus if the system preserves $C_{3z}$ symmetry, these order parameters should satisfy the following relationship:
\begin{eqnarray}
\Delta _{0x}(\widetilde{\mathbf{k}})=\cos(2\theta)\Delta _{0x}(C^{-1}_3\widetilde{\mathbf{k}})+\sin(2\theta)\Delta _{zy}(C^{-1}_3\widetilde{\mathbf{k}})\;\nn
\Delta _{zy}(\widetilde{\mathbf{k}})=-\sin(2\theta)\Delta _{0x}(C^{-1}_3\widetilde{\mathbf{k}})+\cos(2\theta)\Delta _{zy}(C^{-1}_3\widetilde{\mathbf{k}})\;\nn
\Delta _{0y}(\widetilde{\mathbf{k}})=\cos(2\theta)\Delta _{0y}(C^{-1}_3\widetilde{\mathbf{k}})-\sin(2\theta)\Delta _{zx}(C^{-1}_3\widetilde{\mathbf{k}})\;\nn
\Delta _{zx}(\widetilde{\mathbf{k}})=\sin(2\theta)\Delta _{0y}(C^{-1}_3\widetilde{\mathbf{k}})+\cos(2\theta)\Delta _{zx}(C^{-1}_3\widetilde{\mathbf{k}})\;\nn
\label{eq:order-C3z}
\end{eqnarray} 

\begin{figure}[htbp]
\includegraphics[width=3.5in]{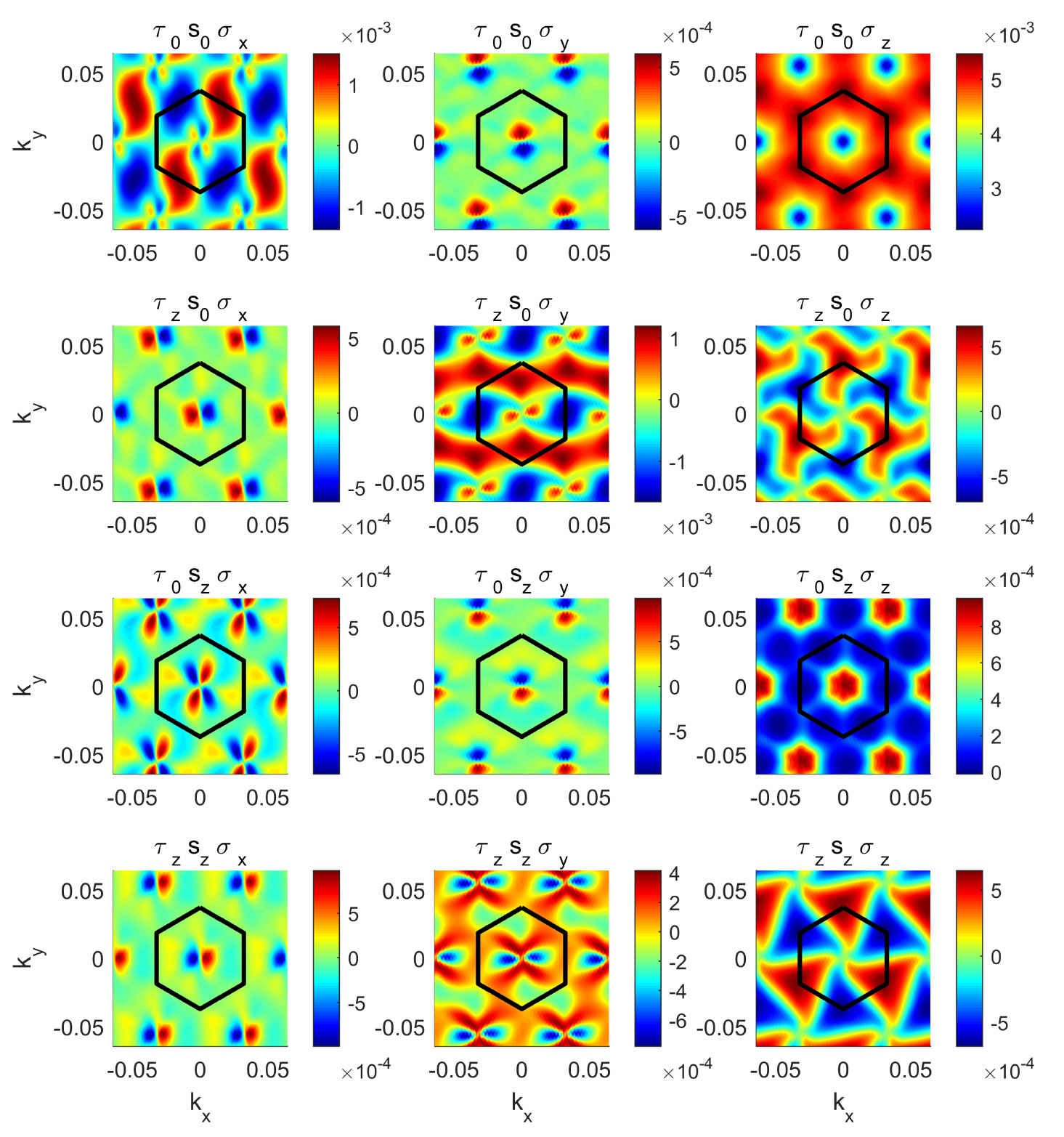}
\caption{~\label{figss1} The reciprocal-space distributions of main order parameters at half filling under --0.4\,V/nm field ($U_d=0.0536$\,eV) with $\epsilon$ = 9.6 and $\kappa$ = 0.005\,\AA{}$^{-1}$. The reciprocal-space coordinates are limited to -0.0644$\sim$ 0.0644\,\AA{}$^{-1}$.}
\end{figure}

In Fig.~\ref{figss1} and Fig.~\ref{figss3} we present the distribution of the dominant order parameters in moir\'e Brillouin zone at 1/2 filling of TBMG under $U_d=0.0536\,$eV and $U_d=-0.0536\,$eV respectively ($\theta\!=\!1.25\,^{\circ}$). The dominant order parameters in the spin polarized, nematic insulator states involve the $\tau^{0,z}s^{0,z}\sigma^{x,y}$ orders, which spontaneously break $C_{3z}$ symmetry and spin rotational symmetry. It is interesting to note that the $\tau^{z}\sigma^x$ and $\sigma^{y}$ order break time-reversal symmetry $\mathcal{T}=\tau^{x}\mathcal{K}$, but preserve a ``Kramers" time-reversal symmetry introduced in Ref.~\onlinecite{zaletel-tbg-hf-prx20} $\widetilde{\mathcal{T}}=\tau^{z}\mathcal{T}$, which is the combination of valley $U(1)$ symmetry operation and the physical time-reversal operation. On the other hand, the operators $\tau^{z}\sigma^{y}$ and $\sigma^{x}$ preserve the physical time-reversal symmetry. Either the $\mathcal{T}$ or $\widetilde{\mathcal{T}}$ symmetry would guarantee that the Chern number  and the net orbital magnetization of the system vanish.

\begin{figure}[htbp]
\includegraphics[width=3.5in]{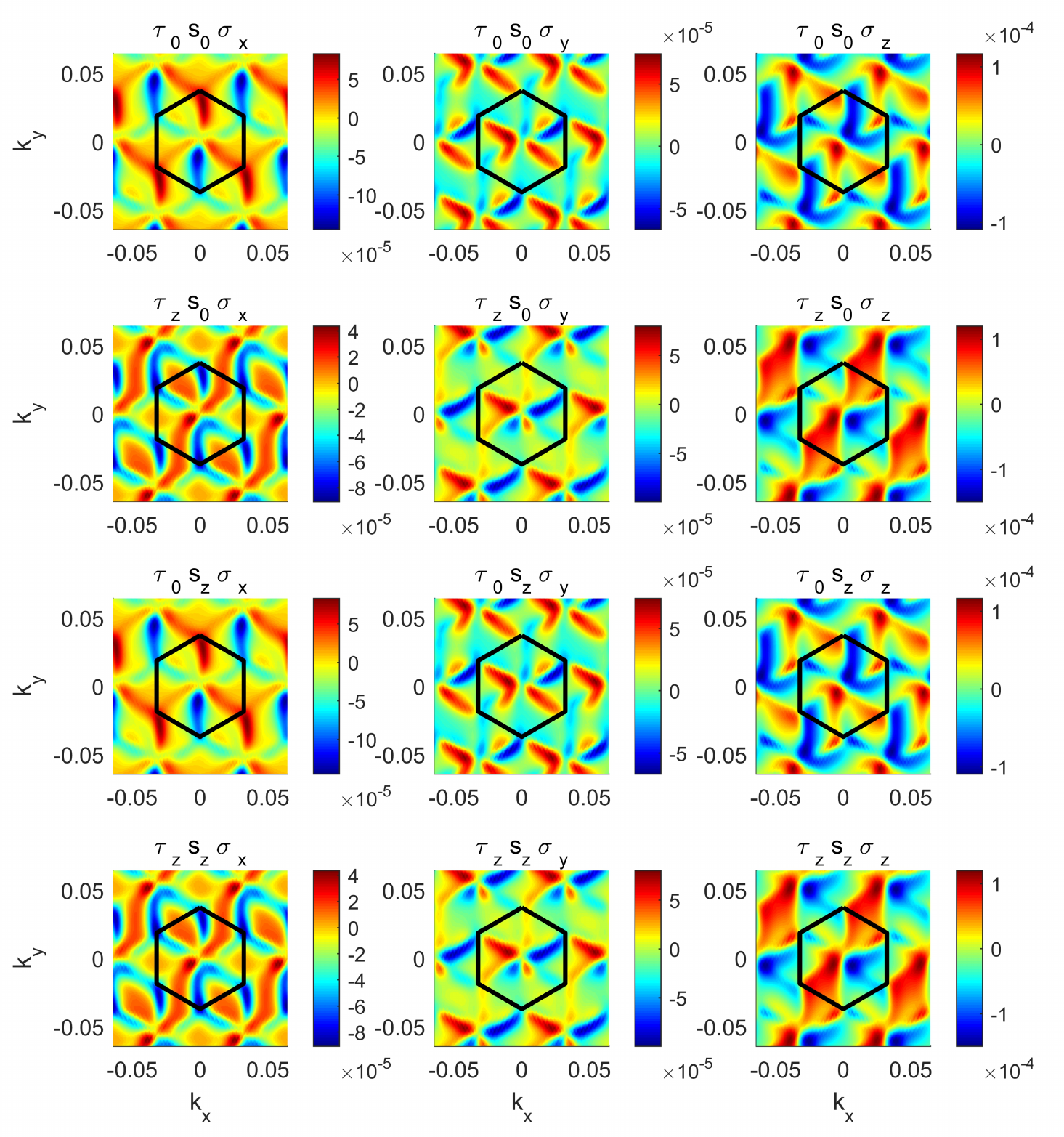}
\caption{~\label{figss2} The reciprocal-space distributions of main order parameter symmetrical deviations ($\Delta _{sym}-\Delta$) at half filling under --0.4\,V/nm field ($U_d=0.0536$\,eV) with $\epsilon$ = 9.6 and $\kappa$ = 0.005\,\AA{}$^{-1}$. The reciprocal-space coordinates are limited to -0.0644$\sim$ 0.0644\,\AA{}$^{-1}$.}
\end{figure}

In  Fig.~\ref{figss2} and Fig.~\ref{figss4} we present the differences between the actual order parameters obtained from self-consistent Hartree-Fock calculations (denoted by $\Delta$) from the ``symmetrized" order parameters that are enforced to obey $C_{3z}$ symmetry (denoted by $\Delta_{symm}$) for $U_d=\pm 0.0536\,$eV at 1/2 filling. For ($\sigma_x$, $\tau_z\sigma_y$) and ($\sigma_y$, $\tau_z\sigma_x$) order parameters, the conditions that these order parameters obey $C_{3z}$ symmetry are given by Eqs.~(\ref{eq:order-C3z}). ($\sigma_x$, $\tau_z\sigma_y$) and  ($\sigma_y$, $\tau_z\sigma_x$) form two pairs of two dimensional irreducible representations of $C_{3z}$ operation.  
From Fig.~\ref{figss2} and Fig.~\ref{figss4} we can see the deviation from $C_{3z}$ symmetry can be as large as 0.1\,meV at some $\kt$ points in the moir\'e Brillouin zone; while  from Fig.~\ref{figss1} and Fig.~\ref{figss3} we see that the maximal magnitudes of the dominant order parameters $\sim 0.5\rm{-}1$\,meV at half filling under both positive and negative displacement fields. Thus we conclude that the order parameters at 1/2 filling of TBMG under finite  displacement fields at $\theta=1.25\,^{\circ}$ strongly break $C_{3z}$ symmetry, which give rise to the nematic insulator state as discussed in main text.

\begin{figure}[htbp]
\includegraphics[width=3.5in]{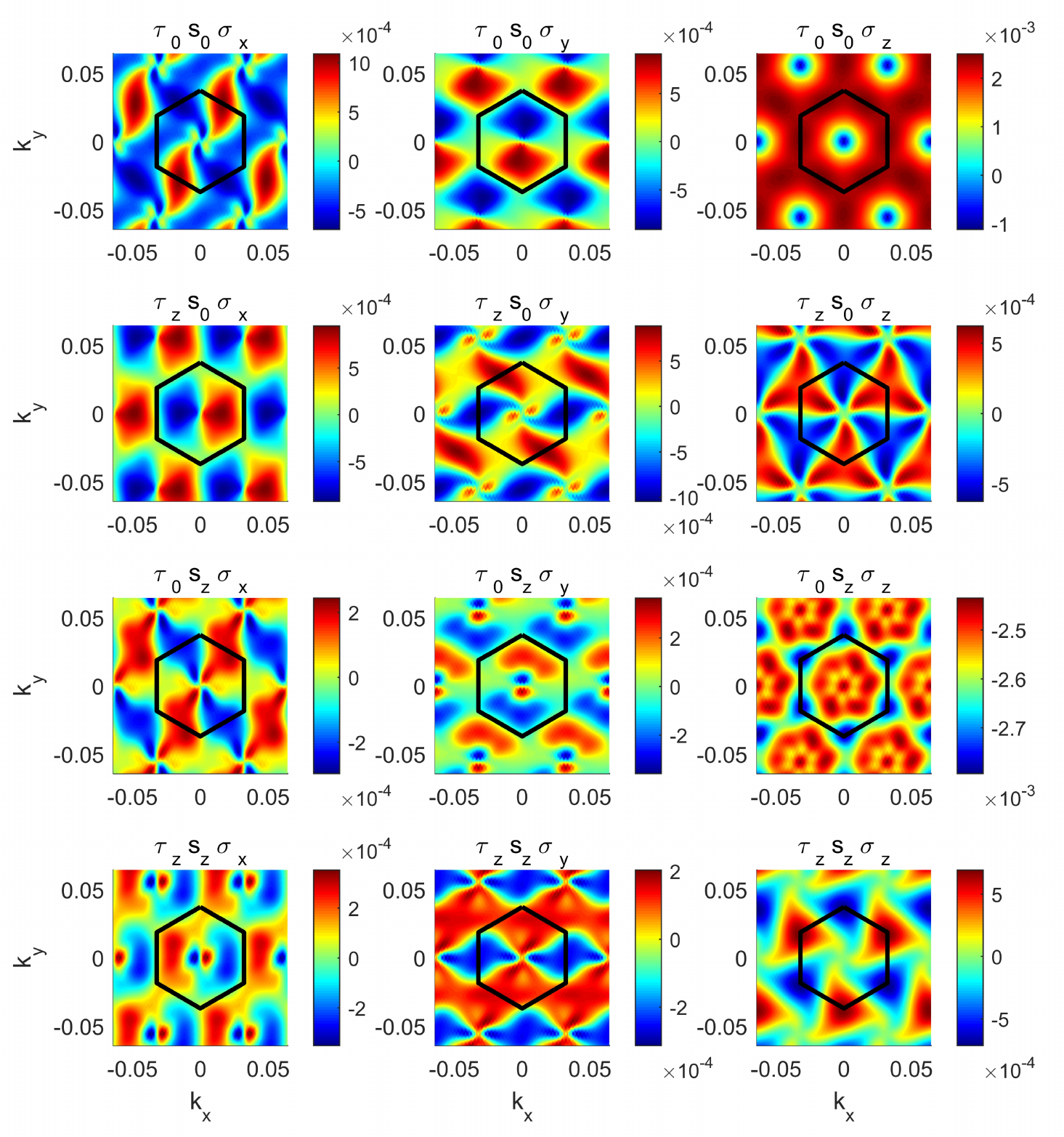}
\caption{~\label{figss3} The reciprocal-space distributions of main order parameters at half filling under 0.4\,V/nm field ($U_d=-0.0536$\,eV) with $\epsilon$ = 9.6 and $\kappa$ = 0.005\,\AA{}$^{-1}$. The reciprocal-space coordinates are limited to -0.0644$\sim$ 0.0644\,\AA{}$^{-1}$.}
\end{figure}

\begin{figure}[htbp]
\includegraphics[width=3.5in]{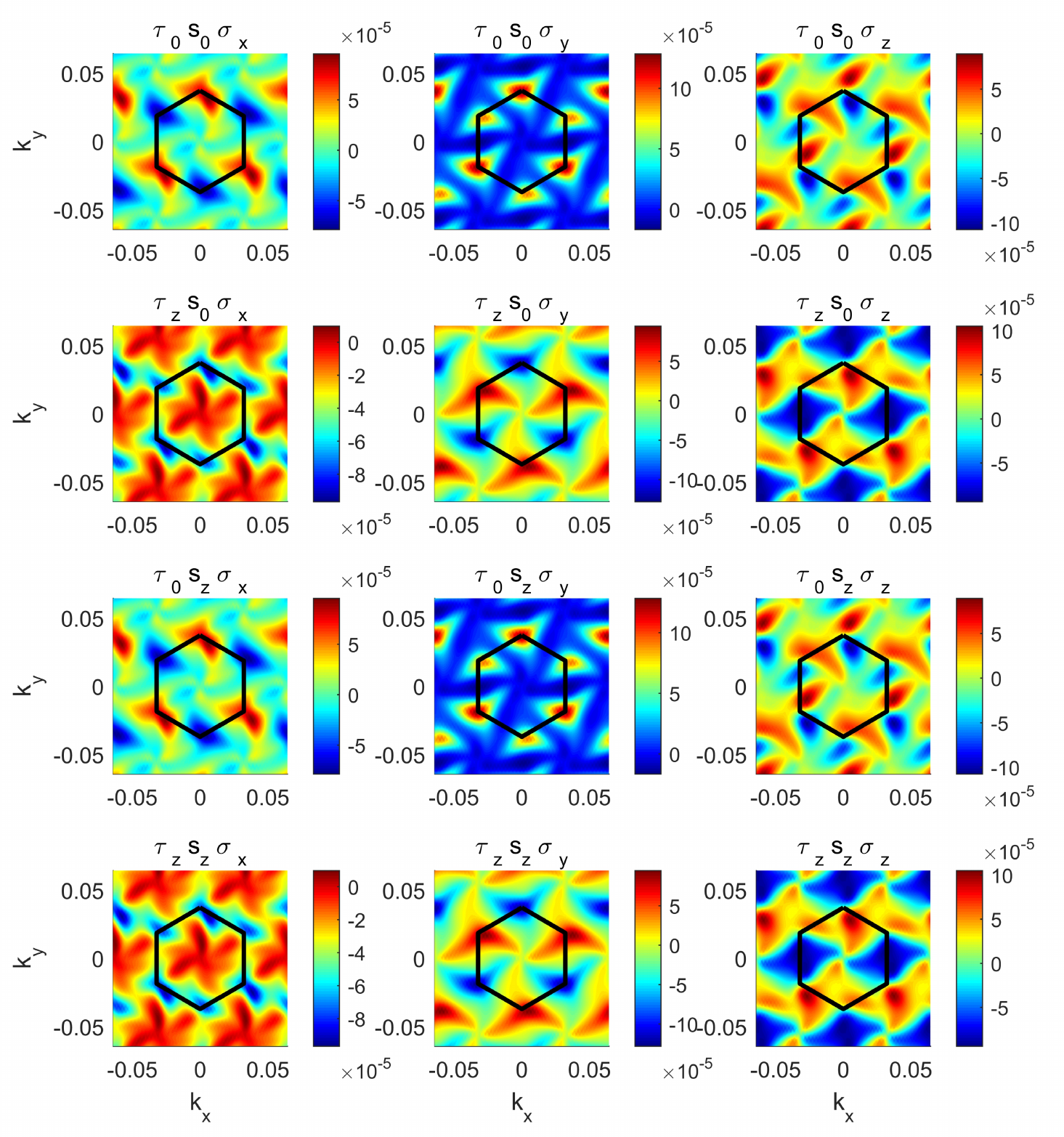}
\caption{~\label{figss4} The reciprocal-space distributions of main order parameter symmetrical deviations ($\Delta _{sym}-\Delta$) at half filling under 0.4\,V/nm field ($U_d=-0.0536$\,eV) with $\epsilon$ = 9.6 and $\kappa$ = 0.005\,\AA{}$^{-1}$. The reciprocal-space coordinates are limited to -0.0644$\sim$ 0.0644\,\AA{}$^{-1}$.}
\end{figure}

\vspace{12pt}
\begin{center}
\textbf{\large \VI\ Competition between spin-polarized and valley-polarized states}
\end{center}
\vspace{12pt}
\begin{center}
\textbf{\large VIA. Quasi-degeneracy between spin-polarized and valley-polarized states}
\end{center}
\label{sec:form}
We first discuss the degeneracy between spin-polarized and valley-polarized states in the twisted multilayer graphene systems under finite displacement fields. First, we define 
\begin{equation}
\lambda _{\mu \sigma m,\mu \sigma n}(\kt,\qt,\Q)=\sum _{\alpha \G}C^*_{\mu \alpha \G+\Q,n\kt + \qt}C_{\mu \alpha \G, m\kt}
\label{eq:lambda}
\end{equation}
where $C_{\mu \alpha \G, m\kt}$ is the non-interacting wavefunction (see Eq.~(17)), and $\mu$, $\sigma$, and $\alpha$ refer to the valley, spin, and layer/sublattice  degrees of freedom. Then the form factor can be re-written as
$\Omega ^{\mu \sigma, \mu '\sigma '}_{nm,n'm'}(\kt,\kt ',\qt,\Q)=\lambda _{\mu\sigma m,\mu \sigma n}(\kt,\qt,\Q)\lambda ^*_{\mu ' \sigma 'n',\mu ' \sigma ' m'}(\kt '-\qt,\qt,\Q).$
It should be noted  that the conduction flat band is usually isolated from other bands in the twisted multilayer graphene under finite displacement fields. Therefore, in such case of isolated flat band, the band index can be dropped, and $\lambda _{\mu\sigma,\mu\sigma}(\kt,\qt,\Q)$ becomes a $4\times 4$ diagonal matrix defined in the valley-spin space at this time. Thus the Coulomb interaction projected onto single conduction band becomes
\begin{equation}
\hat{H}_{single-band}=\frac{1}{N_s}\sum _{\kt,\kt',\qt}\sum_{\mu\mu'\sigma\sigma'}\left(\sum_{\Q}V(\qt+\Q)\lambda _{\mu\sigma ,\mu \sigma }(\kt,\qt,\Q)\lambda ^*_{\mu ' \sigma ',\mu ' \sigma '}(\kt '-\qt,\qt,\Q)\right)\hat{c}^{\dagger}_{\mu\sigma,\kt+\qt}\hat{c}^{\dagger}_{\mu ' \sigma ',\kt '-\qt}\hat{c}_{\mu ' \sigma ' ,\kt '}\hat{c}_{\mu \sigma,\kt}
\end{equation}

Under the Hartree-Fock approximation, the interaction energy can be divided into the Hartree energy $E^H$ and the Fock energy $E^F$,
\begin{equation}
E^H=\frac{1}{2N_s}\sum_{\kt \kt '}\sum_{\Q}\sum_{\mu \mu '\sigma \sigma '}V(\Q)\lambda _{\mu \sigma ,\mu\sigma}(\kt ,0,\Q)\lambda^*_{\mu '\sigma ',\mu '\sigma '}(\kt ' ,0,\Q)\Delta_{\mu\sigma,\mu\sigma}(\kt)\Delta_{\mu '\sigma ',\mu '\sigma '}(\kt '),
\end{equation}
\begin{equation}
E^F=-\frac{1}{2N_s}\sum_{\kt \qt}\sum_{\Q}\sum_{\mu \mu '\sigma \sigma '}V(\qt + \Q)\lambda _{\mu \sigma ,\mu\sigma}(\kt ,\qt, \Q)\lambda^*_{\mu '\sigma ',\mu '\sigma '}(\kt ,\qt, \Q)\Delta_{\mu\sigma,\mu '\sigma '}(\kt+\qt)\Delta_{\mu '\sigma ',\mu \sigma}(\kt).
\end{equation}
Here the density operator $\Delta_{\mu\sigma,\mu '\sigma '}(\kt)$ is defined as 
\begin{equation}
\Delta_{\mu\sigma,\mu '\sigma '}(\kt)=\langle \hat{c}^{\dagger}_{\mu\sigma,\kt}\hat{c}_{\mu '\sigma ',\kt} \rangle\;.
\end{equation}
At 1/2 filling and when the system is in an insulator state, the trace of the density operator at every $\kt$ point equals to 2, thus one can decompose the $4\times 4$ matrix $\hat{\Delta}(\kt)$ defined in the valley-spin space as
\begin{equation}
\hat{\Delta}(\kt)=(\mathbbm{1}+\hat{\Sigma}(\kt))/2\;,
\label{eq:sigma}
\end{equation}
where $\mathbbm{1}$ is the $4\times 4$ identity matrix, and $\hat{\Lambda}(\kt)=\tau^{a}\, s^{b}$ ($a, b =0,x,y,z$) is a traceless $4\times 4$ matrix, which can be written as the tensor product of two Pauli matrices $\tau^{a}$ and $s^{b}$ defined in the valley and spin space respectively ($\hat{\Sigma}(\kt)$ cannot be identity).

We first discuss the Hartree energy $E^H$. The Hartree energy $E^H$ can be re-written in the matrix form
\begin{equation}
E^H=\frac{1}{2N_s}\sum _{\Q}V(\Q)\sum _{\kt}\rm{tr}\left[\hat{\lambda}(\kt,0,\Q)\hat{\Delta}(\kt)\right]\sum _{\kt '}\rm{tr}\left[\hat{\lambda} ^{\dagger}(\kt ',0,\Q)\hat{\Delta}(\kt ')\right]\;,
\end{equation}
where ``$\rm{tr}$" means taking the trace in the valley-spin space.
Clearly the dominant Hartree energy is $\Q=0$ term. In this term, the matrix $\hat{\lambda}(\kt,\qt=0,\Q=0)$ becomes the identity matrix in the valley-spin space. So the dominant Hartree energy is a constant for given filling, which is independent of the flavor symmetry breaking.

\begin{figure*}[!htbp]
\includegraphics[width=5in]{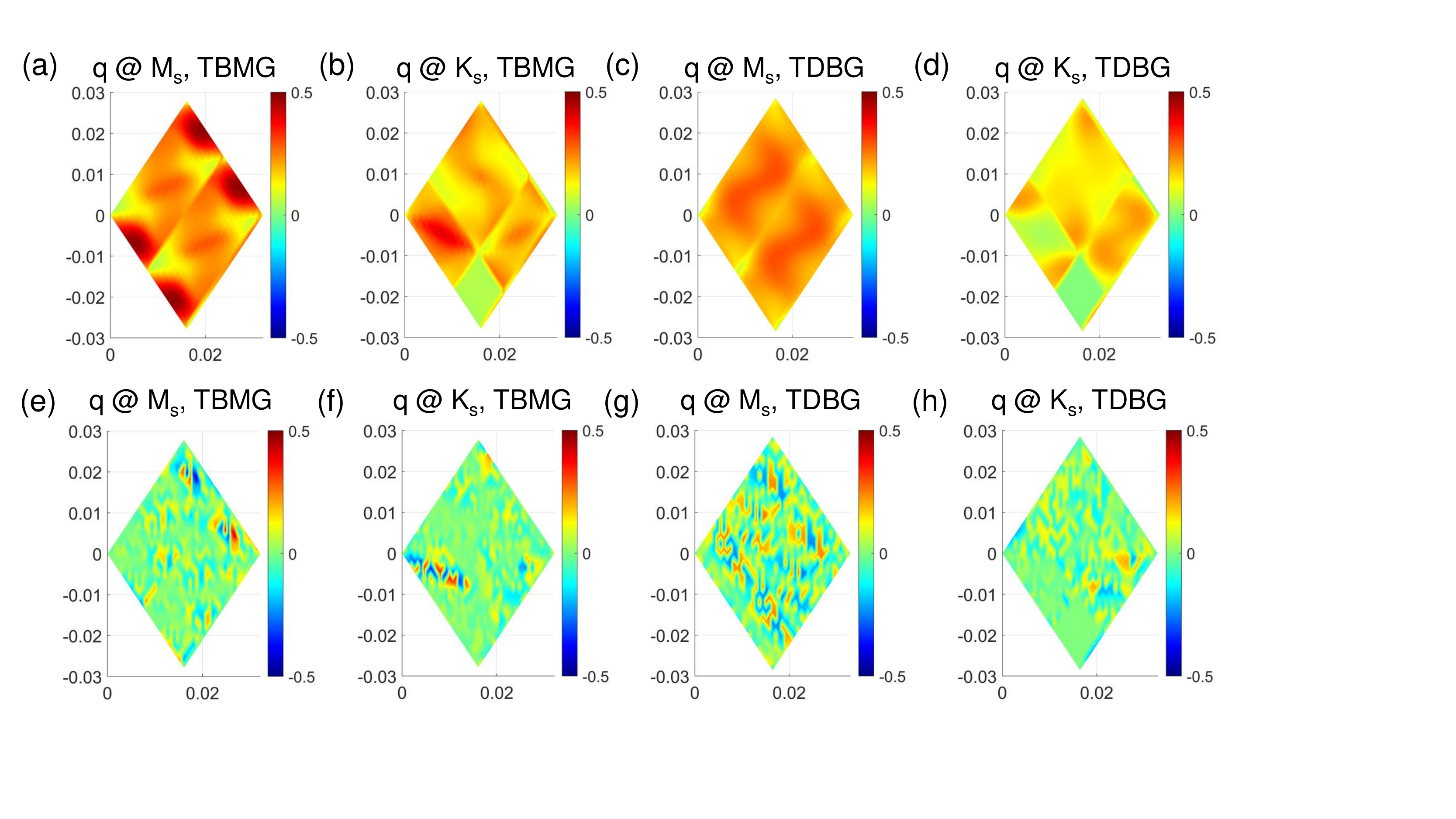}
\caption{~\label{figs-lambda} The reciprocal space distributions of $ |\lambda_0(\kt,\qt,\Q=0)|^2+ |\lambda_z(\kt,\qt,\Q=0)|^2$, (a)-(b) for  twisted bilayer-monolayer graphene (TBMG), and (c)-(d) for twisted double bilayer graphene (TDBG). The reciprocal space distributions of $2$Im$[\lambda_0(\kt,\qt,\Q=0)\lambda^*_z(\kt,\qt,\Q=0)]$, (e)-(f) in  TBMG, and (g)-(h) in TDBG system. In the subfigures (a,c,e,g), $\qt=\G_1/2$,  and in  (b,d,f,h), $\qt=\G_1/3+2\G_2/3$.}
\end{figure*}

As for the Fock energy $E^F$, it can be re-written in the matrix form as follows:
\begin{equation}
E^F=-\frac{1}{2N_s}\sum_{\kt \qt}\sum _{\Q} V(\qt+\Q)\rm{tr}\left[ \hat{\Delta}(\kt)\hat{\lambda}(\kt,\qt,\Q)\hat{\Delta}(\kt+\qt)\hat{\lambda}^{\dagger}(\kt,\qt,\Q) \right]
\end{equation}
The matrix $\hat{\lambda}(\kt,\qt,\Q)$ is explicitly written as 
\begin{equation}
\hat{\lambda}(\kt,\qt,\Q)=\begin{pmatrix}
\lambda _{+,+} & 0 & 0  & 0 \\
 0 & \lambda _{+,+} & 0  & 0 \\
 0 & 0 & \lambda _{-,-}  & 0 \\
 0 & 0  & 0  &  \lambda _{-,-}
\end{pmatrix}
\end{equation}
in which $\mp$ refers to the $K/K'$ valley, $\lambda_{\mu,\mu}$  (with implicit wavevector dependence) is the single-band form factor as defined in Eq.~(\ref{eq:lambda}).  Because  spin-orbit coupling is negligibly weak in graphene, the non-interacting wavefunctions of spin-up and spin-down electrons are identical, thus $\lambda_{\mu,\mu}$ is independent of spin. This matrix $\hat{\lambda}(\kt,\qt,\Q)$ can be further written as 
\begin{equation}
\hat{\lambda}(\kt,\qt,\Q)=\lambda _0(\kt,\qt,\Q)\mathbbm{1} +\lambda _z(\kt,\qt,\Q)\tau _z\;,
\end{equation} 
where
\begin{align}
\lambda _0(\kt,\qt,\Q)&=[\lambda_{+,+}(\kt,\qt,\Q)+\lambda_{-,-}(\kt,\qt,\Q)]/2,\;\nn
\lambda _z(\kt,\qt,\Q)&=[\lambda_{+,+}(\kt,\qt,\Q)-\lambda_{-,-}(\kt,\qt,\Q)]/2.
\end{align} 
With these notations, the Fock energy can be written as
\begin{align}
E^F=& -\frac{1}{2N_s}\sum_{\kt,\qt}\sum_{\Q}V(\qt+\Q)\left[ |\lambda_0(\kt,\qt,\Q)|^2\,\rm{tr}[\hat{\Delta}(\kt)\hat{\Delta}(\kt+\qt)] +\lambda_0(\kt,\qt,\Q)\lambda^*_z(\kt,\qt,\Q)\rm{tr}[\hat{\Delta}(\kt)\hat{\Delta}(\kt+\qt)\tau_z]\right.\;\nn
&+\left.\lambda^*_0(\kt,\qt,\Q)\lambda_z(\kt,\qt,\Q)\,\rm{tr}[\hat{\Delta}(\kt)\tau_z\hat{\Delta}(\kt+\qt)]+ |\lambda_z(\kt,\qt,\Q)|^2\,\rm{tr}[\hat{\Delta}(\kt)\tau_z\hat{\Delta}(\kt+\qt)\tau_z]\right]\;.
\label{fockenergy}
\end{align}

Now we focus on the symmetry-breaking ground states with zero magnetic field. The second and third term in the Eq.~(\ref{fockenergy}) either favors a strongly $\kt$-dependent IVC (intervalley coherent) ordered state with $\hat{\Sigma}(\kt)=\tau_{x,y}$ ($\hat{\Sigma}(\kt)$ defined in Eq.~(\ref{eq:sigma})),  or favors a $\kt$-independent valley polarized state with $\hat{\Sigma}(\kt)=\tau_{z}$. This is because $\rm{tr}[\hat{\Delta}(\kt)\hat{\Delta}(\kt+\qt)\tau_z] \neq 0$ and $\rm{tr}[\hat{\Delta}(\kt)\tau_z\hat{\Delta}(\kt+\qt)] \neq 0$ in two cases: (\i) $\hat{\Sigma}(\kt) \neq \hat{\Sigma}(\kt+\qt) \neq \tau_z$, and (\ii) $\hat{\Sigma}(\kt)=\hat{\Sigma}(\kt+\qt)=\tau_z$.  
However, the form factors in the second and the third terms ($\lambda^*_0(\kt,\qt,\Q)\lambda_z(\kt,\qt,\Q)$  and $\lambda_0(\kt,\qt,\Q)\lambda_z^*(\kt,\qt,\Q)$) are much smaller than those of the first term and fourth terms ($|\lambda_0(\kt,\qt,\Q)|^2$ and $|\lambda_z(\kt,\qt,\Q)|^2$)  in  Eq.\ref{fockenergy}. In Fig.~\ref{figs-lambda}(a)-(b) we present the calculated $|\lambda_0(\kt,\qt,\Q=0)|^2+ |\lambda_z(\kt,\qt,\Q=0)|^2$  for the conduction flat band for twisted bilayer-monolayer graphene at $\theta\!=\!1.25^{\circ}$ and $U_d\!=\!0.0536\,$eV, with  $\qt$ at $M_s$ and $K_s$ points respectively; and in Fig.~\ref{figs-lambda}(e)-(f) we show $2$Im$[\lambda_0(\kt,\qt,\Q=0)\lambda^*_z(\kt,\qt,\Q=0)]$ also for TBMG system with the same choice of parameters. Similarly, in Fig.~\ref{figs-lambda}(c)-(d) we show the calculated $|\lambda_0(\kt,\qt,\Q=0)|^2+ |\lambda_z(\kt,\qt,\Q=0)|^2$ for the isolated conduction flat band of twisted double bilayer graphene with $\theta=1.28^{\circ}$ and $U_d=0.04$\,eV; whereas in Fig.~\ref{figs-lambda}(g)-(h) we present $2$Im$[\lambda_0(\kt,\qt,\Q=0)\lambda^*_z(\kt,\qt,\Q=0)]$ for TDBG with the same parameter choice.  For both systems,
the average value of $ |\lambda_0(\kt,\qt,\Q=0)|^2+ |\lambda_z(\kt,\qt,\Q=0)|^2$ reaches $0.13\!\sim\!0.25$, but the average value of $2$Im$[\lambda_0(\kt,\qt,\Q=0)\lambda^*_z(\kt,\qt,\Q=0)]\sim 10^{-3}$ for $\qt$ at $K_s$ and is zero for $\qt$ at $M_s$. Therefore, we rule out the IVC states as ground-state candidates due to the small form factor Im$[\lambda_0(\kt,\qt,\Q=0)\lambda^*_z(\kt,\qt,\Q=0)]$. We do not rule out the valley polarized state at this moment, since the valley polarized state is favored by the dominant first and fourth terms in Eq.~(\ref{fockenergy}) which we explain below.


Since the form factors in  the first term and fourth terms (($\lambda^*_0(\kt,\qt,\Q)\lambda_z(\kt,\qt,\Q)$  and $\lambda_0(\kt,\qt,\Q)\lambda_z^*(\kt,\qt,\Q)$)) in  Eq.\ref{fockenergy} are dominating, now we only consider these two terms. First, we note that the first term in  Eq.~\ref{fockenergy} favors a $\kt$-independent order parameter, because \rm{tr}$[\hat{\Delta}(\kt)\hat{\Delta}(\kt+\qt)]=2$ (see Eq.~(\ref{eq:sigma})) for all $\kt$ for an insulator state at 1/2 filling of conduction flat band. Thus, for all types of $\kt$-independent flavor ordered states $\hat{\Sigma}(\kt)=\tau^{a} s^{b}$, the term in Eq.~(\ref{fockenergy}) contributes to the same Fock energy $-\sum_{\kt \qt}\sum_\Q V(\qt+\Q)|\lambda_0(\kt,\qt,\Q)|^2/N_s$.

As for the fourth term in the Eq.~\ref{fockenergy}, for convenience, we let $\hat{\Delta}(\kt)=\hat{A}$, $\tau_z\hat{\Delta}(\kt+\qt)\tau_z=\hat{B}$. Then we use the  Cauchy-Schwarz inequality $(\rm{tr}{\hat{A}\hat{B}})^2\leq (\rm{tr} \hat{A})^2 (\rm{tr} \hat{B})^2 $, from which we obtain 
\begin{equation}
\rm{tr}[\hat{\Delta}(\kt)\tau_z\hat{\Delta}(\kt+\qt)\tau_z]\leq\sqrt{(\rm{tr} [\hat{\Delta}(\kt)])^2(\rm{tr} [\tau_z\hat{\Delta}(\kt+\qt)\tau_z])^2}
\end{equation}
the equality condition is satisfied if and only if
\begin{equation}
\hat{\Delta}(\kt)=\tau_z\hat{\Delta}(\kt+\qt)\tau_z.
\end{equation}
We note that similar trick has been used in Ref.~\onlinecite{liu-prr21} in the analysis of ground state at charge neutrality point of TBG.
With this condition, the Fock energy contributed by the fourth term in the Eq.~\ref{fockenergy} reaches maximum magnitude. Therefore, the fourth term favors certain types of $\kt$-independent order parameters that commute with $\tau_z$.  It gives us three order parameters: $\tau_z$ (valley polarized), $s_z$ (spin polarized) and $\tau_zs_z$ (valley spin locking) where we choose the spin-ordering direction to be the ``$z$" direction.  We see that the spin polarized and valley polarized states are degenerate if only considers the dominant component of the Hartree and Fock energies, but the sub-leading terms may lead to slight energy difference.
This explains why we have the quasi-degeneracy between the valley polarized and spin polarized states. Inclusion of atomic Hubbard interactions would split such quasi degeneracy and favors a spin polarized state. Therefore, the ground state at 1/2 filling is spin polarized in twisted multilayer graphene systems under finite displacement fields with isolated conduction flat bands. On the other hand, applying a vertical magnetic field generates nonzero valley polarization thus favors a valley polarized state, which will be explained in detail in the following two subsections.

\begin{figure}[!htbp]
\includegraphics[width=3.5in]{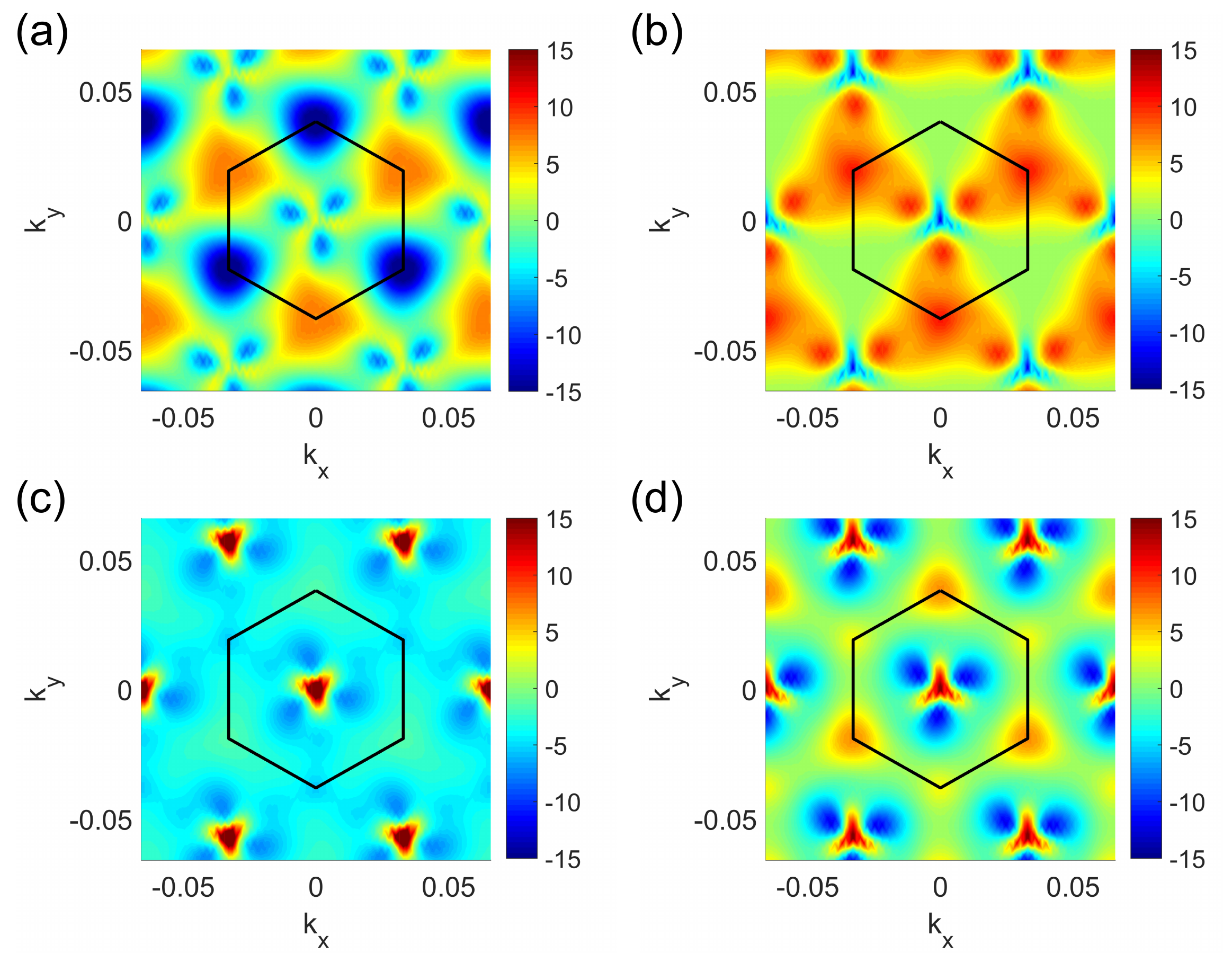}
\caption{~\label{figs9} The orbital g-factors of valence flat band (a) and conduction flat band (b) under $U_d$ = 0.0536\,eV field and those of valence flat band (c) and conduction flat band (d) under $U_d$ = -0.0536\,eV field in  twisted bilayer-monolayer graphene at $\theta=1.25\,^{\circ}$. The reciprocal-space coordinates are limited to -0.032$\sim$ 0.032\,\AA{}$^{-1}$. }
\end{figure}

\begin{figure}[!htbp]
\includegraphics[width=3.5in]{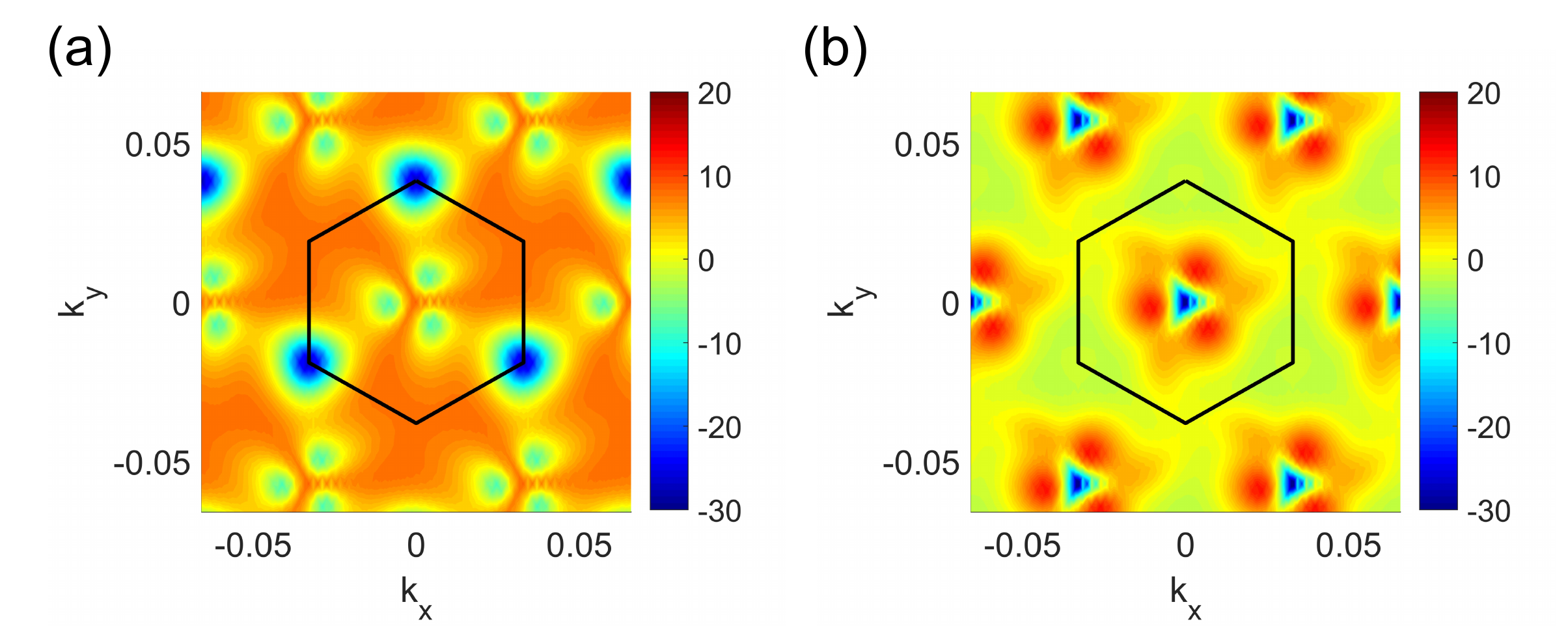}
\caption{~\label{figs10} The orbital g-factors of valence flat band (a) and conduction flat band (b) under $U_d$ = 0.04\,eV field in the twisted double-bilayer graphene with $\theta=1.28\,^{\circ}$ . The reciprocal-space coordinates are limited to -0.032$\sim$ 0.032\,\AA{}$^{-1}$. }
\end{figure}

\vspace{12pt}
\begin{center}
\textbf{\large VIB. Orbital magnetic Zeeman effects in twisted graphene systems}
\label{sec:zeeman}
\end{center}
The effects of magnetic fields can be separated into two parts:  the spin Zeeman effects and the orbital magnetic effects.  The former can be trivially described by the spin Zeeman splitting, $H^{s}_{\rm{Zeeman}}=\mu_B\mathbf{s}\cdot\mathbf{B}$, where $\mathbf{s}$ represents the Pauli matrix in spin space, and $\mathbf{B}$ is the external magnetic field. The orbital magnetic effects deserve careful discussions.
First, the vertical magnetic field tends to recombine the flat bands into a series of recurring Landau levels (LLs), i.e., the  Hofstadter butterfly spectra \cite{hofstadter-prb76},  which are dependent on the number of magnetic fluxes in each moir\'e primitive cell.  Second, aside from the formation of LLs, the magnetic
field also induces splitting between the flat bands from the opposite valleys, because when $C_{2z}$ (and $C_{2z}\mathcal{T}$) symmetry is broken, the states in the two valleys have opposite orbital magnetizations which couple linearly to vertical magnetic fields. Such valley-contrasting orbital magnetizations are giant in twisted graphene systems because the orbital angular momenta of the electrons circulating on the moir\'e length scale are large \cite{jpliu-prx19}. For example, the orbital magnetization contributed by the flat bands of is  on the order of $\pm 10\,\mu_{\textrm{B}}$ per moir\'e cell in TBMG system.     
Therefore, the vertical magnetic field tends to drive the system into a valley polarized, time-reversal broken state even without the necessity of forming LLs. For weak vertical magnetic fields, the magnetic flux for each moir\'e supercell is small in the TBMG, e.g., for $B\!=\!2\,$T, the flux per supercell $\Phi/\Phi_0\!=\!0.053\!\approx\!7/132$ ($\Phi_0\!=\!h/e$), which is far from forming notable Hofstadter bands. Thus we can neglect the effects of LL quantization for $B\!\lessapprox\!2\,$T, and only consider the orbital magnetic Zeeman effect in the subspace of the two flat bands for each valley, which can be conveniently described using the orbital $g$ factor defined in the subspace of the flat bands \cite{song-zeeman-arxiv15,Koshino-orbital-prb11,ashvin-double-bilayer-nc19, wu-tdbg-arxiv20, Sun-zeeman-prb20}, the matrix element of the orbital magnetic $g$ factor in the flat-band basis is expressed as:
the matrix element is expressed as
\begin{align}
\hat{g}^{\mu}_{mm^{\prime}}(\kt)=&\frac{-im_e}{2\hbar ^2}\sum _{l}\left(\frac{1}{E^{\mu}_{m\kt}-E^{\mu}_{l\kt}}+\frac{1}{E^{\mu}_{m'\kt}-E^{\mu}_{l\kt}}\right)\;\nn
&\times (\hat{v}^{x,\mu}_{ml}\hat{v}^{y,\mu}_{lm^{\prime}}-\hat{v}^{y,\mu}_{ml}\hat{v}^{x,\mu}_{lm^{\prime}}),
\label{eq:g-factor}
\end{align}
where $m, m'$ refer to the indices of the flat bands, and $l$ refers to remote band index, $E^{\mu}_{m\kt}$ represents the non-interacting flat-band energy of valley $\mu$ at moir\'e wavevector $\kt$ , and $v^{a,\mu}=\partial H(\k)/(\hbar\partial k_a)$ ($a=x,y$) is the velocity operator for valley $\mu$.  
Then the orbital magnetic Zeeman effects can be described by $(H^{\mu}_{\rm{Zeeman}})_{m m'}(\kt)=\mu_B \hat{g}^{\mu}_{mm^{\prime}}(\kt)\,B_z$, where $B_z$ is the $z$ component of the external magnetic field.   In Fig.~\ref{figs9}(a)-(b) we show the orbital magnetic $g$ factors of the valence flat band and conduction flat band from the $K$ valley in the TBMG at $\theta=1.25\,^{\circ}$ with $U_d=0.0536\,$eV, we see that  the maximal amplitude $\sim 10\rm{-}15$, and they are of opposite signs for the $K'$ valley. Similarly, in Fig.~\ref{figs9}(c)-(d) we show the orbital $g$ factor distributions of the valence flat band and conduction flat band of TBMG with $U_d\!=\!-0.0536\,$eV and $\theta\!=\!1.25\,^{\circ}$, which have maximal amplitudes around $\Gamma_s$ $\sim 15$. In Fig.~\ref{figs10}(a) and (b) we show the orbital $g$ factor distributions of the valence flat band and conduction flat band in TDBG with $U_d=0.04\,$eV and $\theta\!=\!1.28\,^{\circ}$. The orbital $g$ factor in TDBG is as large as -30 around $K_s$ point in the moir\'e for the valence flat band (Fig.~\ref{figs10}(a)), and is maximal around $\Gamma_s$ point ($\sim -35$) for the conduction flat band (Fig.~\ref{figs10}(b)).  
Such giant orbital magnetic $g$ factor corresponds to a significant orbital Zeeman splitting $\sim 1\rm{-}2\,$meV between the flat bands from the $K$ and $K'$ valleys for vertical magnetic field  $B_z=1\,$T, which completely dominates over LL spacing. For example, for TBMG with $U_d=0.0536\,$eV, for the Fermi velocity around 1/2 filling, $\hbar\omega_c\!=\!74\,\mu$\,eV with $B_z=1\,$T, which is much smaller than the orbital magnetic Zeeman splitting.

\vspace{12pt}
\begin{center}
\textbf{\large VIC. Berry-curvature correction to the density of states under vertical magnetic fields}
\end{center}

Magnetic field does not only couples to the electrons through the orbital Zeeman effect, for topological bands with nonzero Chern numbers, magnetic field also changes the density of the Chern bands. To be specific, the change of particle number per cell $\delta n$ is described by the Streda formula $\delta n = B_z \Omega_M C/\Phi_0$, where $\Omega_M$ is the area of the moir\'e primitive cell, $C$ is the Chern number of the occupied bands, and $\Phi_0=h/e$ is the flux quantum. The  change of particle number  in the Chern band induced by magnetic field can be well characterized by introducing a Berry-curvature correction to density of states under vertical magnetic field \cite{xiao-prl05}. Following the semi-classical treatment introduced in Ref.~\onlinecite{xiao-prl05}, when the magnetic field is present, the density matrix  at $\kt$ for each spin species $\rho_{\mu n,\mu m}(\kt)$ ($m$ and $n$ are band indices and $\mu$ is valley index) is multiplied by a factor of $(\mathbbm{1}+eB_z\hat{\Omega}(\kt)/\hbar)$, 
i.e.
\begin{equation}
\hat{\rho}(\kt,B_z)=\hat{\rho}(\kt,0)\cdot (\mathbbm{1}+eB_z\hat{\Omega}(\kt)/\hbar)
\label{eq:rho-B}
\end{equation}
where $\mathbbm{1}$ is the identity matrix defined in the valley-band space, and $\hat{\Omega}(\kt)$ is the Berry curvature matrix at $\kt$ defined in the valley-band basis, and $\hat{\rho}(\kt,B_z)$ is the Berry-curvature-corrected density matrix at $\kt$ with magnetic field $B_z$. The ``$\cdot$" symbol means a matrix product. Here we only the density matrices of the valley and/or spin polarized states since the IVC states are energetically unfavored as argued in Sec.~\ref{sec:form}.
From Eq.~(\ref{eq:rho-B}) it follows that the density under magnetic field $B_z$ is
\begin{align}
&\rho(B_z)\;\nn
=&\frac{1}{N_k \Omega_M}\sum_{\kt}\rm{tr}[\hat{\rho}(\kt,0)\cdot(\mathbbm{1}+eB_z\hat{\Omega}(\kt)/\hbar)]\;\nn
=&\rho(0)+ \frac{e}{\hbar}B_z\,\frac{1}{N_k\Omega_M}\sum_{\kt}\rm{tr}[\hat{\rho}(\kt,0)\hat{\Omega}(\kt)]\;\nn
=&\rho(0)+\frac{e B_z}{h} C
\end{align}
where $\rho(0)$ is the density with zero magnetic field, and $C$ is the Chern number of the occupied bands, which is expressed as 
\begin{align}
C&=\frac{1}{2\pi}\int d^2\kt \,\rm{tr}[\hat{\Omega}(\kt) \cdot \hat{\rho}(\kt)]\;\nn
&=\frac{1}{2\pi}\sum_{\kt} \frac{(2\pi)^2}{N_k\Omega _M}\, \rm{tr}[\hat{\Omega}(\kt) \cdot \hat{\rho}(\kt)]\;\nn
&=\frac{2\pi}{\Omega _M N_k}\sum_{\kt} \,\rm{tr}[\hat{\Omega}(\kt) \cdot \hat{\rho}(\kt)].
\end{align}
Therefore, with the Berry-corrected density matrix as given in Eq.~(\ref{eq:rho-B}),  Streda formula $\delta\rho(B_z)=\rho(B_z)-\rho(0)=eB_z C/h$ immediately follows. All the self-consistent Hartree-Fock calculations under finite magnetic fields reported in main text are performed with such a Berry-curvature correction to density operator. Since the Chern numbers of the flat bands for the $K$ and $K'$ valleys are opposite, the field-induced change of densities for the $K$  and $K'$ valleys have opposite sign, thus favors a valley polarized state over a spin polarized state. 
In particular, if the valley polarized state around 1/2 filling has a Chern number of the same sign as the magnetic field, it gains more Fock energy and is more easily to open up a gap and becomes a Chern-insulator state. 
Our calculations indicate that such effects is even more dramatic than the valley polarization induced by the orbital Zeeman splitting. The results presented in Fig. 4 of main text are obtained through self-consistent Hartree-Fock calculations including both orbital magnetic Zeeman effects and the Berry-curvature correction to the density of Chern bands.  Since the charge density would be changed under magnetic fields for states with nonzero Chern numbers, we have performed self-consistent Hartree-Fock calculations at three filling factors around filling 2: (a) $n(C\!=\!0)=2$, (b) $n(C\!=\!2)=2+2\Omega_M e B/h$, and (c) $n(C\!=\!-2)=2-2\Omega_M e B/h$, where $\Omega_M$ is the area of the moir\'e primitive cell. For each magnetic field $B$, we have performed self-consistent Hartree-Fock calculations for both spin polarized and valley polarized for all the three filling factors $n(C\!=\!0$), $n(C\!=\!2)$, and $n(C\!=\!-2)$, and compare the energies of the different symmetry-breaking states for each filling with increasing magnetic field. 



\vspace{12pt}
\begin{center}
\textbf{\large \VII\ More results of Hartree-Fock calculations for TBMG and TDBG systems}
\end{center}
\begin{figure}[htbp]
\includegraphics[width=3.5in]{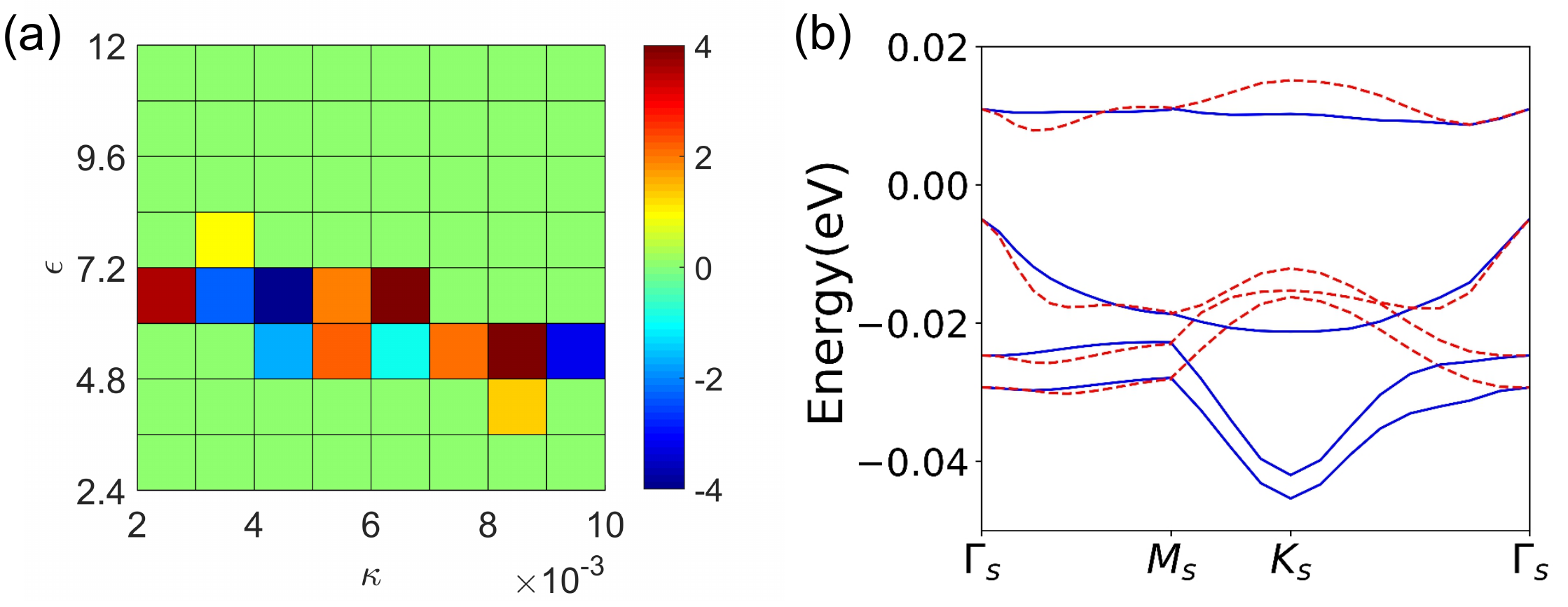}
\caption{~\label{figssHF} The HF energy bands of twisted double bilayer graphene with $U_d\!=\!0.04$\,eV at 1/2 filling ($\epsilon\!=\!9.6$ and $\kappa\!=\!0.005$\,\AA{}$^{-1}$). The blue and red lines represent the energy bands from different valleys.}
\end{figure}

\vspace{12pt}
\begin{center}
\textbf{\large VIIA. Correlated insulators at 1/2 filling}
\end{center}
In this section we present more results for the Hartree-Fock calculations of TBMG and TDBG systems. First we present the Hartree-Fock phase diagram of TDBG at 1/2 filling with $U_d\!=\!0.04\,$eV and $\theta\!=\!1.28\,^{\circ}$  in Fig.~\ref{figssHF}(a), where $\epsilon$ is the background dielectric constant, and $\kappa$ is the inverse screening length. The color coding indicates the Chern number of the Hartree-Fock ground states. We see that in most of the parameter space the system stays in a correlated insulator state with Chern number zero. A more detailed analysis reveals that these  states are spin polarized, $C_{3z}$-breaking, and quantum valley Hall insulator states with valley Chern numbers $\pm 2$. The nonzero valley Chern numbers also result from the nontrivial band topology as shown in Fig.~\ref{figstdbg}(a)-(b): the Chern numbers ($C$) for the valence flat band and the conduction flat band from the $K^{\prime}$ valley are $+2$ and $-2$ for respectively for $U_d\!=\!0.04\,$eV and $\theta\!=\!1.28\,^{\circ}$. As a result, in a spin polarized state at 1/2 filling with six of the eight flat bands being occupied, there would be one $C=-2$ ($C=2$) conduction flat band from the $K'$ ($K$) valley being unoccupied, giving rise to nonzero valley Chern numbers $\pm 2$ of the occupied bands. Such spin polarized, nematic, and quantum valley Hall insulator states are also 
characterized by the order parameters $\tau^{0,z}s^{0,z}\sigma^{x,y}$, which are similar to those found at 1/2 filling of TBMG. The energy bands of the Hartree-Fock ground state at 1/2 filling of TDBG under $U_d\!=\!0.04$\,eV at $\theta\!=\!1.28\,^{\circ}$ are shown in Fig.~\ref{figssHF}(b), with the dielectric constant $\epsilon=9.6$, and the inverse screening length $\kappa=0.005\,\angstrom^{-1}$. We see that the band gaps are on the order of 10\,meV.  

\begin{figure}[!htbp]
\includegraphics[width=3.5in]{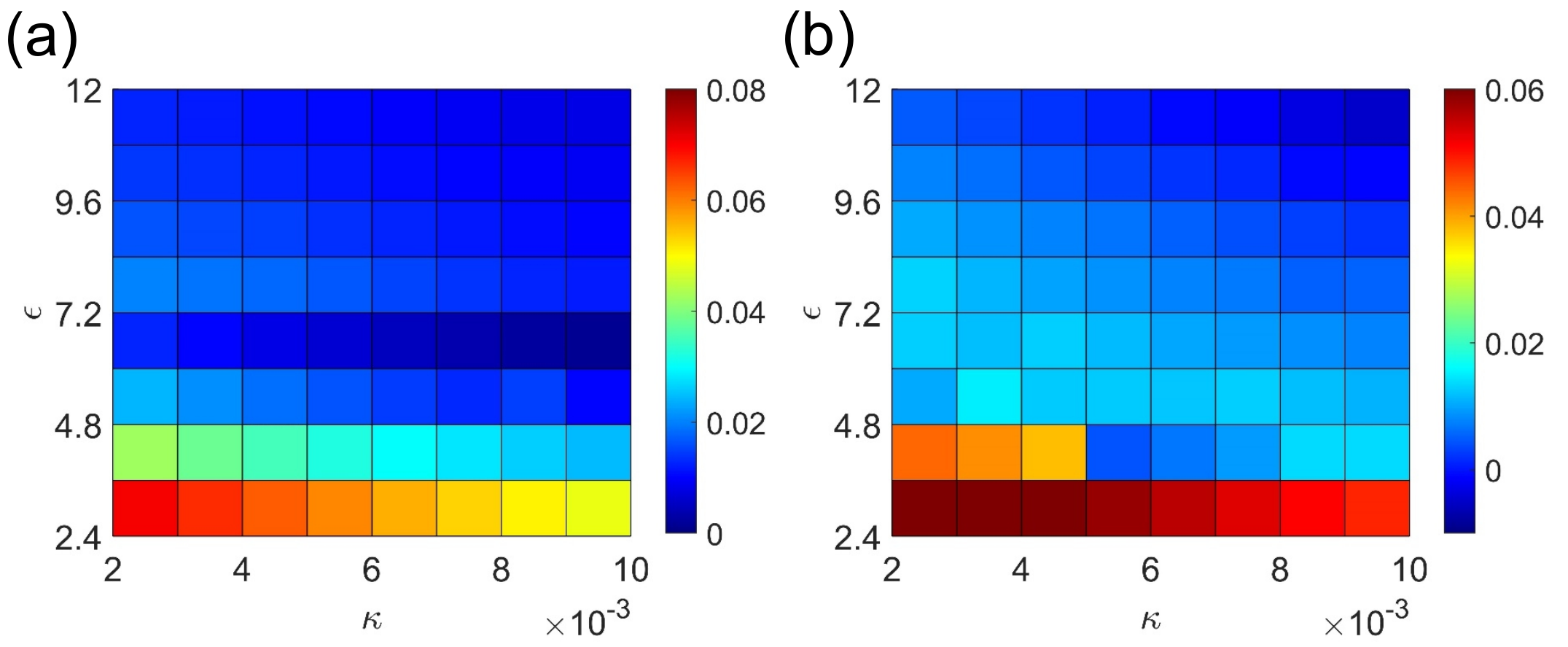}
\caption{~\label{figs2} The indirect gaps of the Hartree-Fock ground states including atomic Hubbard interactions at 1/2 filling of twisted bilayer-monolayer graphene at $\theta=1.25\,^{\circ}$: (a) $U_d=-0.0536\,$eV, and (b) $U_d=0.0536\,$eV.} 
\end{figure}

In Fig.~\ref{figs2}(a)-(b) we show the indirect gaps of the Hartree-Fock ground states in the parameter space of $\epsilon$ and $\kappa$ at 1/2 filling of the TBMG system with $U_d=\mp0.0536\,$eV at $\theta=1.25\,^{\circ}$. This figure should be compared with Fig.~2 of the main text which shows the Chern numbers of the gapped states.  Moreover, we note that the $C\!=\!-1$ phases in Fig.~2(a) and (b) of main text are states with one $C\!=\!2$ band and one $C\!=\!-1$ band being occupied from the majority spin of the $K'$ valley, and two $C\!=\!-2$ bands and two $C\!=\!1$ bands being occupied from both spins of the $K$ valley. Such a state is a valley- and spin-polarized insulator state at 1/2 filling, and the resulted Chern number -1 is inconsistent with the experimental observation with zero Chern number  \cite{young-monobi-nature20}.

\begin{figure}[!htbp]
\includegraphics[width=3.5in]{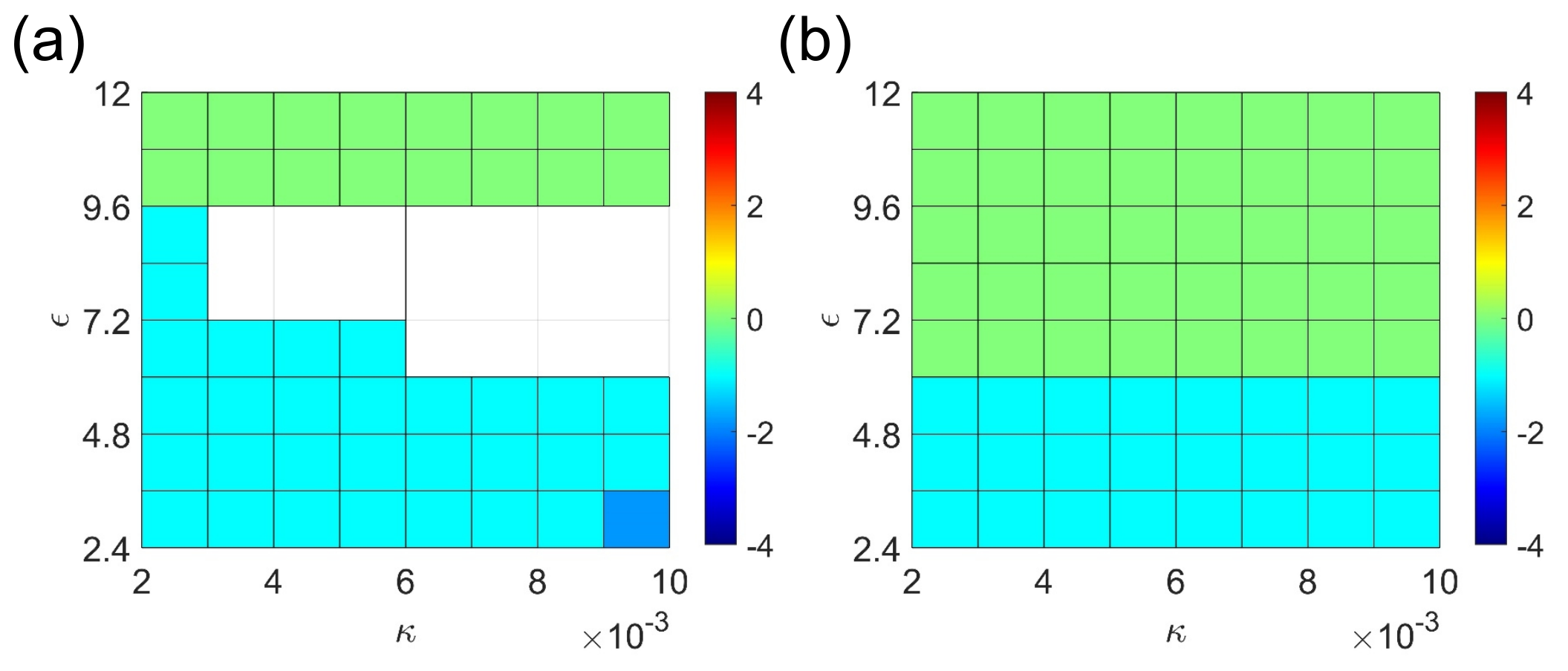}
\caption{~\label{figs3} The Chern numbers and indirect gaps of the Hartree-Fock (HF) ground states at 1/2 filling of TBMG at $\theta=1.25\,^{\circ}$ under different displacement fields: (a) Chern numbers of the HF ground states, $D=0.3\,$V/nm ($U_d=-0.0402\,$eV), and (b) Chern numbers of the HF ground states, $D=0.5\,$V/nm ($U_d=-0.067\,$eV). } 
\end{figure}

In Fig.~\ref{figs3} (a) and (b) we show the Chern numbers of the Hartree-Fock ground states at $U_d\!=\!-0.0402\,$eV ($D\!=\!0.3\,$V/nm) and $U_d\!=\!-0.067\,$eV ($D\!=\!0.5\,$V/nm) at 1/2 filling. The Chern-number-zero states in Fig.~\ref{figs3} are the same state as that explained in the main text: they are spin polarized, nematic states stabilized by atomic Hubbard interactions.  Such zero-Chern-number states are more robust for larger displacement fields  as they occupy larger area in the phase diagram shown in Fig.~\ref{figs3}(b); while such a zero-Chern-number state is less robust for weaker displacement fields, which only survives for relatively weak interactions as shown in Fig.~\ref{figs3}(a). The blank in Fig.~\ref{figs3}(a) indicates that at these points the Hartree-Fock ground states are metallic such that Chern numbers of occupied states are ill defined.

\vspace{12pt}
\begin{center}
\textbf{\large VIIB. Including more active bands into the Hartree-Fock calculations}
\end{center}

\begin{figure}[!htbp]
\includegraphics[width=3.5in]{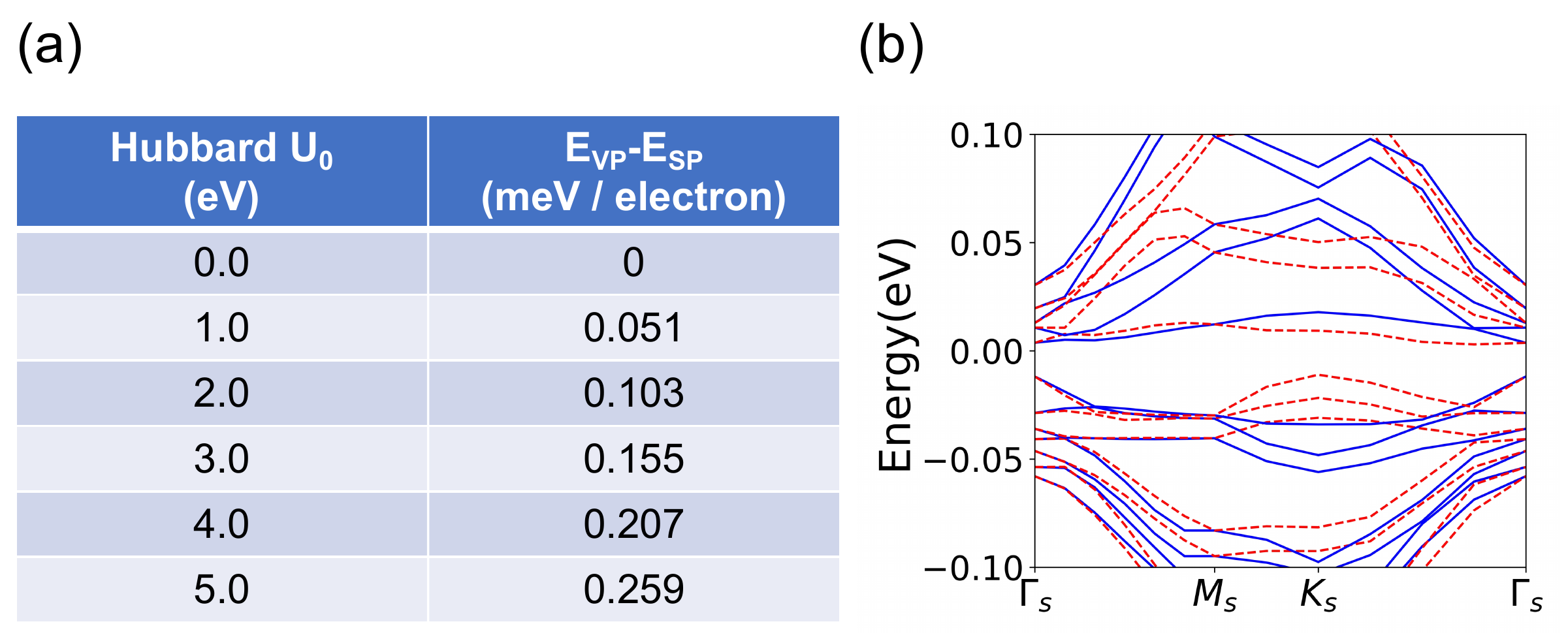}
\caption{~\label{nflat3} (a) The energy difference between VP and SP states at the half filling in the TBMG with six-band-projected interactions. (b) The Hartree-Fock energy bands with $\epsilon$ = 9.6 and $\kappa$ = 0.005\,\AA{}$^{-1}$. } 
\end{figure}

To verity our conclusion, we also project the interactions onto six active bands (including two flat bands and four remote bands) for each valley each spin in the TBMG, and perform Hartree-Fock calculations at 1/2 filling of the flat bands with $U_d\!=\!0.0536\,$eV and $\theta\!=\!1.25\,^{\circ}$. 
We find that the ground state is still the spin polarized, nematic, and quantum valley Hall insulator state, which is consistent with the calculations with interactions only projected to the two flat bands. The energy difference between the VP and SP states at the half filling for such six-band (per spin per valley) Hartree-Fock calculations are shown in the Fig.~\ref{nflat3} (a), from which  we can see that the energy difference becomes larger while the on-site Hubbard interaction $U_0$ is increasing. This is consistent with the results obtained from the calculations with interactions only projected to the two flat bands.

\begin{figure}[!htbp]
\includegraphics[width=3.5in]{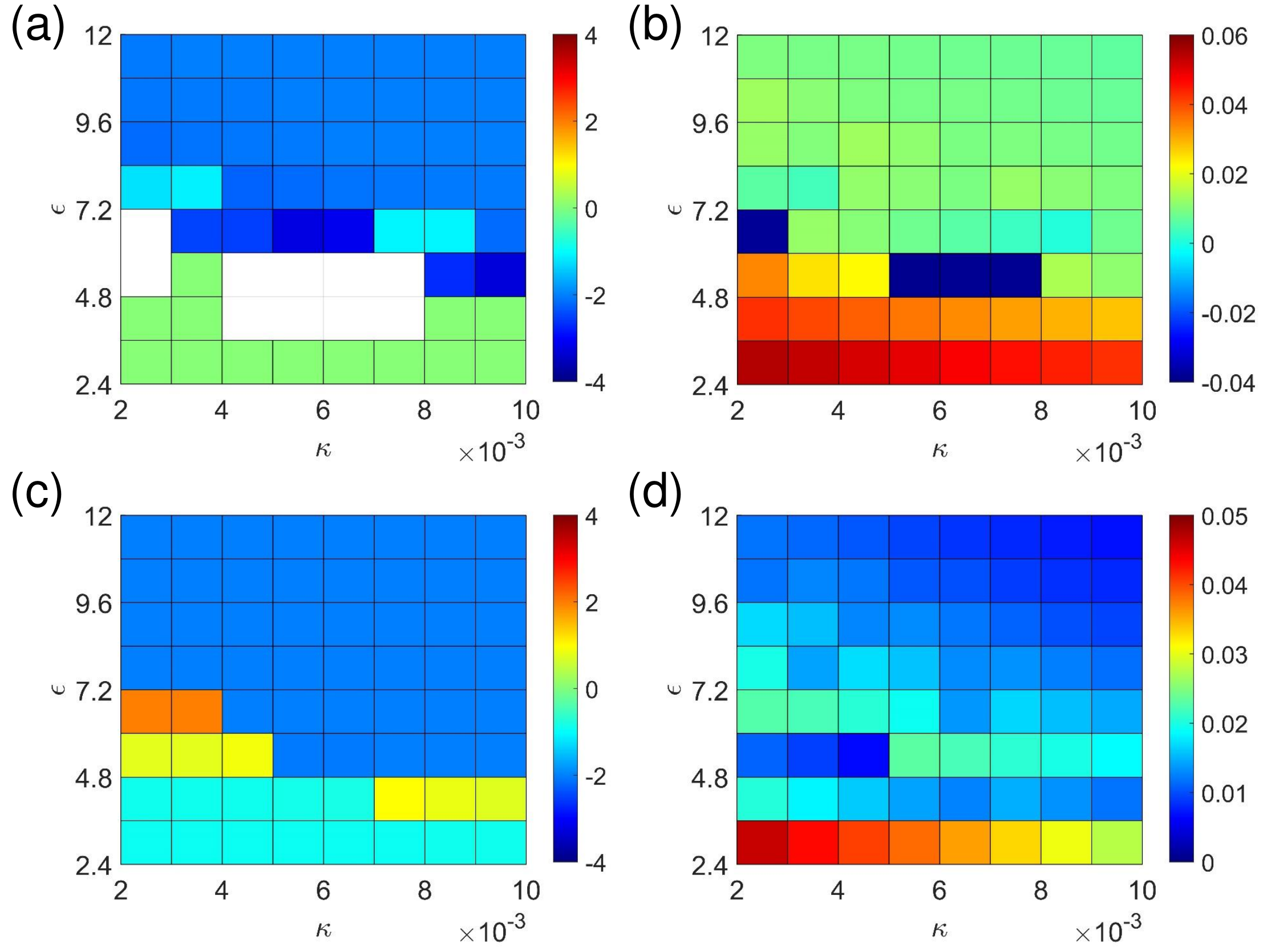}
\caption{~\label{figs6} The Hartree-Fock phase diagram of Chern number (a) and bandgap (b) at 1/4 filling under 0.5\,V/nm field ($U_d = -0.0670$\,eV) for twisted bilayer-monolayer graphene at $\theta=1.25\,^{\circ}$. The ground-state phase diagram of Chern number (c) and bandgap (d) at 3/4 filling under 0.3\,V/nm field ($U_d = -0.0402$\,eV).}
\end{figure}

\vspace{12pt}
\begin{center}
\textbf{\large VIIC. Quantum anomalous Hall states at 1/4 and 3/4 fillings}
\end{center}
Now we discuss the quantum anomalous Hall states at  1/4 filling and 3/4 filling of TBMG. In Fig.~3(a) and (b) of the main text we have presented the Chern numbers of the Hartree-Fock ground states at 1/4 filling of TBMG with $D\!=\!0.5\,$V/nm ($U_d\!=\!-0.067\,$eV) and at 1/4 filling with $D\!=\!0.3\,$V/m ($U_d=-0.0402\,$eV), from which we see that in most regions of the phase diagrams, the system stays in the $\vert C\vert=2$ quantum anomalous Hall states. Here In Fig.~\ref{figs6} (a)-(b) we show the indirect gaps of the Hartree-Fock ground states at 1/4 filling with $D\!=\!0.5\,$V/nm (Fig.~\ref{figs6}(a)) and at 3/4 filling with $D\!=\!0.3\,$V/nm (Fig.~\ref{figs6}(b)).  

It is interesting to note that the valley Chern numbers of the valence flat band and conduction flat band are interchanged under opposite displacement fields (see Fig.~1(c)-(d)) in the main text): when $U_d>0$, for $\theta\gtrapprox 1.05\,^{\circ}$, the Chern numbers of the conduction (valence) flat band from the $K'$ valley is -1 (2); for $U_d<0$, the Chern numbers of the conduction (valence) flat band from the $K'$ valley becomes 2 (-1). Therefore, when $U_d<0$ ($D>0$ in our definition), the system favors a spin and valley polarized state with one  conduction flat band with $\vert C\vert=2$ being occupied (unoccupied) at 1/4 (3/4) filling, which gives rise to the $\vert C\vert=2$ quantum anomalous Hall effects observed in experiments \cite{young-monobi-nature20}.   

\begin{figure}[!htbp]
\includegraphics[width=3.5in]{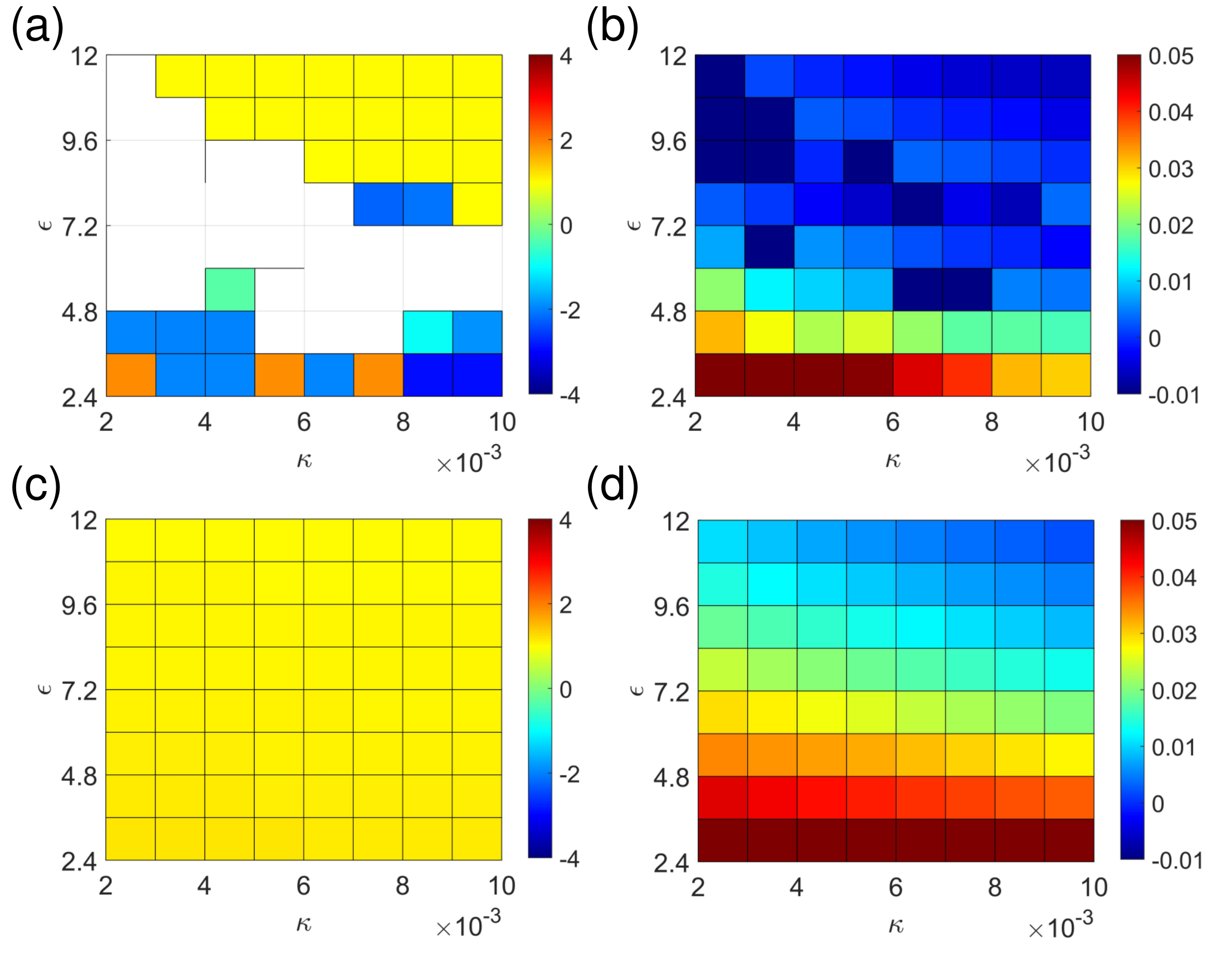}
\caption{~\label{figs8} The ground-state phase diagram of Chern number (a) and bandgap (b) at 1/4 filling under -0.5\,V/nm field ($U_d = 0.0670$\,eV) for twisted bilayer-monolayer graphene at $\theta=1.25\,^{\circ}$. The ground-state phase diagram of Chern number (c) and bandgap (d) at 3/4 filling under -0.3\,V/nm field ($U_d = 0.0402$\,eV).}
\end{figure}

Following the above argument, it is naturally expected that when the displacement field is flipped, i.e., when $U_d>0$ ($D<0$), the ground state at 1/4 (3/4) filling would be a valley and spin polarized state with $\vert C\vert=1$, since the Chern number of the conduction flat band of the $K$ valley has been changed from $2$ to $-1$ due to the flip of displacement field. This has been verified by Hartree-Fock calculations at 1/4 filling with $D=-0.5$\,V/nm and 3/4 filling with $D=-0.3$\,V/nm  as shown by the calculated Chern numbers of the Hartree-Fock ground states in Fig.~\ref{figs8}(a) and (c) respectively. We see tha the $C=1$ quantum anomalous Hall state is extremely robust at 3/4 filling with $D=-0.3\,$V/nm. Basically  the system stays at the $C=1$ QAH state for the entire parameter space we have explored.   The $C=1$ QAH state survives for weaker interactions at 1/4 filling with $D=-0.5\,$V/nm ($U_d=0.0536\,$eV) as shown in Fig.~\ref{figs6}(a).  

\vspace{12pt}
\begin{center}
\textbf{\large \VIII\ Hartree-Fock calculations for hBN-aligned twisted bilayer graphene system}
\end{center}

\begin{figure}[!htbp]
\includegraphics[width=3.5in]{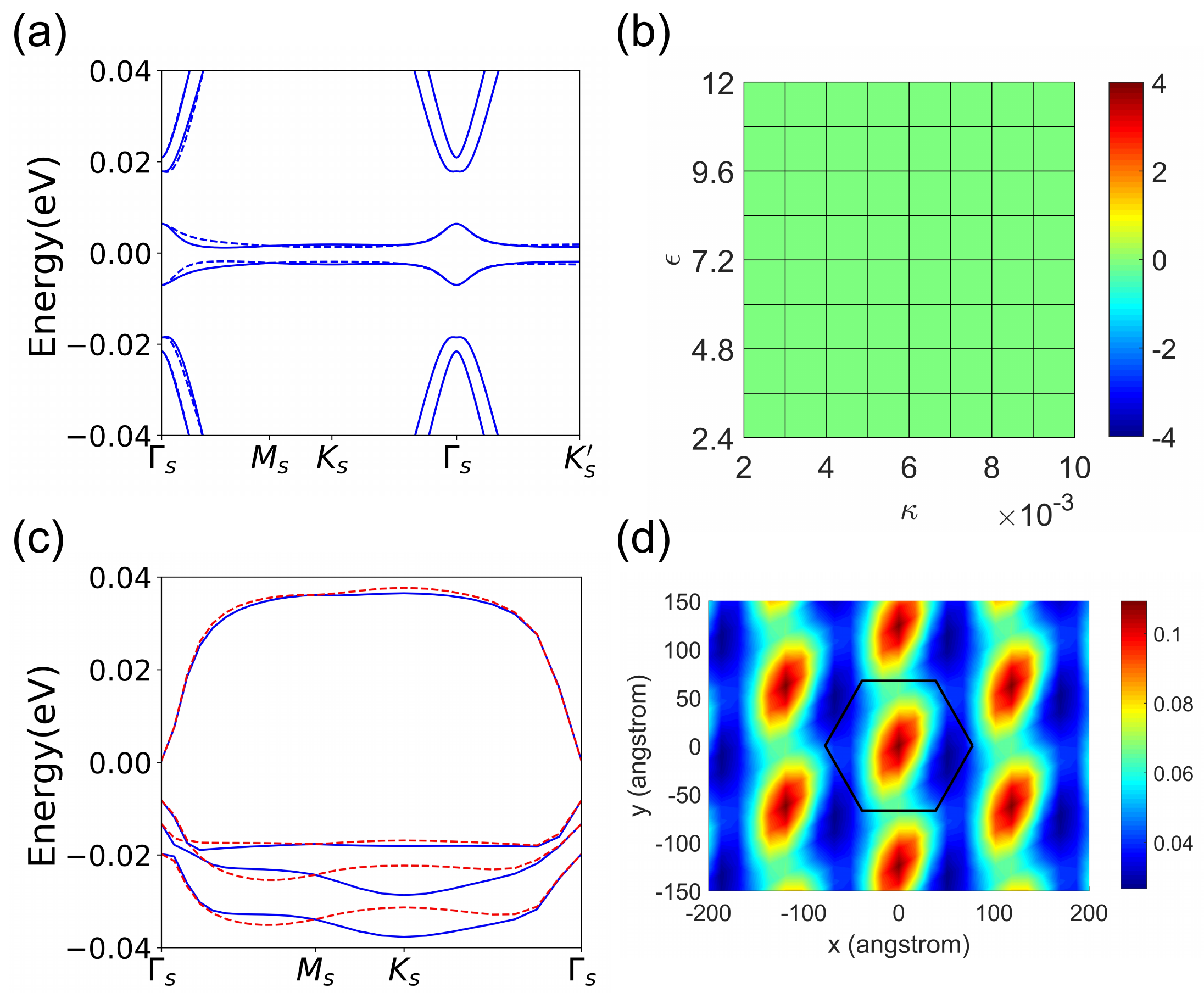}
\caption{~\label{fig-tbg-BN} (a) The non-interacting energy bands of hBN-aligned twisted bilayer graphene. The solid (dashed) lines represent the energy bands of K (K$^{\prime}$) valley. (b) The Hartree-Fock phase diagram of hBN-aligned twisted bilayer graphene at the half filling. (c) The Hartree-Fock energy bands at the half filling in the hBN-aligned twisted bilayer graphene. (d) The nematic charge density of spin-polarized phase at the half filling.}
\end{figure}

Now we consider the hBN-aligned magic-angle twisted bilayer graphene (TBG) system where we assume the aligned hBN substrate introduces a staggered sublattice potential $\approx\!15\,$meV to the bottom graphene layer. The staggered sublattice potential breaks the $C_{2z}$ symmetry, thus opens a gap at the Dirac point.  The non-interacting energy bands of hBN-aligned TBG system are present in the Fig.~\ref{fig-tbg-BN} (a). We can see that a gap $\sim 4\,$meV opens up at the Dirac points.

We present Hartree-Fock phase diagram of this system at  half filling in the Fig.~\ref{fig-tbg-BN} (b). Similar to TBMG and TDBG, there are two degenerate states of VP and SP state at the half filling if we only consider the inter-site Coulomb interaction. However, further calculations including both intersite Coulomb interaction and on-site Hubbard interaction reveal that the SP state becomes the only ground state with zero Chern number as shown in Fig.~\ref{fig-tbg-BN} (b). Here, the atomic Hubbard interactions lowers the energy of the SP state by about 0.6\,meV per electron.  The order parameters of the correlated insulator state at 1/2 filling of hBN-aligned TBG are also $\tau^{0,z}s^{0,z}\sigma^{x,y}$, which are qualitatively the same as those found in TBMG and TDBG. Similarly, the local charge density of flat bands at the half filling as shown in Fig.~\ref{fig-tbg-BN} (d) also exhibits the nematicity. We also show the Hartree-Fock energy bands at the half filling  with $\epsilon$ = 9.6 and $\kappa$ = 0.005\,\AA{}$^{-1}$ in Fig.~\ref{fig-tbg-BN} (c). 

\vspace{12pt}
\begin{center}
\textbf{\large \IX\ Hartree-Fock calculations for other twisted multilayer graphene systems}
\end{center}

\begin{figure}[!htbp]
\includegraphics[width=3.5in]{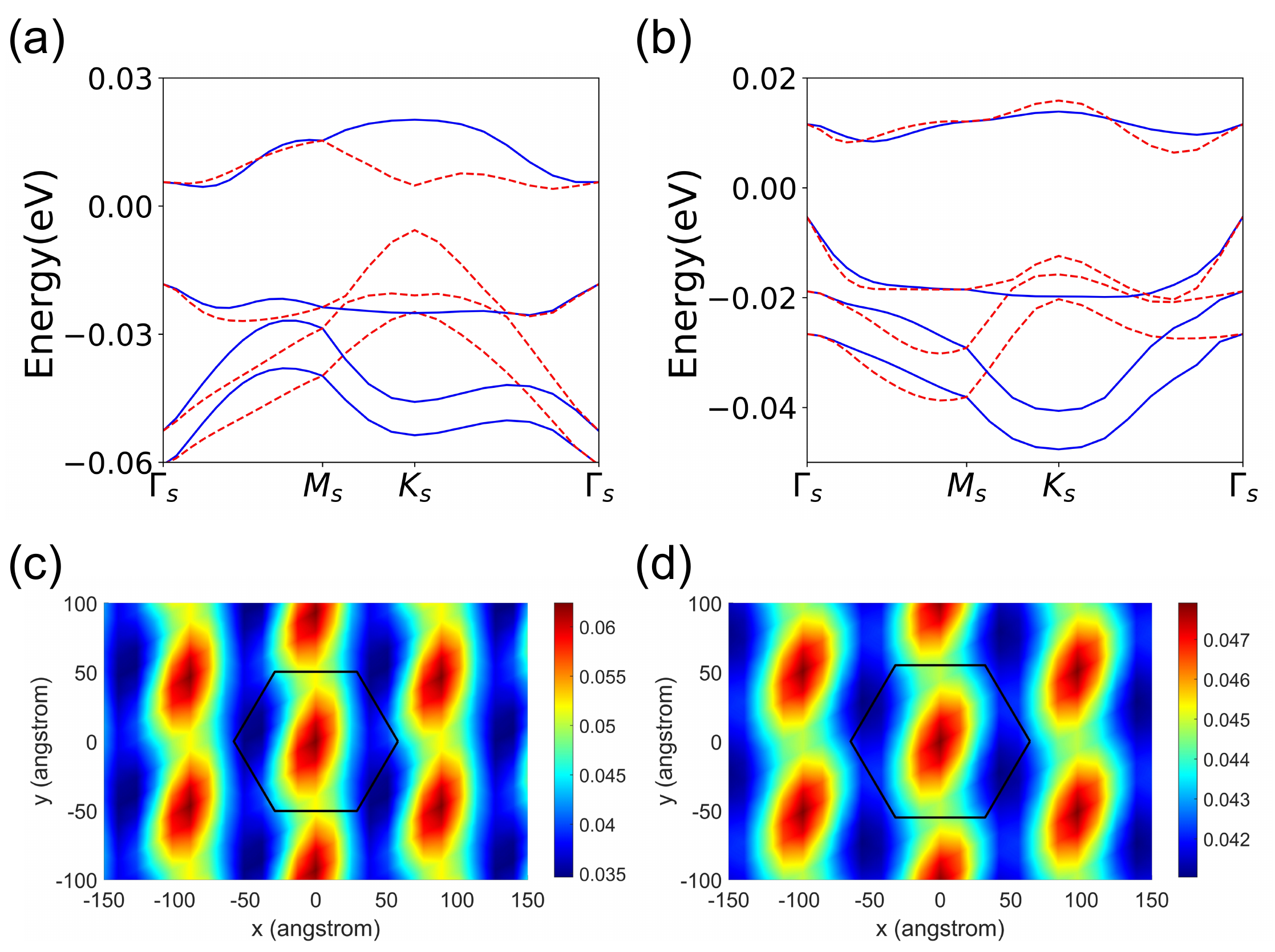}
\caption{~\label{fig3and} The Hartree-Fock energy bands at the half filling: (a) twisted $(3+1)$-layer graphene with $\theta\!=\!1.4^{\circ}$,  and $U_d=-0.06\,$eV, and (b) twisted $(3+2)$-layer graphene with $\theta\!=\!1.28\,^{\circ}$ and $U_d=0.04\,$eV. The local charge densities of twisted $(3+1)$-layer graphene (c) and twisted $(3+2)$-layer graphene (d) are also present.}
\end{figure}

In this section, we present the Hartree-Fock energy bands of other twisted multilayer graphene system, for example, the twisted $(3+1)$-layer graphene with $\theta\!=\!1.4^{\circ}$,  and $U_d=-0.06\,$eV, and twisted $(3+2)$-layer graphene with $\theta\!=\!1.28\,^{\circ}$ and $U_d=0.04\,$eV in Fig.~\ref{fig3and} (a,b). 

The ground states at half filling in these systems are also spin-polarized (SP) states which are stabilized by atomic Hubbard interactions. The inclusion of atomic Hubbard interactions lowers the energy of the SP state by about 0.5\,meV per electron in the twisted $(3+1)$-layer graphene or  0.4\,meV per electron in the twisted $(3+2)$-layer graphene. The order parameters are $\tau^{0,z}s^{0,z}\sigma^{x,y}$, which are same with those at 1/2 filling of the TBMG and TDBG systems. Because the distributions of these order parameters in the moir\'e Brillouin zone break $C_{3z}$ symmetry, the local charge density contributed by the low-energy bands at half filling also exhibits nematicity as shown in Fig.~\ref{fig3and} (d).

\end{document}